\documentclass[12pt]{article}
\pdfoutput=1
\usepackage{float}  
\usepackage{cite}
\usepackage{booktabs}
\usepackage[english]{babel}
\usepackage{amsmath,amssymb,amsbsy,amstext, amsthm, simplewick}
\usepackage{hyperref}
\usepackage{graphicx}
\usepackage{amsfonts}
\usepackage{amssymb}
\usepackage[small]{caption}
\usepackage{upgreek}
\usepackage[svgnames,dvipsnames,x11names,table]{xcolor}
\usepackage{multirow}
\usepackage{geometry}
\usepackage[hang,flushmargin]{footmisc}
\usepackage{bm}
\usepackage{braket}
\usepackage{subcaption}
\usepackage{mathtools}
\usepackage{setspace}
\usepackage{cleveref}
\usepackage{comment}
\usepackage{scalerel}
\usepackage[normalem]{ulem}
\usepackage{slashed}
\usepackage{enumitem}
\usepackage{dsfont}
\usepackage{tikz}
\usetikzlibrary{decorations.markings}
\usetikzlibrary{shapes.misc}

\makeatletter
\g@addto@macro\bfseries{\boldmath}
\makeatother

\hypersetup{
    colorlinks=true,
    linkcolor={red!50!black},
    citecolor={blue!50!black},
    urlcolor={blue!80!black}
}

\usepackage{colortbl}

\makeatletter
\newlength{\apb@width}
\newcommand{\autoparbox}[2][c]{\settowidth{\apb@width}{#2}\parbox[#1]{\apb@width}{#2}}

\makeatother

\definecolor{lightgray}{gray}{0.9}

\usepackage[framemethod=default]{mdframed}
\newmdenv[skipabove=7pt,
skipbelow=7pt,
rightline=false,
leftline=false,
topline=false,
bottomline=false,
backgroundcolor=gray!10,
linecolor=gray,
innerleftmargin=5pt,
innerrightmargin=5pt,
innertopmargin=5pt,
innerbottommargin=5pt,
leftmargin=0cm,
rightmargin=0cm,
linewidth=4pt]{eBox}

\usepackage[most]{tcolorbox}
\tcbset{colback=white, colframe=black,
        highlight math style= {enhanced, 
            colframe=red,colback=red!10!white,boxsep=0pt}
        }
\definecolor{light-gray}{gray}{0.95}
\newcommand\apj{\ref@jnl{ApJ}}
\newcommand\apjl{\ref@jnl{ApJL}}     
\newcommand\apjs{\ref@jnl{ApJS}}
\crefname{table}{Table}{Tables}
\crefname{equation}{Eq.}{Eqs.}
\crefname{appendix}{App.}{Apps.}
\crefname{section}{Sec.}{Secs.}
\crefname{figure}{Fig.}{Figs.}

\numberwithin{equation}{section}

\def\beq{\begin{equation}}
\def\eeq{\end{equation}}

\def\bea{\begin{eqnarray}}
\def\eea{\end{eqnarray}}

\def\keq{k_{\mathrm{eq}}}
\def\kmax{k_{\mathrm{max}}}
\def\kmin{k_{\mathrm{min}}}

\def\beq{\begin{equation}}
\def\eeq{\end{equation}}
\def\bea{\begin{eqnarray}}
\def\eea{\end{eqnarray}}

\def\L{{\cal L}}

\def\fnl{f_{\rm NL}}

\def\H{{\cal H}}

\def\k{{\vec{k}}}

\def\x{{\vec x}}

\def\u{{\vec u}}
\def\vn{{\vec \nabla}}

\def\Bt{B_{\rm template}}

\def\dL{{\delta_{\rm lin}}}
\def\PhiL{{\Phi_{\rm in}}}
\def\hdL{{{\hat \delta}_{\rm lin}}}
\def\hdLa{{{\hat \delta}_{\rm lin}^{(1)}}}
\def\hdLb{{{\hat \delta}_{\rm lin}^{(2)}}}
\def\hdLc{{{\hat \delta}_{\rm lin}^{(3)}}}
\def\dgo{\delta_g^{\rm obs}}

\def\PL{{P_{\rm lin}}}
\def\fnl{f_{\rm NL}}
\def\fnlloc{f_{\rm NL}^{\rm loc}}
\def\fnleq{f_{\rm NL}^{\rm eq}}

\def\blin{b_{\rm lin}}
\def\bquad{b_{\rm quad}}

\DeclareRobustCommand{\SkipTocEntry}[4]{}

\definecolor{colorTC}{rgb}{.2,.7,.2}

\definecolor{amethyst}{rgb}{0.6, 0.4, 0.8}

\definecolor{acolor}{rgb}{0.4, 0.2, 0.4}

\definecolor{blue3}{RGB}{31, 119, 180}
\definecolor{red3}{RGB}{	214, 39, 40}
\definecolor{orange3}{RGB}{255, 127, 14}
\definecolor{green3}{RGB}{44, 160, 44}

\begin{document}

\begin{titlepage}
\setcounter{page}{1} \baselineskip=15.5pt
\thispagestyle{empty}
$\quad$
\vskip 70 pt

\begin{center}
{\LARGE \selectfont \bf Primordial Non-Gaussianity and  \\[8pt]
the Field-Level Cram\'er-Rao Bound}
\end{center}

\vskip 20pt
\begin{center}
\noindent
{\fontsize{12}{18}\selectfont Eugene Chen$^1$, Daniel Green$^1$, and Vincent S. H. Lee$^{1,2}$}
\end{center}

\begin{center}
\vskip 4pt
\textit{$^1${\small Department of Physics, University of California at San Diego,  La Jolla, CA 92093, USA}}
\vskip 4pt
\textit{$^2${\small Department of Physics, University of California Berkeley, Berkeley, CA 94720, USA}}

\end{center}

\vspace{0.4cm}
 \begin{center}{\bf Abstract}
 \end{center}

\noindent
Primordial non-Gaussianity is one of the most powerful probes of the inflationary epoch. The particle spectrum relevant to inflation, including masses and spins, is encoded in the precise form of statistical correlations of the adiabatic modes. Yet, in the presence of nonlinear structure formation, the optimal approach to measuring these signals remains unclear. 
Accurate modeling becomes crucial as late-time non-Gaussianty can become degenerate with primordial physics. Moreover, scale-dependent bias shows that information can move from non-Gaussian initial conditions to the amplitude of the Gaussian fluctuations. In this paper, we aim to clarify how primordial information is encoded in maps of galaxies. We use the field-level Cram\'er-Rao bound to investigate the ultimate limit of what can be extracted from realistic maps of the Universe. For local non-Gaussianity, we show that multi-tracer scale-dependent bias can exceed the sensitivity of conservative higher-point analyses. However, as expected, the multi-tracer analysis falls short of the optimal constraint when all the modes at the scale of the dark matter halos are included. We then forecast the potential reach of future surveys for equilateral and local non-Gaussianity. Equilateral in particular is highly sensitive to priors and modeling assumptions and can benefit dramatically from theoretical input such as the redshift evolution of the bias.

\end{titlepage}
\setcounter{page}{2}

\restoregeometry

\begin{spacing}{1.2}
\newpage
\setcounter{tocdepth}{2}
\tableofcontents
\end{spacing}

\setstretch{1.1}
\newpage

\section{Introduction}

One of the great challenges in fundamental physics is to understand the microphysics of the inflationary epoch directly from observations~\cite{Baumann:2009ds,Achucarro:2022qrl}. Inflation is responsible for the formation of structure in our Universe and may provide a window into the highest-energy physical processes accessible through observation. It is essential to our goal of understanding the Universe, and the fundamental laws that govern it, that we extract as much information about this period as we can from current and future cosmic maps~\cite{Chang:2022lrw,Green:2022bre}.

While the information stored in the two-point statistics is generally the easiest to measure, the form is so heavily constrained by statistical homogeneity and isotropy that only specific dynamical processes leave a measurable signal~\cite{Slosar:2019gvt}. In contrast, higher-point statistics, such as the bispectrum or trispectrum, contain a wealth of information~\cite{Babich:2004gb} about the dynamics of the inflaton~\cite{Creminelli:2003iq} and its coupling to any additional particles~\cite{Creminelli:2004yq,Chen:2009zp}. For this reason, the search for primordial non-Gaussianity (PNG) has emerged as one of the key signals for fundamental physics in the 2020s and beyond~\cite{Meerburg:2019qqi,Chang:2022lrw}. 

Unfortunately, the ultimate reach of future searches for primordial non-Gaussianity is unknown. To date, the strongest constraints arise from the cosmic microwave background (CMB)~\cite{Planck:2019kim}. The CMB is essentially a linear system, ensuring that PNG can be measured directly via higher-point statistics of temperature and/or polarization~\cite{Babich:2004yc}. Yet, the CMB is effectively two-dimensional and therefore contains vastly less information than is available in the three-dimensional structure of the observable universe. Exactly how much of this information is accessible to future large-scale structure (LSS) surveys is far more uncertain~\cite{Baldauf:2016sjb}. The late Universe is nonlinear on all but the largest scales and therefore mixes primordial and late-time non-Gaussianity. For most observables, maximizing the amount of information requires modeling~\cite{McDonald:2009dh} and removing correlations between primordial non-Gaussianity and non-linear evolution. As a result, the ultimate limits of our knowledge of the inflationary era are tied to the unknown limits of our modeling techniques.

The field-level Cram\'er-Rao (CR) bound offers one window into our potential to isolate inflationary physics~\cite{Baumann:2021ykm,Cabass:2023nyo}. Assuming that all the information about the Universe is encoded in maps of its contents, any specific measurement is limited by the total information in the underlying fields~\cite{Schmittfull:2018yuk}. From the field-level likelihood function~\cite{Jasche:2009hx,Jasche:2009hz,Kitaura:2012tu,Wang:2013ep,Leclercq:2018who,Ramanah:2018eed,Schmidt:2018bkr,Cabass:2019lqx,Cabass:2020jqo,Cabass:2020nwf}, one can calculate the Fisher-matrix of the fields themselves to determine the ultimate limit of any particular measurement~\cite{Seljak:2017rmr,Elsner:2019rql,Schmidt:2020viy,Andrews:2022nvv,Nguyen:2024yth}. This well-known observation is the basis for analyses of the CMB~\cite{Creminelli:2006gc}. Conventional CMB bispectrum estimators saturate the field-level CR bound and thus represent an optimal measurement of the amplitude of a given template~\cite{Smith:2006ud}. In principle, the same is true of bispectrum estimators used in galaxy surveys~\cite{Schmittfull:2014tca,Akitsu:2025boy}, except that practical measurements require marginalizing over a large number of additional parameters~\cite{Spezzati:2025zsb}. Under these circumstances, the field-level CR bound may differ significantly from the CR of just the low-point correlators (e.g.~power spectrum+bispectrum) as information about these nuisance parameters must ultimately reside the maps as well~\cite{Baumann:2021ykm}. 

One manifestation of our uncertainty over the limits of these analyses is the wide range of forecasts in the literature for PNG using higher-point correlators. Forecasts involving the CMB correlators or galaxies power spectra are sufficiently mature that most published results will agree within 10-20\%~\cite{Wu:2014hta}. In contrast, forecasts for equilateral non-Gaussianity for a single galaxy survey, such as DESI~\cite{DESI:2013agm}, Euclid~\cite{Amendola:2016saw} and Spec-S5~\cite{Spec-S5:2025uom}/MegaMapper~\cite{Schlegel:2022vrv}, can vary by up to an order of magnitude~\cite{Amendola:2016saw,Cabass:2022epm,Euclid:2025hlc,Braganca:2023pcp}. These differences can be attributed to the varying assumptions and choices of nuisance parameters, but even the determination of which parameters are most important are not well-known. 

The need to address this uncertainty is made more imperative given that PNG is one of the key targets for current and future surveys. For local non-Gaussianity, the path to surpassing the sensitivity of the CMB is more plausible with the help of scale-dependent bias, but the precise sensitivity is uncertain. For equilateral non-Gaussianity, published results with BOSS~\cite{Cabass:2022wjy,DAmico:2022gki} and DESI~\cite{Chudaykin:2025vdh} are a factor of five away from being competitive with the CMB. For a target of such importance to future cosmic surveys, this situation is far from desirable; charting a course towards reliable and competitive measurements is therefore essential.

In this paper, we will address some of these issues by connecting the forecasts for PNG in galaxy surveys with the underlying field-level information content of the Universe. Assuming adiabatic initial conditions, all cosmological information is encoded in the initial conditions and time evolution. As such, we define the field-level Cram\'er-Rao bound in terms of Gaussian probability distribution for the initial conditions and a forward model. Galaxy biasing presents one acute challenge to our understanding of the information content of the Universe because it can move information from the power spectrum to higher-point correlators and, in the case of scale-dependent bias, from higher-point correlators to the power spectrum. This is particularly confusing in the case of local non-Gaussianity where the relative importance of the power spectrum~\cite{Hamaus:2011dq} and bispectrum~\cite{Baldauf:2010vn} is not manifest. 

After defining the field-level Fisher matrix, including scale-dependent bias, we will directly evaluate the relative importance of various measures of local PNG at the field level. We will review the intuition and technical reasons why the matter bispectrum should contain all of the relevant information. We will then show how this explains the difference in the dependence on the smallest wavenumber, $\kmin$, between the single and multi-tracer analyses. Finally, we will see that for realistic maximum wavenumbers, $\kmax$, the multi-tracer power spectrum analysis is effectively optimal. However, this suggests that novel uses of small-scale bispectrum information (see e.g.~\cite{Goldstein:2022hgr}) can exceed the sensitivity of even an ``optimal" multi-tracer analysis~\cite{Seljak:2008xr,Hamaus:2011dq}. 

In addition, we evaluate the potential for maps of the Universe to increase our understanding of non-local PNG, represented in terms of the equilateral template\footnote{Previous studies~\cite{Babich:2004gb,Green:2023uyz} have shown that a wide range of shapes are well-approximated by the equilateral template. While equilateral specifically presents a compelling theoretical target, it also effectively describes the sensitivity to a wide range of other compelling targets. Oscillatory signals are the primary exception that requires special treatment~\cite{Beutler:2019ojk,Green:2026yev}.}. Using realistic forecasts in redshift space and marginalizing over biasing, cosmological parameters, and redshift space distortions, we assess the long term potential for LSS to exceed the sensitivity of the CMB to PNG of both the local and non-local varieties. At the field level, these forecasts are only sensitive to linear and quadratic biases. These, nevertheless, are the dominant obstacles to improving measurements of $\fnleq$. We discuss strategies, including simulation-based priors, to improve the long term prospects of the surveys. 

Our goal throughout the paper is to provide a more intuitive understanding of how the various assumptions and analysis strategies affect overall sensitivity. While our forecasts are consistent with previous analyses, we also show that for $\fnleq$, the forecasts can vary by orders of magnitude for the same survey and $\kmax$, only by changing how we marginalize over nuisance parameters. This helps explain the wild variation of forecasts in the literature, and why the situation with $\fnleq$ is so different from the $\fnlloc$ forecasts. While, in principle, the prospects for a survey like Spec-S5~\cite{Spec-S5:2025uom}/MegaMapper~\cite{Schlegel:2022vrv} to deliver a transformative measurement of PNG beyond the local type, the challenge of controlling biasing remains a necessary challenge if we are to meet these ambitions.

The paper is organized as follows: In Section~\ref{sec:likelihood}, we introduce the model of initial conditions and nonlinear evolution that underlie our field level analysis. In Section~\ref{sec:CR}, we use these ingredients to determine the field level likelihood and Fisher matrix. In Section~\ref{sec:local}, we apply this to $\fnlloc$ to understand the relative importance of the power spectrum and bispectrum. In Section~\ref{sec:fnleq}, we perform realistic forecasts and the conclude in Section~\ref{sec:conclusions}. Details of the calculations are contained in four Appendices.

\subsection{Conventions and notation}
\label{subsec:notation}

Throughout this work we adopt standard large-scale structure notation, but
explicitly fix several conventions that may vary across the literature. Shorthand notations used for momentum integrals and Dirac delta functions
are collected in Table~\ref{tab:shorthand} for convenience.
We specify our normalization of correlation functions,
the definition of linear and nonlinear density fields, and the treatment
of growth factors and transfer functions.
These conventions are summarized in Table~\ref{tab:conventions}. Finally, we collect parameters relevant to the formation of halos in Table~\ref{tab:parameters}.

\begin{table}[H]
\centering
\caption{Shorthand notation used throughout this work. 
}
\label{tab:shorthand}
\renewcommand{\arraystretch}{1.2}
\setlength{\tabcolsep}{8pt}
\begin{tabular}{ll}
\hline\hline
 $\displaystyle \int d^3x$, \,  $\displaystyle \int \frac{d^3\vec{k}}{(2\pi)^3}$ & $\displaystyle \int_x$ \, , \, $\displaystyle \int_k$ \\
$\displaystyle \int \frac{d^3\vec{k}_1}{(2\pi)^3}\cdots \frac{d^3\vec{k}_n}{(2\pi)^3}\,(2\pi)^3\delta_D(\vec{k}_1+\cdots \vec{k}_n)$ & $\displaystyle \int_{k_1,\cdots ,k_n}$ \\
 
$(2\pi)^3\delta_D(\mathbf{0})$& $V_{\rm survey}$ \\
\hline\hline
\end{tabular}
\end{table}

\begin{table}[H]
\centering
\caption{Notation and conventions used in this work.
The table fixes definitions that may differ across the large-scale structure literature.}
\label{tab:conventions}
\renewcommand{\arraystretch}{1.25}
\setlength{\tabcolsep}{6pt}
\begin{tabular}{p{4.3cm} p{3.7cm} p{6.0cm}}
\hline\hline
Quantity & Symbol & Convention used here \\
\hline

Dirac delta function
& $\delta_D(\k)$& $(2\pi)^3\delta_D(\k)$ enforces momentum conservation \\

Initial matter density contrast
& $\delta(\k,z_0)$
& Defined at reference redshift $z_0$ \\

Linear matter density contrast
& $\dL(\k,z)$
& $\dL(\k,z)=\dfrac{D(z)}{D(z_0)}\,\delta(\k,z_0)$ \\

Nonlinear matter density contrast
& $\delta(\k,z)$
& Real-space matter over-density field \\

Galaxy density contrast
& $\delta_g(\k)$
& Real-space galaxy over-density field \\

Primed correlator
& $\langle\cdots\rangle'$
& Overall momentum-conserving Dirac delta removed \\

Linear power spectrum
& $\PL (k,z)$
& $\langle\dL(\k,z)\dL(\k',z)\rangle'=\PL(k,z)$ \\

Nonlinear matter power spectrum
& $P(k)$
& Defined from $\langle\delta(\k)\delta(\k')\rangle'$ \\

Galaxy power spectrum
& $P_g(k)$
& Defined from $\langle\delta_g(\k)\delta_g(\k')\rangle'$ \\

Shot noise
& $N^2$& Poisson shot-noise contribution to $P_g(k)$ \\

Initial Newtonian potential
& $\Phi(\k)$
& Gaussian primordial potential \\

Primordial potential power spectrum
& $P_\Phi(k)$
& $\langle\Phi(\k)\Phi(\k')\rangle'=P_\Phi(k)$ \\

Transfer function
& $\mathcal{T}(k,z)$
& $\mathcal{T}(k,z)\equiv\sqrt{\PL(k,z)/P_\Phi(k)}$ \\

Observed galaxy density contrast
& $\delta_g^{\rm obs}(\k)$
& Includes survey window and observational effects \\

Observed galaxy power spectrum
& $P_g^{\rm obs}(k)$
& Angular-averaged spectrum derived from $\delta_g^{\rm obs}$ \\

\hline\hline
\end{tabular}
\end{table}

\begin{table}[H]
\centering
\caption{Primordial and large-scale structure parameters used in this work.}
\label{tab:parameters}
\renewcommand{\arraystretch}{1.25}
\setlength{\tabcolsep}{6pt}
\begin{tabular}{p{4.3cm} p{1.5cm} p{8cm}}
\hline\hline
Quantity & Symbol & Definition / meaning \\
\hline

Primordial potential amplitude
& $\Delta_\Phi$
& $\Delta_\Phi \equiv k^3 P_\Phi(k)$ \\
\hline

Characteristic smoothing scale
& $R_*$
& Lagrangian smoothing scale associated with galaxy formation, setting the natural size of higher-derivative (gradient) bias operators. We take $R_*=2.66\,h^{-1}\mathrm{Mpc}$. \\
\hline

Critical collapse threshold
& $\delta_c$
& Critical linear over-density for spherical collapse. We fix $\delta_c=1.686$~\cite{Gunn:1972sv}. \\
\hline

Primordial scalar power spectrum amplitude
& $A_s$
& Size of the initial density perturbations. We set $A_s=2\times 10^{-9}$~\cite{Planck:2018vyg}. \\
\hline

Matter density Parameter
& $\Omega_m$
& Fraction of cold dark matter denstiy today. We fix $\Omega_m h^2=0.12$, where the Hubble constant to day is denoted by $H_0=100\,h$\,km/s/Mpc~\cite{Planck:2018vyg}. \\
\hline

Turnover scale of the matter power spectrum
& $k_{\mathrm{eq}}$
& Comoving wavenumber corresponding to the horizon scale at matter–radiation equality. We fix $k_{\mathrm{eq}}=0.01$~Mpc$^{-1}$\cite{Eisenstein:1997ik}. \\
\hline

Size of halo mass bin in log scale
& $\Delta (\log M)$& Size of a mass bin in log scale, where $M$ is the halo mass. We take $\Delta (\log M)=1$ as a benchmark. \\
\hline

\hline\hline

\end{tabular}
\end{table}

\section{From Initial Conditions to Galaxies }\label{sec:likelihood}

In this section, we will 
introduce the field-level forward models that describe the distribution for galaxies derived from a map of the non-Gaussian initial conditions. These models contain free parameters that are ultimately determined by data. It is through this forward modeling that we will later derive likelihoods for both PNG and these nuisance parameters.

Like any model of galaxy biasing, we will start by describing the initial conditions, and then move to the nonlinear evolution of matter. Finally, we will translate the matter distribution to a distribution of galaxies using a local (Lagrangian) biasing model. Our goal here is to be explicit about these assumptions so that they can be easily improved upon or extended.

\subsection{Initial Conditions}

We define the initial density field in terms of the totals matter density fluctuation, 
\beq
\delta(\x,z_0) \equiv \frac{\rho_m(\x,z_0) - \bar\rho_m(z_0)}{\bar \rho_m(z_0)} \approx \dL(\x,z_0)
\eeq
at some large redshift after recombination, $1100 > z_0\gg 1$, where $\rho_m(\x,z_0)$ is the local matter density and $\bar \rho_m(z_0)$ is the spatial average. We set the initial conditions at this early time so that the non-linear field, $\delta$, is indistinguishable from linear density fluctuation, $\dL$. The initial conditions are defined in the matter dominated era and are therefore more easily expressed in terms of nearly\footnote{In linearized gravity, $\Phi$ is constant during matter domination and therefore $\Phi_{\rm in}(\k)$ can be specified in the linear regime at any $z\in [2,10^3]$. }
 constant initial Newtonian potential $\PhiL(\x)$,
\begin{equation}\label{eqn:DM_density_field}
	\dL(\vec{k}) = \frac{2k^2T(k)D(z)}{3\Omega_mH_0^2}\PhiL(\vec{k}) \equiv \mathcal{T}(k,z)\PhiL(\vec{k}) \, ,
\end{equation}
with the transfer function $T(k)$ satisfying $T(k)\to 1$ for small $k \to 0$. Here $D(z)$ is the linear growth factor, $\Omega_m$ is the cold dark matter abundance today, and $H_0\equiv 100\,h$~km/s/Mpc is the Hubble rate today. We will denote the (equal-time) nonlinear matter power spectrum as $P(k)$, defined by
\beq
\langle \delta(\vec{k},z)\delta(\vec{k}',z)\rangle = (2\pi)^{3}\,\delta_{D}(\vec{k}+\vec{k}')P(k,z)
\eeq
and similarly for the linear power spectrum $\PL (k,z)$. From here onward, we will drop the redshift dependence to avoid clutter, unless required for clarity. Note that in a matter-dominated universe, $\Phi(\k)$ is constant in linear evolution, so there is no meaningful distinction between the initial conditions and the linearized $\Phi$. Of course, the Poisson equation, $\nabla^2 \Phi = \frac{3}{2} \Omega_m \H^2 \delta$, where $\H\equiv aH$ is the conformal Hubble parameter, is a non-perturbative statement in Newtonian gravity so it will be important to distinguish $\PhiL$ and $\Phi$.

The initial conditions for $\Phi(\k)$ are defined statistically. The minimal assumption is that each adiabatic fourier mode, $\zeta(\k)$, is drawn from a Gaussian distribution with a power spectrum 
\beq
\Delta_{\zeta}^2(k)=(k^3/2\pi^2)P_{\zeta}(k)=A_s(k/k_0)^{n_s-1} \, ,
\eeq
where $A_s$ is the primordial power spectrum amplitude at some scale $k_0$, and $n_s$ is the spectral index~\cite{Larson:2010gs}. In terms of the (linear) Newtonian potential during matter domination, $\Phi = -3 \zeta / 5$, the potential power spectrum is
\begin{align}\label{eqn:Gaussian_power_spectrum}
    P_{\Phi}(k) = \frac{9}{25}\frac{2\pi^2}{k^3}A_s\left(\frac{k}{k_0}\right)^{n_s-1} &\approx \frac{18\pi^2}{25}\frac{A_s}{k^3} \nonumber \\
    \Delta_{\Phi} \equiv k^3P_{\Phi}(k) &\approx \frac{18\pi^2}{25}A_s \, ,
\end{align}
where we have taken $n_s\approx 1$ for simplicity.

Our goal here is to investigate the possibility that the initial conditions are, in fact, non-Gaussian. For the simplest perturbative models of non-Gaussian statistics, the leading non-Gaussian correlator is the bispectrum, defined as
\begin{equation}\label{eqn:bispectrum_def}
	\langle\PhiL(\vec{k}_1)\PhiL(\vec{k}_2)\PhiL(\vec{k}_3)\rangle  =B_{\Phi}(k_1,k_2,k_3)(2\pi)^3\delta_D(\vec{k}_1+\vec{k}_2+\vec{k}_3) \, .
\end{equation}
We can learn a lot of about the detectability of a wide range of scale invariant signals from considering two template: the local and equilateral shapes, with the bispectrum given by~\cite{Komatsu:2001rj, Babich:2004gb}
\begin{align}\label{eqn:loc_bispectrum}
	B^{\mathrm{loc}}_{\Phi}(k_1,k_2,k_3) &= 2f_{\mathrm{NL}}^{\mathrm{loc}}P_{\Phi}(k_1)P_{\Phi}(k_2)+\mathrm{perms} \\
	B^{\mathrm{eq}}_{\Phi}(k_1,k_2,k_3) &= 162f_{\mathrm{NL}}^{\mathrm{eq}}\frac{\Delta_{\Phi}^2}{k_1k_2k_3(k_1+k_2+k_3)^3}  \label{eqn:eq_bispectrum}\, .
\end{align}
Much more complex signals can arise from additional fields, particle production, and time-dependent couplings~\cite{Achucarro:2022qrl}. Nevertheless, our interest here is understanding the limitations due to degeneracies with local physics and these two examples are sufficient to underscore the litany of potential issues.  

In practice, we will not observe the bispectra of the Newtonian potential or the adiabatic mode directly. Instead, even in the linear description, the observables derived from the matter over-density involve the transfer function so that 
\begin{align}\label{eqn:DM_spectrum}
	P(k) &\equiv \langle \delta(\vec{k})\delta(\vec{k}')\rangle' = \mathcal{T}^2(k)P_{\Phi}(k)  = \frac{8\pi^2}{25}\frac{A_sD^2(z)T^2(k)}{\Omega_m^2H_0^4}k  \nonumber \\
	B(k_1,k_2,k_3) &\equiv \langle \delta(\vec{k}_1)\delta(\vec{k}_2)\delta(\vec{k}_3)\rangle' = \mathcal{T}(k_1)\mathcal{T}(k_2)\mathcal{T}(k_3)B_{\Phi}(k_1,k_2,k_3) \, .
\end{align}
Here $P(k)$ is really the linear solution to the matter power spectrum. We expect $P_m(k)\approx P(k)$ on large scale, $k<k_{\mathrm{NL}}\approx 0.1h\,\mathrm{Mpc}^{-1}$.

In order to maintain consistency with CMB normalization, it will be useful for us to define the template matter bispectrum as
\begin{align}\label{eqn:template_bispectrum}
	B(k_1, k_2, k_3) &= f_{\mathrm{NL}}\Bt(k_1, k_2, k_3) \nonumber \\
    &= f_{\mathrm{NL}}B_{\Phi,{\rm template}}(k_1, k_2, k_3)\mathcal{T}(k_1)\mathcal{T}(k_2)\mathcal{T}(k_3)\, ,
\end{align}
where, from Eqs.~\eqref{eqn:loc_bispectrum}-\eqref{eqn:eq_bispectrum}
\begin{align}\label{eqn:B_template}
	B_{\Phi,{\rm template}}^{\mathrm{loc}}(k_1, k_2, k_3) &= \frac{648\pi^4A_s^2}{625}\frac{k_1^3+k_2^3+k_3^3}{k_1^3k_2^3k_3^3} \nonumber \\
	B_{\Phi,{\rm template}}^{\mathrm{eq}}(k_1, k_2, k_3) &=\frac{52488\pi^4A_s^2}{625}\frac{1}{k_1 k_2 k_3(k_1+k_2+k_3)^3} \, ,
\end{align}
In principle, we should be able to use the factorizable bispectrum templates in place of these analytic ones. However, given that we are anticipating significant degeneracies between $\fnl$ and the bispectra associated with non-linear evolution, we do need to worry that small differences between PNG templates could have a significant impact on the final results for $\fnl$.

\subsection{Nonlinear Evolution}

The central challenge in the search for PNG in galaxy surveys is the nonlinearity of the galaxy distribution. This comes from two effects, non-linearity of the matter and non-linearity of halo/galaxy formation. We will begin by discussing the former, where the rules are well understood. The later will be modeled using the bias expansion described in the next section. However, in both cases it is important that these are models for the changes to the maps themselves and not merely the summary statistics.

In principle, the nonlinear evolution of dark matter density field is governed by the collisionless Boltzmann (Vlasov) equation. Solving the equation exactly is intractable, and even perturbative solutions are challenging at the full phase space level. Two standard simplifications are widely used (see Ref.~\cite{Bernardeau:2001qr} for a reference). The first is the Eulerian formulation, standard perturbation theory (SPT), which takes the lowest two moments of the Boltzmann equation, yielding the continuity and Euler equations for a pressure-less fluid. The second is the Lagrangian formulation, Lagrangian perturbation theory (LPT), which follows individual fluid elements by solving Newton's equation of motion in comoving coordinates. In this work, we adopt the Eulerian/SPT framework because it provides a more natural setting for introducing the effective field theory of large scale structure and for incorporating the associated counter-terms. We will be able to account for some aspects of the Lagrangian formulation using IR resummation (the Zeldovich approximation).

In standard perturbation theory (SPT), one treats the matter as a cold fluid coupled to gravity, 
\begin{align}
\frac{\partial \delta(\x, \tau)}{\partial \tau}+\vn \cdot\{[1+\delta(\x, \tau)] \u(\x, \tau)\}&=0 \ , \\
\frac{\partial \u(\x, \tau)}{\partial \tau}+\mathcal{H}(\tau) \u(\x, \tau)+\u(\x, \tau) \cdot  \vn \u(\x, \tau)&= -\vn \Phi(\x, \tau) \ ,
\end{align}
where $\H(\tau) = a'/a$. Solving the fluid equations order by order yields an expansion of the late-time density contrast. In an Einstein-de Sitter universe ($\Omega_m=1$), the time evolution can be determined easily order by order, to that the resulting solution takes the form 
\begin{align}
    \delta_{\rm SPT}(\k,\tau)
    = \sum_{n=1}^{\infty} a^n(\tau)\, \delta^{(n)}(\k) , 
\end{align}
where $\delta^{(n)} = {\cal O}(\delta^n)$ is the $n$-th order solution. The general solution, written as 
\begin{align}
    \delta^{(n)}(\k)
    = \int_{k_1 \cdots k_n}
      \delta_D(\k_1 + \cdots + \k_n - \k)\,
      F_n(\k_1,\ldots,\k_n)\,
      \delta_L(\k_1)\cdots\delta_L(\k_n)\ ,
\end{align}
can be solved using a difference equation that determines the kernels $F_n$. Here $\int_{k_1\ldots k_n}\equiv \int \frac{d^3 k_1}{(2\pi)^3}\ldots\frac{d^3 k_n}{(2\pi)^3}$. 
These kernels are the standard SPT mode-coupling functions, whose explicit forms can be found in the literature~\cite{Crocce:2005xy }. 

In order to model data accurately, it is important that the matter is not literally a perfect fluid, especially on small scales. On large scales, the impact of the physics on small scales\footnote{Summary statistics such as the power spectrum computed from SPT are not, in general, ultraviolet finite, since short-wavelength modes contribute spuriously when integrated perturbatively. These ultraviolet contributions are absorbed into counter-terms that arise from an effective stress tensor.} is encoded in changes to the fluid equations via a non-zero stress tensor, $\sigma^{ij}$, so that 
\beq
\frac{\partial \u(\x, \tau)}{\partial \tau}+\mathcal{H}(\tau) \u(\x, \tau)+\u(\x, \tau) \cdot  \vn \u(\x, \tau)=  -\vn \Phi(\x, \tau)-\frac{1}{\rho} \left(\vn_j\sigma^{ij} \right) \ .
\eeq
The effective field theory~\cite{Baumann:2010tm,Carrasco:2012cv,Carrasco:2013mua,Pajer:2013jj,Porto:2013qua} of large-scale structure (EFTofLSS) accounts for this stress using the approximate locality (in space) of the non-zero stress. To leading order, in the EFT power counting, 
\begin{align}
  \partial_i( \rho^{-1} \partial_j\tau^{ij} ) \approx c_s^2 \partial^2 \delta \to \delta(\k,\tau)
    = \delta_{\rm SPT}(\k,\tau)
      - c_s^2(a)\frac{k^2}{k_{NL}^2} \, D(a)^3\, \delta_L(\k) \ ,
\end{align}
where the coefficient $c_s^2(a)$ encodes the impact of unresolved short-scale physics~\cite{Carrasco:2012cv}. In general, the coefficients are non-local in time~\cite{Carrasco:2013mua,Carroll:2013oxa}, which is important at higher orders.

The full details of the EFT of LSS are reviewed elsewhere in far greater detail~\cite{Cabass:2022avo,Ivanov:2022mrd}. The key observations we need here is only that the equations of motion of SPT or the EFT of LSS give rise to a relationship between $\delta(\k)$ and the map of the initial conditions. The big difference between STP and the EFT of LSS is that the EFT allows for stochastic contributions to the stress tensor and therefore is not completely determined by the initial map. As we will see, the same thing arises due to shot noise in the galaxy density contrast and therefore we will absorb this issue into the definition of the galaxies.

\subsection{Biasing}
\label{subsection:biasing}
The formation of galaxies from the underlying distribution of matter is complex and difficult to model, even numerically. However, because the formation process should be governed by local physics, the density contrast of galaxies should be a local function of the underlying distribution of matter and associated gravitational fields. For initial conditions that obey the single field consistency conditions~\cite{Creminelli:2013mca,Creminelli:2013poa}, this expression should obey~\cite{McDonald:2009dh}
\beq\label{eq:bias_general}
\delta_g(\x) = F[\nabla_i \nabla_j \Phi] \approx b_1 \delta + b_2 \delta^2 + \ldots\ , 
\eeq
where $\nabla_i \nabla_j \Phi$, the tidal tensor, is the building block for any observable consistent with the equivalence principle. As usual, we have expanded this function as series in powers of $\Phi$, or equivalently $\delta$, and derivatives thereof. While this formula intuitively captures the local nature of galaxy formation, it over-simplifies the relationship between the matter density that is observable, at late times and large distances, and the density contrast of galaxies in the same regime (see~\cite{Desjacques:2016bnm} and references therein for review). 

For our purposes, the most significant departures from this simple picture arise at linear order in the fluctuations in the presence of 
PNG, in the form of scale-dependent bias~\cite{Grinstein:1986en,Dalal:2007cu}. The scale-dependent bias is defined by the behavior of the bispectrum template in the squeezed limit, 
\beq
\Bt(k_1,k_2,k_3)|_{k_1 \ll k_2,k_3} \to 4\lambda\fnl \left(\frac{k_1}{k_2}\right)^\Delta P(k_1) P(k_2) \ ,
\eeq
where $\lambda$ defines the relative amplitude in the squeezed limit and $\Delta =0,2$ for local and equilateral non-Gaussianity respectively. For simplicity\footnote{For $\fnlloc$, this expression is defined so that $\lambda=1$. For single field inflation, where $\Delta=2$, $\lambda$ depends on the specific interactions and the precise form of the template~\cite{Creminelli:2013cga}. For the purpose of our paper, $\fnlloc$ is the only case where the scale-dependent bias dominants the signal, although there are now additional examples~\cite{Green:2026yev}.}, we will define $\lambda=1$ to avoid clutter. For the equilateral template, we adopt the standard normalization in which the bispectrum in the squeezed limit carries an overall factor of 3, so that our definition of $f_{\rm NL}^{\rm eq}$ is consistent with existing surveys. The squeezed limit generates correlations between the variance on small scales with the density contrast on large scales, giving rise to a correction to the linear bias of the form~\cite{Dalal:2007cu,Baumann:2012bc,Assassi:2015fma} 
\begin{equation}\label{eqn:bias_expansion}
	\delta_g(x) = b_1\delta_x(x) + \fnl \frac{b_{\Phi}}{\mathcal{T}(k)}(kR_*)^{\Delta}\delta(x) + \epsilon(x) +\cdots \, ,
\end{equation}
where $b_\Phi$ is the scale-dependent bias coefficient, and $\epsilon(x)$ is the stochastic bias. These two terms represent the influence of of modes we don't observe on the formation of galaxies. The scale-dependent bias, $b_\Phi$, arises in the presence of primordial non-Gaussianity that correlates these short distance fluctuations with a long-wavelength Newtonian potential. 
The stochastic bias represents information encoded in those short wavelength modes that is uncorrelated with the long wavelength $\delta(\x)$ and thus is effectively a random variable. In practice, the stochastic bias is well approximated as shot noise with variance $\langle \epsilon(x)\epsilon(x')\rangle  = N^2\,\delta(x-x') $ where $N^2 = 1/\bar n_g$ where $\bar n_g$ is the average density of galaxies in the survey.

Even before getting to higher-order biases, or the relationship to the EFT of LSS, it is critical that we are trying to model $\delta_g(\x)$ after having integrated out (i.e.~marginalized over) the short distance fluctuations that we cannot resolve. As we are trying to reconstruct the underlying density field from the galaxy distribution, this loss of information is ever-present in our attempt to understand what primordial information is accessible in galaxy surveys.  

Keeping only these linear terms, the power spectrum of the long wavelength galaxy fluctuations can be described by
\begin{equation}\label{eqn:galaxy_power_spectrum}
	P_g(k) = \left[b_1+f_{\mathrm{NL}}\frac{b_{\Phi}}{\mathcal{T}(k)}(kR_*)^{\Delta}\right]^2P(k) + N^2  \, .
\end{equation}
It will be important that galaxies are binned into groups with different properties (e.g.~halo mass) for understanding how information is mapped from primordial initial conditions to late time statistics. For example, if we split the sample into two subsets of galaxies, each tracer can be labeled by $i=1,2$ with its own bias expansion
\begin{equation}\label{eqn:bias_expansion_multi}
	\delta_{g}^{(i)}(\vec{k})=\left[b_{1}^{(i)} + f_{\rm NL}\,\frac{b_{\Phi}^{(i)}}{\mathcal T(k)}\,(kR_*)^{\Delta}\right]\delta(\vec{k})+ \epsilon^{(i)}(\vec{k}),
\end{equation}
with $\langle \epsilon^{(i)}(\vec{k})\epsilon^{(j)*}(\vec{k}')\rangle = (2\pi)^{3}\,\delta_{D}(\vec{k}+\vec{k}')\,\delta_{ij}N_i^2$.
The covariance matrix is given by
\begin{align}\label{eqn:covariance_matrix}
	C_{ij}(k)&\equiv\langle \delta_{g}^{(i)}(\vec{k})\,\delta_{g}^{(j)}(\vec{k})\rangle' \nonumber \\
	&=\left[b_{1}^{(i)} + f_{\rm NL}\,\frac{b_{\Phi}^{(i)}}{\mathcal T(k)}\,(kR_*)^{\Delta}\right]\left[b_{1}^{(j)} + f_{\rm NL}\,\frac{b_{\Phi}^{(j)}}{\mathcal T(k)}\,(kR_*)^{\Delta}\right]P(k)+ \delta_{ij}\,N_i^2\, .
\end{align}
To avoid clutter, we will largely work with a single sample except for Section~\ref{sec:local}, where we will focus on local non-Gaussianity, in which case multi-tracer analyses are crucial.
\medskip

It is important galaxy formation is a nonlinear phenomenon and therefore requires higher-order biasing, as in Eq.~(\ref{eq:bias_general}). This is covered in great detail in reviews such as Refs.~\cite{McDonald:2009dh,Desjacques:2016bnm}. For our purposes, the quadratic bias terms will play the most significant role and can be expressed in Fourier space as 
\begin{equation}\label{eqn:bias_expansion2}
	\begin{aligned}
	    &\delta_g(\k) = b_{\rm lin}(k)\dL(\k)\\
        &+ \int_{k_1,k_2}\delta_D(\k_1+\k_2-\k)\Big[b_{\rm quad}(\k_1,\k_2)+b_{\rm lin}(k)F_2(\k_1,\k_2)\Big]\dL(\k_1)\dL(\k_2) +\ldots \ ,
	\end{aligned}
\end{equation}
where 
\begin{align}
        & b_{\rm lin}(k) = b_1 +  b_{k^2} (kR_{*})^2+ b_{k^4} (kR_{*})^4
        + f_{\mathrm{NL}}\frac{b_{\Phi}}{\mathcal{T}(k)}(kR_*)^{\Delta} \\
        & b_{\rm quad}(\k_1,\k_2) = b_2
        + b_{\rm s^2}\Big(\frac{(\vec{k}_1\cdot \vec{k}_2)^2}{k_1^2k_2^2}-\frac{1}{3}\Big)
        + f_{\rm NL} \frac{\Bt(k_1,k_2,|\k_1+\k_2|)}{2\PL(k_1)\PL(k_2)} \; .
\end{align}
In principle $\Bt$ could be any bispectrum of interest, but for our purposes we will consider only local and equilateral so that $\fnl=\fnlloc$  or $\fnleq$ with the associated templates from Eqs.~\eqref{eqn:loc_bispectrum}-\eqref{eqn:eq_bispectrum} respectively. Here we have introduced the $k$-dependent bias parameters $b_{\rm lin}(k)$ and $b_{\rm quad}(k_1,k_2)$ to simplify our expressions~\cite{Schmittfull:2018yuk}, at the cost of making the expressions non-local. Specifically, $b_{\rm lin}(k)$ collects all the linear bias contributions, including derivative terms and the scale-dependent bias. All quadratic bias terms are collected in $b_{\rm quad}(k_1,k_2)$, including the map from purely Gaussian to weakly non-Gaussian initial conditions. The scale $R_{*}$ denotes a characteristic (Lagrangian) size of the tracer. It is common to take
$b_\Phi = \delta_c(b_1-1)$ as the PNG bias coefficient
and $\delta_c \simeq 1.686$ as critical over-density based on the critical threshold model~\cite{Press:1973iz}. 
The bias parameters, $b_1$, $b_{k^2}$, $b_{k^4}$, $b_2$, and $b_{s^2}$, are often take to be free parameters, but can be determined from simulations or from a specific biasing model (see e.g.~\cite{Sharma:2025xss} for discussion in the context of $\fnleq$). 

Finally, we note that $\fnl$ appears in both $b_{\rm lin}$ and $b_{\rm quad}$ even though they are different orders in $\delta$. At late times, $T(k,z) \gg 1$ because of the growth of structure, and therefore the contributions from linear and quadratic bias are similar. As we will see, the dominant contribution will depend in detail on the shape of the bispectrum and the scales that can be resolved.

\subsection{Redshift-Space Distortions}
\label{subsection:rsd}

The bias expansion described in Sec.~\ref{subsection:biasing} specifies the galaxy density field in three-dimensional comoving coordinates. Observationally, however, galaxy positions are inferred from redshifts, so that peculiar velocities along the line of sight induce anisotropic distortions in the observed clustering pattern. These redshift-space distortions (RSD) arise from the velocity-dependent mapping between real-space galaxy positions and observed redshift-space coordinate~\cite{Bernardeau:2001qr}, and are typically modeled in order to relate the biased galaxy density field to measured clustering statistics~\cite{Schmittfull:2020trd}.

At linear order, RSD is described by the Kaiser effect~\cite{Kaiser:1987qv}, such that the redshift-space galaxy over-density is given by
\begin{equation}
\delta_g^{(\rm rsd)}(\vec{k})
=
\delta_g^{(r)}(\vec{k}) + f\,\mu^2\,\dL(\vec{k}),
\end{equation}
where $\delta_g^{(\rm rsd)}$ denotes the redshift-space over-density, $\delta_g^{(r)}$ denotes the over-density in real space, 
$f \equiv \partial \log D/\partial \log a$ is the linear growth rate, and $\mu$ denotes the cosine of the angle between the wave-vector $\vec{k}$ and the line of sight. This mapping introduces anisotropic $\mu$-dependence in the observed clustering statistics, even when the underlying real-space density field is isotropic.

Beyond linear order, the real-to-redshift-space mapping enhances the sensitivity of clustering observables to short-scale velocity and displacement modes. In practice, these modes cannot be fully resolved in a perturbative description~\cite{Jackson:1971sky,Scoccimarro:2004tg} and are integrated out. Their impact on large-scale redshift-space observables is therefore modeled phenomenologically.

In this work, we account for small-scale random motions using a Gaussian Finger-of-God~\cite{Jackson:1971sky} (FoG) damping model, following standard practice (see, e.g., Ref.~\cite{SaitoRSDNotes}). At a schematic level, FoG effects suppress power along the line of sight and can be represented by multiplicative damping factors acting on the redshift-space galaxy power spectrum and bispectrum,
\begin{equation}
    P^{\rm FoG}_g(k,\mu)
    \equiv
    e^{-\sigma_{v,P}^2\,\mu^2\,k^2}\,P_g(k,\mu),
\end{equation}
\begin{equation}
    B^{\rm FoG}_g(k_1,k_2,k_3;\mu_1,\mu_2,\mu_3)
    \equiv
    e^{-\sigma_{v,B}^2\,\sum_i \mu_i^2 k_i^2}\,
    B_g(k_1,k_2,k_3),
\end{equation}
where $\mu$ (and $\mu_i$) denote the cosine of the angle between each wavevector and the line of sight, and $\sigma_{v,P}$ and $\sigma_{v,B}$ parametrize the effective velocity dispersion relevant for the power spectrum and bispectrum, respectively. These expressions are meant to indicate the structure of the FoG suppression used in the forecast, rather than to define new observables.

In addition to FoG suppression, the coarse-graining of short-scale velocity and displacement modes induces corrections to the truncated perturbative description of the redshift-space mapping. At the level of the power spectrum, these corrections can be parametrized by effective counterterms with distinct $\mu$-dependence. Following the notation of Ref.~\cite{Chen:2020fxs}, we include the leading such contributions,
\begin{equation}
    P_g^{\rm RSD}(k,\mu)
    \;\ni\;
    \alpha_2\,(R_* k)^2\,\mu^2
    + \alpha_4\,(R_* k)^2\,\mu^4 ,
\end{equation}
where $\alpha_2$ and $\alpha_4$ are treated as nuisance parameters and marginalized over in the forecast.

Throughout this analysis, we neglect the Alcock--Paczynski effect~\cite{Alcock:1979mp}, assuming that the background cosmology is sufficiently well constrained by external data such that geometric distortions do not significantly affect the inferred constraints on primordial non-Gaussianity.

\section{Cram\'er-Rao at the Field Level} \label{sec:CR}

The central goal of this paper is to understand the limits of cosmological information available in maps of the Universe. The Cram\'er-Rao (CR) bound of the underlying map provides a robust lower limit on the variance with which we can measure any cosmological parameter. In this section, we will review the field-level likelihood function for a distribution of matter $\delta_m(\x,z)$ or galaxies $\delta_g(\x,z)$ and derive the expressions for the Cram\'er-Rao bound for the amplitudes of primordial non-Gaussianity.

\subsection{Galaxy Surveys and Field-Level Inference}

Field-level inference\footnote{Here we will make no distinction between field-level inference, simulation-based inference, and likelihood-free inference. For our purposes, field-level inference means that we define the likelihood directly in terms of the initial conditions at the field level.} for an LSS survey is a Bayesian inference for both the cosmological parameters and initial conditions, applied directly at the level of the observed distribution of galaxies and/or matter~\cite{Jasche:2009hx,Jasche:2009hz}. The cosmological parameters are assumed to be deterministic; they take on a single value in our Universe and dictate the evolution of structure from a fixed set of initial conditions. The initial conditions are the only fundamental source of randomness. The form of the probability distribution for the initial conditions therefore dictates the posterior for both the reconstruction of the initial conditions and the cosmological parameters. In practice, however, we do not observe the universe to arbitrarily small scales and therefore some processes will appear to be stochastic. For our purposes, galaxy formation is the most significant such process.

We will make the fundamental assumption that our initial conditions for the matter field, $\delta_{\rm lin}(\x,z=z_0)$, in the Universe can be derived from drawing our initial density field from a purely Gaussian distribution. We can then allow for weakly non-Gaussian initial conditions as a small (deterministic) non-linear transformation of this map~\cite{Schmidt:2010gw}. We will additionally assume that the observed density contrast of galaxies, $\delta_g^{\rm obs}$, is determined by the matter density field (via a map $\delta_g(\x)$), up to shot noise (\textit{i.e.} randomness from sampling a discrete set of objects). From these two core assumptions, the likelihood of an observed density contrast of galaxies, given a set of biases, $b_i$, and cosmological parameters, $\theta_i$, is given by~\cite{Cabass:2019lqx,Baumann:2021ykm}
\begin{equation}
\mathcal{P}[\delta_g^{\rm obs}\mid b_i,\theta_i]
=
\int \mathcal{D}\dL\,
\exp\!\left[
-\int_k \frac{\dL(\k)\dL(-\k)}{2\PL(k)}
-\int_k \frac{|\delta_g(\k)-\delta_g^{\rm obs}(\k)|^2}{2N^2}
\right] \ .
\label{eq:field_level_likelihood}
\end{equation}
Here $N^2$ denotes
the (assumed scale-independent) shot-noise variance.

Although only a single realization of the Universe is observed, we integrate over the Gaussian
prior for $\dL$ because it is the observed distribution of galaxies that is fixed by our measurements.
Statistical homogeneity and isotropy imply that the ensemble distribution of
linear modes can be inferred from angular averaging over Fourier modes, which
motivates the use of the functional integral in
Eq.~\eqref{eq:field_level_likelihood} as the field-level likelihood.

The functional integral in Eq.~\eqref{eq:field_level_likelihood} can be evaluated
systematically by expanding the action around its saddle-point (maximum-likelihood) configuration~\cite{Baumann:2021ykm},
$\hdL$, satisfying
\beq
\frac{\partial}{\partial \dL(-\k)}\mathcal{P}[\delta_g^{\rm obs}\mid b_i,\theta_i]\,|_{\dL(\k)=\hdL(\k)} = 0 \ .
\eeq
The maximum-likelihood solution admits a transparent power counting, obtained iteratively in powers of the observed
field,
\beq
\hdL = \hdLa+\hdLb+\hdLc+\ldots\ , 
\eeq
where $\hdL^{(n)} ={\cal O}((\dgo){}^n)$. Solving the for the maximum-likelihood linear field in this way, one finds the leading terms
\begin{align}
     &\hdLa(\k) = \frac{b_{\rm lin}\PL(k)}{\PL b_{\rm lin}^2+N^2}\delta_g^{\rm obs}(\k)\\
    &\hdLb(\k) = -\int_{k_1}\frac{b_{\rm lin}\PL(k)b_{\rm quad}(\k_1,\k\text{-}\k_1)}{\PL(k)b_{\rm lin}^2+N^2} \hdLa(\k_1) \hdLa(\k\text{-}\k_1) \ .
\end{align}
Higher-order terms involving additional nonlinear bias insertions are neglected.
Details of the saddle-point solution are presented in Appendix \ref{app:Marginal likelihood}. 

At this stage, the most important feature to notice is that linear maximum-likelihood initial density field, $\hdLa$, is effectively a Wiener filtered copy of $\dgo$. The second order correction, $\hdLb$, is effectively removing the first nonlinear contribution to $\dgo$ from the quadratic bias using the maximum-likelihood linear condition. At high signal to noise, we are simply inverting the map from $\dL$ to $\dgo$ perturbatively. However, the filtering that occurs because of the noise is crucial to reproducing the Fisher information associated with individual summary statistics~\cite{Baumann:2021ykm}. In fact, in this formulation, expanding the integrand at the saddle point already
reproduces the likelihoods associated with the standard hierarchy of summary
statistics, including the power spectrum, bispectrum, and higher-point
correlators generated by nonlinear bias.
Genuine loop corrections arise only in the presence of finite shot noise, through the integration over deviations from the saddle point.

Expanding the action around the saddle-point solution
$\dL=\hdL+\dL'$ leads to a Gaussian integral over the residual
field $\dL'$ with curvature
$R_2(k)=(\PL(k)b_{\rm lin}^2(k)+N^2)/(\PL(k)N^2)$, together with perturbative
corrections $\Delta R_2$ generated by nonlinear bias operators.
Evaluating the resulting expansion yields a finite set of contributions
to the marginalized likelihood, which can be organized unambiguously into
saddle-point (tree-level) terms and curvature-induced corrections.
Both classes of terms inherit explicit factors of
$[\PL b_{\rm lin}^2+N^2]^{-1}$ through the saddle-point solution and the inverse
curvature, as shown explicitly in Appendix \ref{app:Marginal likelihood}.

\subsection{Cramér–Rao Bound at the Field Level}\label{CR bound at the field level}

We will be particularly interested in the field-level likelihood, Eq.~\eqref{eq:field_level_likelihood}, marginalized over $\dL$. Our focus will be on measuring the cosmological parameters, $\theta_i$ and bias parameters, $b_i$, after marginalizing over the uncertainty in $\dL$ given a fixed map $\dgo$.

A cosmological analysis with the field level likelihood is, of course, conventional Bayesian inference applied to $\dL$ and all the cosmological parameters simultaneously. As such, the likelihood will define the posterior for a given observed map $\delta_g^{\rm obs}(\x)$. The Fisher matrix associated with this posterior is therefore
\beq
F_{ij} =- \left\langle \frac{\partial^2}{\partial \vartheta_i \partial \vartheta_j}\log \mathcal{P}[\{\vartheta\} |\delta_g^{\rm obs}] \right \rangle \ ,
\eeq 
where $\vartheta_i, \vartheta_j \in \{\vartheta \} \equiv \{ \vec b, \vec \theta \}$ are specific cosmological or bias parameters. It is important that this is the posterior given a fixed observed map $\dgo(\x)$ even though we have marginalized over $\dL(\x)$. This should be distinguished from the Fisher information for an observed summary statistic, such as $P_g^{\rm obs}(k)$, which is only given by the observed value of the summary statistic. We calculate this posterior by marginalizing over $\dL$ using the maximum likelihood reconstruction of the initial conditions described in the previous section. In this case, the field-level log likelihood is determined in
Eq.~\eqref{eq:marginalized likelihood}. Even if we are going to analyze our maps via summary statistics, this Fisher matrix is still useful as the Cramér–Rao bound state determines the minimum error measurement that can arise from any manipulation of $\dgo$. For any parameter $\theta_i$ in the model, the
variance of any unbiased estimator is bounded from below by the inverse of the
Fisher matrix,
\begin{equation}
\mathrm{Var}(\vartheta_i) \ge (F^{-1})_{ii}.
\end{equation}
Any unbiased estimator that saturates this bound can be defined as ``optimal" as no other estimator can provide a more precise measurement statistically.

Below we present representative blocks of the Fisher matrix that illustrate the
structure of the field-level Cramér–Rao bound. In particular, we show the
cosmological–cosmological sector, which reduces to the familiar Gaussian
power-spectrum Fisher matrix in the appropriate limit, and the linear
bias–linear bias sector, which illustrates how bias parameters enter on the same
footing as cosmological parameters. The remaining Fisher-matrix blocks are
discussed in Appendix \ref{app:Fisher matrix}.

\subsubsection{Linear Evolution}

One of the first checks of this approach is that it reproduces the expected results for linear evolution. Therefore, as a warm-up, we will set $\bquad=0$ and determine the Fisher matrix elements.

First, we will differentiate the marginalized log-likelihood with respect to cosmological
parameters $\theta_i$ and $\theta_j$. Treating $\blin$ as an independent parameter then yields 
\begin{equation}
\begin{aligned}
F_{\theta_i\theta_j}
&= -\int_k \frac{(2\pi)^3\delta_D(0)}{2}
\frac{b_{\rm lin}^4\,\partial_{\theta_i}\PL(k)\,\partial_{\theta_j}\PL(k)}
{[\PL(k)b_{\rm lin}^2+N^2]^2} \\
&\quad
-\int_k \frac{2b_{\rm lin}^4\,\partial_{\theta_i}\PL(k)\,\partial_{\theta_j}\PL(k)}
{[\PL(k)b_{\rm lin}^2+N^2]^3}\,
P_g^{\rm obs}(k)\,(2\pi)^3\delta_D(0) \\
&\quad
+\int_k \frac{3b_{\rm lin}^6\PL(k)\,
\partial_{\theta_i}\PL(k)\,\partial_{\theta_j}\PL(k)}
{[\PL(k)b_{\rm lin}^2+N^2]^4}\,
P_g^{\rm obs}(k)\,(2\pi)^3\delta_D(0) \ .
\end{aligned}
\end{equation}
In the large signal-to-noise limit, this
expression simplifies to
\begin{equation}
F_{\theta_i\theta_j}
\;\approx\;
\int_k \frac{V_{\rm survey}}{2}\,
\frac{b_{\rm lin}^4\,\partial_{\theta_i}\PL(k)\,
\partial_{\theta_j}\PL(k)}
{[\PL(k)b_{\rm lin}^2+N^2]^2},
\end{equation}
which coincides with the standard Gaussian power-spectrum Fisher matrix.

We can similarly consider the Fisher matrix elements associated with the linear bias. To simplify the functional form, we will vary the amplitude, holding the $k$-dependence fixed via
\beq
b_{\rm lin}(k)=\bar b_{\rm lin}K_1(k) \ . 
\eeq
The corresponding Fisher matrix element is
\begin{equation}
\begin{aligned}
F_{\rm lin,lin}
&=
-V_{\rm survey}\int_k
\frac{b_{\rm lin}^2 \PL(k)^2}{[\PL(k)b_{\rm lin}^2+N^2]^2}\,K_1(k)^2 \\
&\quad
+3V_{\rm survey}\int_k
\frac{b_{\rm lin}^2 \PL(k)^2}{[\PL(k)b_{\rm lin}^2+N^2]^3}\,
P_g^{\rm obs}(k)\,K_1(k)^2 
\end{aligned}
\end{equation}
Assuming $P_g^{\rm obs}(k)=b_{\rm lin}^2\PL(k)+N^2$, this simplifies to
\begin{equation}
F_{\rm lin,lin}
=
2V_{\rm survey}\int_k
\frac{b_{\rm lin}^2\PL(k)^2}{[\PL(k)b_{\rm lin}^2+N^2]^2}\,
K_1(k)^2 
\label{linear Fisher}
\end{equation}
We again reproduce the Fisher matrix for the power spectrum at high signal to noise.

The general conclusion here should have been anticipated: in a linear and Gaussian universe, the cosmic information is all contained in the power spectrum. At high signal to noise, the Cram\'er-Rao bound can therefore be saturated with an appropriate power spectrum estimator. 

\subsubsection{The Fisher Matrix in a Nonlinear Universe}

Nonlinear evolution is the primary reason that Fisher information for $\fnl$ in galaxy surveys is not literally equivalent to the Fisher information of the CMB or directly in the initial conditions.

At quadratic order in $\dL$ (or $\dgo$), most of the nonlinear information has been encoded in $\bquad$. At this stage there is very little difference between $b_2$ and $\fnl$, as they appear in $\bquad$ in the same way. The additional complication of $\fnl$ is that it appears both in $\blin$ and $\bquad$.

\paragraph{Mildly non-linear regime}
When $\delta_g(\k)\approx b_{\rm lin}(k)\dL(\k)$, or equivalently for the cutoff $k_{\rm max}$ is sufficiently small such that
$$
\int_{k'<k_{\rm max}}b_{\rm quad}(\k',\k-\k')\dL(\k')\dL(\k-\k')\ll b_{\rm lin}(k)\dL(k)
$$
the information of the quadratic bias goes to the bispectrum sector.

By differentiating the marginal likelihood eq.~\ref{eq:marginalized likelihood} with respect to the coefficient of the quadratic term, and then plugging in $P_g^{\rm obs}(k) = P_{\rm lin}(k) b_{\rm lin}^2(k)+N^2$ (we expect that the observation will agree with the deterministic model plus the noise), we obtain the Fisher matrix for the quadratic order as the following,
\begin{equation}
    \begin{aligned}
        &F_{\rm quad,quad}\\
        &=2\int_{k,k'} \frac{K_2(\k',\k)^2[\PL(k)b_{\rm lin}]^2[\PL(k')b_{\rm lin}]^2}{[\PL(k) b_{\rm lin}^2+N^2][\PL(k') b_{\rm lin}^2+N^2][\PL(k-k') b_{\rm lin}^2+N^2]}\\
      & +4V_{\rm survey}\int_{k,k'}\frac{K_2(\k\text{-}\k',\k')K_2(\k'\text{-}\k,\k)[\PL(k)b_{\rm lin}][\PL(k')b_{\rm lin}][\PL(|\k\text{-}\k'|)b_{\rm lin}]^2}{[\PL(k) b_{\rm lin}^2+N^2][\PL(k') b_{\rm lin}^2+N^2][\PL(|\k\text{-}\k'|) b_{\rm lin}^2+N^2]}
    \end{aligned}
\end{equation}

This expression is equivalent to the standard tree-level bispectrum Fisher matrix,
\begin{equation}
    \begin{aligned}
        F_{\rm quad, quad} = V_{\rm survey}\int_{k,k'} \frac{\frac{\partial }{\partial b_{\rm quad}}B(k,k',|\k-\k'|)\frac{\partial }{\partial b_{\rm quad}}B(\k,\k',|\k-\k'|)}{6[\PL(k) b_{\rm lin}^2+N^2][\PL(k') b_{\rm lin}^2+N^2][\PL(|\k\text{-}\k'|) b_{\rm lin}^2+N^2]}
    \end{aligned}
\end{equation}
Thus quadratic operators contribute Fisher matrix through its imprint on the bispectrum.

Even if fiducial quadratic biases are small, nonlinear gravitational evolution generates an effective quadratic operator through the SPT kernel $F_2$. Since this term is multiplied by the nonzero linear bias coefficient $b_{\rm lin}$. It induces mixing between linear and quadratic sectors. Physically, this corresponds to correlations between linear fluctuations and nonlinear mode-coupling contributions.
Schematically,
\begin{equation}
    \begin{aligned}
        &F_{\rm lin,quad}\\
        &=2V_{\rm survey}\int_{k,k'}\frac{K_2(\k',\k)K_1(|\k'\text{-}\k|)F_2(\k',\k) [\PL(k)b_{\rm lin}]^2[\PL(k')b_{\rm lin}]^2}{[\PL(k) b_{\rm lin}^2+N^2][\PL(k') b_{\rm lin}^2+N^2][\PL(|\k\text{-}\k'|) b_{\rm lin}^2+N^2]}\\
      & +4V_{\rm survey}\int_{k,k'}\frac{K_2(\k\text{-}\k',\k')K_1(k')F_2(\k'\text{-}\k,\k)[\PL(k)b_{\rm lin}][\PL(k')b_{\rm lin}][\PL(|\k\text{-}\k'|)b_{\rm lin}]^2}{[\PL(k) b_{\rm lin}^2+N^2][\PL(k') b_{\rm lin}^2+N^2][\PL(|\k\text{-}\k'|) b_{\rm lin}^2+N^2]}
    \end{aligned}
\end{equation}

Primordial non-Gaussianity contributes to both linear and quadratic operators. It modifies the linear bias through scale-dependent terms and generates a primordial bispectrum contribution. Consequently, the Fisher matrix element $F_{\rm f_{NL}, f_{NL}}$ receives contributions from: the linear-linear (power spectrum) block, the quadratic-quadratic (bispectrum) block, and the linear-quadratic(also bispectrum) block,
\begin{equation}
    \begin{aligned}
        &F_{f_{\rm NL},f_{\rm NL}} \\
        &= \Bigg\{\int_{k,k'} \frac{\Big[b_{\rm lin}(k)b_{\rm lin}(k')b_{\rm lin}(|\k\text{-}\k'|)\Big(B_t(k,k',|\k\text{-}\k'|)+2\frac{b_{\Phi}(|\k\text{-}\k'|R_*)^{\Delta}}{\mathcal{T}(k)}F_2(\k,\k')\PL(k)\PL(k')\Big)\Big]^2}{6[\PL(k) b_{\rm lin}^2+N^2][\PL(k') b_{\rm lin}^2+N^2][\PL(k-k') b_{\rm lin}^2+N^2]}\\
        &\qquad+2\int_k
\frac{b_{\rm lin}^2\PL(k)^2}{[\PL(k)b_{\rm lin}^2+N^2]^2}\,
\Big[\frac{b_{\Phi}}{\mathcal{T}(k)}(kR_*)^{\Delta}\Big]^2 \Bigg\}V_{\rm survey} \ .
    \end{aligned}
\end{equation}
The expression above corresponds to tree-level statistics evaluated around the linear fiducial model $\delta_g(\k)\approx b_{\rm lin}(k)\dL(\k)$. Including loop corrections consistently would require incorporating higher order correlation functions and non-Gaussian covariance contributions. For sufficiently small $k_{\rm max}$, loop terms remain subdominant relative to the tree-level contribution and do not significantly modify the Fisher matrix. We therefore restrict to tree level expression in this work. Details of loop-level structure within the saddle point approximation are discussed in Appendix~\ref{app:loop_term}.

For precision data, modeling to higher order in $\dL$ will ultimately involve higher-order bias parameters. For example, at cubic order, we should include
\beq
\delta_g(\k) \supset \int_{\k_1,\k_2} b_{\rm cub}(k_1,k_2,\k\text{-}\k_1\text{-}\k_2) \delta(\k_1)\delta(\k_2) \delta(\k\text{-}\k_1\text{-}\k_2) \ ,
\eeq
and would find the diagonal Fisher element for $b_{\rm cub}$ is a six-point function. In fact, the Fisher matrix elements involving an $n$th order bias parameters are the connected $n+1$ through $2n$ correlation functions~\cite{Baumann:2021ykm}. This ensures that for $n>2$ ($b_{n>2}$), biasing is not degenerate with the primordial bispectrum ($\fnl$).

\paragraph{Deeply non-linear regime}
The saddle point expansion is not parametrically controlled in the deeply non-linear regime 
$$
\int_{k'<k_{\rm max}}b_{\rm quad}(\k',\k-\k')\dL(\k')\dL(\k-\k')\gg b_{\rm lin}(k)\dL(k)
$$
which marks the regime where non-linear contributions dominate the likelihood and the power-counting underlying the saddle-point expansion, and hence the truncation of higher-order terms adopted in this work, is no longer well controlled.

And in this regime the information associated with lower-order bias parameters is redistributed across higher-order correlation functions, which are not captured within the present saddle-point treatment. For example, if $\delta_g(\k)=b_1\dL(\k)+b_2'\Lambda\int_{k'}\dL(\k')\dL(\k\text{-}\k')$ with $\Lambda \gg 1$ and fiducial value of $b_2'=1$, the Fisher matrix contribution from a Gaussian distributed n-point correlation function $S_n(\k_1,\ldots,\k_n)$ will be
\begin{equation}
    \begin{aligned}
       &F_{2,2}^{(\rm n\,point \;correlator)} \propto V_{\rm survey}\int_{k_1,\ldots, k_n}\frac{\Big[\partial_{b_2'}S_n(\k_1,\ldots,\k_n)\Big]^2}{[P_g(k_1)+N^2]\ldots[P_g(k_n)+N^2]}\delta_D(\sum_i\k_i)\\
        &\approx  V_{\rm survey}\int_{k_1,\ldots, k_n}\frac{\Big[ n\Lambda^n \int_{k'}\PL(k')\PL(|\k'+\k_1|)\ldots\PL(|\k'+\ldots+\k_{n\text{-}1}|) \Big]^2\delta_D(\sum_i\k_i)}{[2\Lambda^2\int_{k_1'}\PL(k_1')\PL(|\k_1\text{-}\k_1'|)+N^2]\ldots[2\Lambda^2\int_{k_n'}\PL(k_n')\PL(|\k_n\text{-}\k_n'|)+N^2]}\\
        &\approx V_{\rm survey}\int_{k_1,\ldots, k_n}\frac{\Big[ n \int_{k'}\PL(k')\PL(|\k'+\k_1|)\ldots\PL(|\k'+\ldots+\k_{n\text{-}1}|) \Big]^2}{[2\int_{k_1'}\PL(k_1')\PL(|\k_1\text{-}\k_1'|)]\ldots[2\int_{k_n'}\PL(k_n')\PL(|\k_n\text{-}\k_n'|)]}\delta_D(\sum_i\k_i)
    \end{aligned}
\end{equation}
up to the symmetrization factor. In other words, in the deeply non-linear regime, the possibility of extracting  additional information from higher-order correlation function cannot be excluded within the field level analysis based on saddle point approximation. A consistent treatment of this regime would require extending the analysis beyond the saddle point expansion.

\subsection{Summary}

There are several important takeaways from this section that will be important for the rest of the paper. Most significantly, the field-level CR bound is determined by the low-point correlation functions. For $\fnl$, it is a combination of the power spectrum and bispectrum. At high signal to noise, the field level Fisher matrix elements for $\fnl$ agrees with the power spectrum and bispectrum estimator forecasts~\cite{Scoccimarro:2003wn}. In this regard, the field-level inference is not addition information beyond what is already included in analyses for the power spectrum and bispectrum of LSS~\cite{Slosar:2008hx,Cabass:2022wjy,Cabass:2022ymb,DAmico:2022gki,DAmico:2022osl,Chudaykin:2025vdh}. The second critical feature is that the bias parameters, $b_n$ (and generalization therefore), that appear at order $\delta^n$, appear in the Fisher matrix primarily through their contribution to the connected $(n+1)$- to $2n$-point functions.

These two statements imply that field-level inference for $\fnl$ is effectively equivalent to the power spectrum and bispectrum measurements that include $\blin$ and $\bquad$ as free parameters, but hold $b_{n>2}$ fixed. However, a realistic analysis using only these two summary statistics will ultimately find weakened constraints due to marginalizing over higher-order terms.  One can recover the field level sensitivity by then including additional higher-point summary statistics which break these degeneracies.

It is worth observing that many approaches to combined power spectrum and bispectrum analyses use different choices for $\kmax$ across the different correlators~\cite{Beyond-2pt:2024mqz}. In principle, this is compatible with the CR bound, indicating that the constraints will be maximized if all the correlators use the largest value of $\kmax$. However, in practice, comparison with simulations~\cite{Ivanov:2021kcd,Eggemeier:2021cam} would suggest that unbiased results require the field-level analysis to use the most conservative $\kmax$, which would be equivalent to a smaller $\kmax$ across all correlators. This would reduce the total information available to field-level inference compared to optimizing $\kmax$ for each summary statistic. While this tension remains unresolved, it seems reasonable to assume that the CR bound from field-level inference and the combined sum over all the individual correlators should give the same result.

\section{Scale-Dependent Bias and Local Non-Gaussianity}\label{sec:local}

The realization that local non-Gaussanity generates scale-dependent bias~\cite{Dalal:2007cu} has greatly simplified the use of galaxy survey data for constraining inflation~\cite{Slosar:2008hx,BOSS:2012vpn,eBOSS:2021jbt,Chaussidon:2024qni}. It moves information from the bispectrum of the initial conditions into the power spectrum of the galaxy distribution.

The question that is less clear is whether the power spectrum analysis is optimal~\cite{Hamaus:2011dq,dePutter:2016trg}. It would be surprising if the scale-dependent bias was a perfect measurement of the local shape; yet, if it isn't, it would imply a more precise measurement is possible with the same data. In this section, we will address the optimality of scale-dependent bias and how that informs the interpretation of realistic forecasts.

The key difference between local and equilateral non-Gaussianity is that the local signals peaks on large scale (small $k$) while the equilateral signals peaks on small scale (large $k$)~\cite{Gleyzes:2016tdh,Green:2022bre,Green:2023uyz}. As a result, when discussing constraints on $\fnlloc$, we can largely ignore the nonlinearity of structure formation that impacts small scales. In contrast, for $\fnleq$, the degeneracy with nonlinear biasing is dominant factor in determining the sensitivity~\cite{Baldauf:2016sjb}.

\subsection{Signal and Noise}

Although $\fnlloc$ can be measured in the galaxy power spectrum, at a fundamental level, all the information about primordial non-Gaussianity is encoded in the bispectrum of the initial conditions~\cite{Creminelli:2006gc}. It is the subsequent evolution under the force of gravity, starting from these initial conditions, that gives rise to non-linear structures, and leads to the formation of galaxies. A complete analysis of the matter field, although not practically feasible, can still serve as the theoretical upper bound on the amount of information that one can extract from galaxy surveys.

Our goal here is to calculate and compare the Fisher information for $\fnlloc$ associated with three types of measurements:
\begin{enumerate}
    \item Field-level measurement of the matter field $\delta(\vec{x})$,
    \item Galaxy power spectrum analysis on a single type of galaxy, $\delta_g(\vec{x})$,
    \item Galaxy power spectrum analysis on multiple types of galaxy, $\delta_{g,i}(\vec{x})$ \ .
\end{enumerate}
While we expect that the matter field ultimately contains more information than the tracers alone, we wish to understand if this is true for practical measurements. Specifically, cosmic variance cancellation through multiple tracers~\cite{Seljak:2008xr} enables measurements that are not limited by the usual mode counting estimates~\cite{Hamaus:2011dq,Abramo:2013awa}. As a result, it is not clear if or how the CR bound in the presence of multiple tracers (or cross-correlations with other maps~\cite{Schmittfull:2017ffw,Munchmeyer:2018eey,Smith:2018bpn}) is related to the CR bound for the underlying map.

The quantity that we are interested in is the Fisher information for $\fnlloc$,
\begin{equation}\label{eqn:sigma_fNL}
	\left(\frac{S}{N}\right)^2 \equiv F_{\fnlloc,\fnlloc} = -\left\langle\frac{\partial^2}{\partial {\fnlloc}^2} \log\mathcal{L}\right\rangle\Bigg|_{\fnlloc=0}\geq \frac{1}{\sigma^2(\fnlloc)}  \, .
\end{equation}
For the local shape, the signal from scale-dependent bias peaks at small-$k$, far from the nonlinear regime. As a result, the expressions for the Fisher matrix simplify as we can largely ignore the bias expansion and nonlinearity, up to marginalizing over $b_1$. The expressions for the Fisher information are derived in App.~\ref{app:fisher_information_derivation} for the three types of analyses outlined above in details, which we summarize here
\begin{align}\label{eqn:SNR_expression_quote}
	\left(\frac{S}{N}\right)_m^2 & = V_{\rm survey}\left(\frac{12A_sk_{\max}^3}{25\pi^2}\right) \log\left(\frac{k_{\max}}{k_{\min}}\right) \nonumber \\
    \left(\frac{S}{N}\right)_{g,\,s}^2& = V_{\rm survey}\left(\frac{2b_{\Phi}^2}{b_1^2}\right)\int\frac{d^{3}\vec{k}}{(2\pi)^{3}}\frac{1}
	{\mathcal T^{2}(k)} \nonumber \\
    \left(\frac{S}{N}\right)_{g,\,d}^2 &=  V_{\rm survey}\frac{\left[b_1^{(1)}b_{\Phi}^{(2)}-b_1^{(2)}b_{\Phi}^{(1)}\right]^2}{\left[\left(b_1^{(1)}\right)^2N_2^2+\left(b_1^{(2)}\right)^2N_1^2\right]} \int \frac{d^3\vec{k}}{(2\pi)^3}P_{\Phi}(k)
    \, ,
\end{align}
where $N^2$ abstractly denotes the noise level in the galaxy survey, and we have assumed $P\gg N^2$. For matter or single tracers, at high signal to noise, these expressions are cosmic variance limited. In contrast, multi-tracers Fisher information is limited by shot noise for each galaxy.

From these expressions, one can already see the need for a careful analysis of the Cram\'er-Rao bound for these types of measurements. Using the analytic expressions in Eq.~\eqref{eqn:DM_density_field} and Eq.~\eqref{eqn:Gaussian_power_spectrum}, we see that for $P \gg N^2$, $(S/N)_m^2$ and $(S/N)_{g,\,d}^2$ both scale as $\log(k_{\max}/k_{\min})$, while $(S/N)_{g,\,s}^2$ scales as $(k_{\max}/k_{\min})$. The log scaling is the familiar behavior of the information directly in the initial conditions~\cite{Babich:2004gb}. Yet, if it were possible to take the limit $k_{\min} \to 0$ within the regime of validity of these expressions, it would suggest the single tracer analysis has most/all of the information about $\fnlloc$. Of course, more information must be contained in the multi-tracer analysis (and the full matter distribution), and therefore the resolution is that (a) the single tracer measurement must be far from optimal and (b) the $k_{\min} \to 0$ limit is not attainable holding the other parameters fixed. The latter is not necessarily obvious, as future surveys measure progressively more $k$-modes as we increase the volume. However, what is crucial is that the range of measurable $k$-modes is bounded by the requirement that $P \gg N^2$, which does not hold as $k\to 0$ in the presence of shot noise (which is unavoidable for galaxies). It is also true that the bias coefficients are not free parameters in principle, but are related to the mass of the galaxies in a given survey (and hence the total number density). To account for the potential correlation between $b_1$ and $\bar n$, we will use a concrete halo formation model to show that $(S/N)^2_m$ dominates over the Fisher information from the two other surveys.

To illustrate this idea, throughout this section we use simplified analytic expressions to estimate ratios of Fisher information between the different analyses outlined above, which capture the correct qualitative behavior. In parallel, we also compute these ratios using exact expressions for the halo statistics and numerical integration over $k$ when evaluating the Fisher information, as detailed below. The figures presented in the remainder of this section (\textit{cf.} Figs.~\ref{fig:bias_parameters}–\ref{fig:fisher_comparison}) are based on these more accurate numerical results, which are in good qualitative agreement with the behaviors from the analytic estimates, and thus serve as a useful consistency check.

\subsection{Halo Bias Models}

The signal of scale-dependent bias is proportional to $\fnlloc b_\Phi$. The sensitivity to $\fnlloc$ is therefore determined by $b_\Phi$, including our ability to predict $b_\Phi$ independently of the data. In addition, the range of $b_\Phi$ that can be achieved with realistic samples is also critical to determining the ultimate limits of the measurement of $\fnlloc$ from a given map. To address these issues, we will use the Press-Schechter / excursion set formalism to determine values number density and bias of halos~\cite{Press:1973iz, Bond:1990iw}. Since galaxies are found in halos, to our level of approximation, this will be sufficient to determine the values of $\bar n_g$ and $b_\Phi$ that arise within a physical model, and we will use the terms ``halos" and ``galaxies" interchangeably. As we will see, the conclusions will not dependent crucially the specifics\footnote{The reason for choosing a model is primarily because we do want to allow $b_\Phi$ and $\bar n$ to be free and independent parameters in a way that could lead to unphysical conclusions. E.g.~if $b_\Phi \to \infty$, then the Fisher information in the galaxies will exceed the information in the natter distribution.} of the model.

A common perspective of halo formation is that the number density of halos is determined from at the locations of peaks in the initial density field, i.e.~a local process in Lagrangian coordinates. The location of halos at low redshifts then include the evolution from their locations in the initial density field through the force of gravity, i.e.~mapping from Lagrangian to Eulerian positions. This leads to the standard relations between the Lagrangian biases, $b_{1,\Phi}^L$, and Eulerian bias, $b_{1,\Phi}^E$, namely~\cite{Dalal:2007cu} \begin{align}\label{eqn:b_Lagrangian_vs_Eulerian}
	b_1^E &= b_1^L + 1 \, , \nonumber \\
    b_{\Phi}^E &= b_{\Phi}^L \, .
\end{align}
The derivation of these relationships are reviewed in Appendix~\ref{app:exact_expressions}.
These bias parameters represent the relationship between the density of galaxies and the underlying matter density contrast, or Newtonian potential, meaning that
\begin{align}\label{eqn:bias_derivatives}
	b^L_1 &= \frac{\partial \delta_h}{\partial \delta}\Bigg|_{\delta=0,\Phi=0} = \frac{1}{\bar{n}_h^L} \frac{\partial n_h^L}{\partial \delta}\Bigg|_{\delta=0,\Phi=0} = \frac{\partial \log n_h^L}{\partial \delta}\Bigg|_{\delta=0,\Phi=0}\nonumber \\
	f_{\mathrm{NL}}b^L_{\Phi} &= \frac{\partial \delta_h}{\partial \Phi}\Bigg|_{\delta=0,\Phi=0} = \frac{1}{\bar{n}_h^L} \frac{\partial n_h^L}{\partial \Phi}\Bigg|_{\delta=0,\Phi=0}= \frac{\partial \log n_h^L}{\partial \Phi}\Bigg|_{\delta=0,\Phi=0}\, ,
\end{align}
and similarly for the Eulerian frame. In this precise sense, the determination of these parameters is therefore directly tied to the number density of galaxies, which also impacts the minimum\footnote{In realistic surveys, shot noise is determined by target selection, but here we are interested in the ultimate limits of sensitivity.} shot noise amplitude. To better understand the range of possible values of these parameters, we will need a model for $n_h$.

In the Press-Schechter model, the mean halo number density is related to the matter field by a theoretical model. In the Lagrangian frame, when there is no over-density ($\delta=0$), the halo number density per mass is given by
\begin{equation}\label{eqn:Press-Schechter}
	\bar{n}_h^L(M)= \sqrt{\frac{2}{\pi}}\frac{\bar{\rho}}{M}\frac{\delta_c}{\sigma(M)}\frac{d\,\log\sigma(M)}{dM}\exp\left(-\frac{\delta_c^2}{2\sigma(M)^2}\right) \, ,
\end{equation}
where $\bar{\rho}$ is the mean matter mass density, and $\sigma(M)$ is the variance of density perturbations smoothed over a sphere of size $R(M)=(3M/4\pi\bar{\rho})^{1/3}$ and $\bar{\rho}=\Omega_m(3H_0^2/8\pi G)$, given by
\begin{equation}\label{eqn:sigma_def}
	\sigma^2(M) = \int \frac{d^3\vec{k}}{(2\pi)^3} P(k)W^2\left(kR(M)\right)  \, ,
\end{equation}
and we have defined the top-hat filter function
\begin{equation}\label{eqn:top_hat}
	W(kR) \equiv \frac{3\left(\sin(kR)-(kR)\cos(kR)\right)}{(kR)^3}  \, .
\end{equation}
Given a region with an average overall over-density $\delta$, the spherical-collapse criterion for fluctuations is modified to $\delta_c\to \delta_c-\delta$. In other words, the average over-density brings all the fluctuation closer to the barrier. As a result, one finds
\begin{equation}\label{eqn:halo_deltac_modulation}
	n_h^L(M;\delta) = \bar{n}_h^L(M)\Bigg|_{\delta_c\to \delta_c-\delta} \, .
\end{equation}
Combining Eq.~\eqref{eqn:bias_derivatives}, Eq.~\eqref{eqn:Press-Schechter}, and Eq.~\eqref{eqn:halo_deltac_modulation}, we can now derive the bias parameters
\begin{align}\label{eqn:b1_deriv}
	b_1^E = 1+b_1^L = 1+ \frac{1}{\delta_c}\left[\frac{\delta_c^2}{\sigma^2(M)}-1\right] \, .
\end{align}
Scale-dependent bias arises in a similar way, however, the main effect is to increase the variance of the small fluctuations in a region of high $\Phi$, $\sigma^2(M) \to \sigma^2(M) (1 + 2 \fnlloc \Phi)$~\cite{Dalal:2007cu}. As a result, $\Phi$ directly biases the formation of halos so that 
\begin{align}\label{eqn:bphi_deriv}
	b_{\Phi}^L = b_{\Phi}^E = 2\left[\frac{\delta_c^2}{\sigma^2(M)}-1\right]\, .
\end{align}
This formula is particularly useful because it correlates $b_\Phi$ with the parameters that control the formation of halos, rather than being an independent parameter\footnote{There has been some recent concern that for galaxies, rather than halos, the relationship between $b_\Phi$ and $b_1$ may differ from this simple formula~\cite{Barreira:2020kvh,Barreira:2020ekm,Barreira:2021ueb,Barreira:2022sey,Lazeyras:2022koc}. Using this formula could then bias the results, while marginalizing over $b_\Phi$ would lead to much weaker constraints on $\fnlloc$. While there is reason to think these concerns are overstated~\cite{Dalal:2025eve}, our results do not require a resolution to this question. For our purposes, we only need a model for the fiducial value of $b_\Phi$ in realistic settings. Our results will depend only weakly on the exact model for $b_\Phi$.}.

At this point, one could simply forecast the sensitivity of various probes using the Press-Schechter bias parameters as the fiducial values. However, in the interest of providing a more analytic understanding of the signal-to-noise ratio, we will approximate the shape of the matter spectrum using the following broken power law approximation:
\begin{equation}\label{eqn:Pk_3}
	T(k) \approx \begin{dcases*}
		1 & if $k<k_{\mathrm{eq}}$ \\
		\left(\frac{k_{\mathrm{eq}}}{k}\right)^{3/2} & if $k>k_{\mathrm{eq}}$
	\end{dcases*}	 \, ,
\end{equation}
where the  pivot scale $k_{\mathrm{eq}}$ is the scale that enters the horizon at matter radiation equality. In this model, $P(k)\sim k$ in large scale when $k<k_{\mathrm{eq}}$, and $P(k)\sim k^{-2}$ in small scale when $k>k_{\mathrm{eq}}$. We anticipate that the statistics for halos with size $R$ will depend on the dimensionless quantity $k_{\mathrm{eq}}R$, which takes the typical value of
\begin{align}\label{eqn:kR}
	k_{\mathrm{eq}}R = \left(\frac{2Mk_{\mathrm{eq}}^3}{\Omega_mH_0^2M_p^2}\right)^{1/3} \sim 0.02 \, \left(\frac{k_{\mathrm{eq}}}{0.01\,\mathrm{Mpc}^{-1}}\right)\left(\frac{M}{10^{12}\,M_{\odot}}\right)^{1/3}\left(\frac{0.12}{\Omega_mh^2}\right)^{1/3}     \, ,
\end{align}
where $M_p\equiv G^{-1/2}$ is the Planck mass with $G$ being the Gravitational constant, $\Omega_m$ is the cold dark matter abundance today, and we denote the Hubble constant today as $H_0=100\,h$\,km/s/Mpc. Therefore, for Milky way sized halo, $k_{\mathrm{eq}}R\ll 1$. We can also define the halo mass where $\sigma$'s scaling behavior turnovers as $M_{\mathrm{eq}}$, which is the mass of a halo with radius comparable to the transition scale of $P(k)$, \textit{i.e.} $1/k_{\mathrm{eq}}$
\begin{align}\label{eqn:M_eq}
	M_{\mathrm{eq}} \sim \frac{4\pi\bar{\rho}}{3k_{\mathrm{eq}}^{3}} 
    \sim 1.4\times 10^{17}\,M_{\odot}\left(\frac{0.01\,\mathrm{Mpc}^{-1}}{k_{\mathrm{eq}}}\right)^{3}\left(\frac{\Omega_mh^2}{0.12}\right)  \, .
\end{align}
For the rest of the paper, we will focus on halos with mass smaller than this transition mass scale, \textit{i.e.} $M\ll M_{\mathrm{eq}}$. 

With a particular form of $P(k)$ in Eq.~\eqref{eqn:DM_density_field}, Eq.~\eqref{eqn:Gaussian_power_spectrum}, and Eq.~\eqref{eqn:Pk_3}, We can now compute $\sigma(M)$ in Eq.~\eqref{eqn:sigma_def}. The integral can be computed exactly, which is done in App.~\ref{app:exact_expressions}. The expression for $\sigma^2(M)$, in the $M\ll M_{\mathrm{eq}}$, is given by a power series expansion on Eq.~\eqref{eqn:sigma2_exact} 
\begin{align}\label{eqn:sigma_def_small}
	\sigma^2(M)
	\sim 1.8\,\left(\frac{10^{12}\,M_{\odot}}{M}\right)^{1/3}\left(\frac{k_{\mathrm{eq}}}{0.01\,\mathrm{Mpc}^{-1}}\right)^{3}\left(\frac{A_s}{2.1\times 10^{-9}}\right) \left(\frac{0.12}{\Omega_mh^2}\right)^{5/3}D^2(z)  \, .
\end{align}
Putting this back to the Press-Schechter formalism in Eq.~\eqref{eqn:Press-Schechter} allows us to estimate the halo mass function. To do so, it is convenient for us to first derive a cutoff mass, $M_*$, by $\sigma^2(M_*)=2\delta_c^2$, which evaluates to
\begin{align}\label{eqn:M_star}
	M_* 
	\sim 2.2\times 10^{12}\,M_{\odot}\,\left(\frac{k_{\mathrm{eq}}}{0.01\,\mathrm{Mpc}^{-1}}\right)^{9}\left(\frac{A_s}{2\times 10^{-9}}\right)^3\left(\frac{0.12}{\Omega_mh^2}\right)^{5} D^6(z)\, .
\end{align}
Physically, $M_*$ corresponds to the maximum size of halo before the halo abundance is exponentially suppressed as in Eq.~\eqref{eqn:power_spectrum_ansatz}. Then, we define the mean halo number density per log-mass as $f(M) \equiv \frac{dN_h}{d\log MdV}= \bar{n}_h^L(M)M$, which can be computed exactly using Eq.~\eqref{eqn:Press-Schechter}. This is also done in App.~\ref{app:exact_expressions}, and an asymptotic form can be obtained by a power series expansion on Eq.~\eqref{eqn:hmf_differential_exact}
\begin{align}\label{eqn:nbar_halo}
	f(M) 
    &\sim 5.5\times 10^{-3}\,\mathrm{Mpc}^{-3}\, \left(\frac{10^{12}\,M_{\odot}}{M}\right)^{5/6}\left(\frac{2\times 10^{-9}}{A_s}\right)^{1/2} \nonumber \\
    &\quad\times \left(\frac{0.01\,\mathrm{Mpc}^{-1}}{k_{\mathrm{eq}}}\right)^{3/2}\left(\frac{\Omega_mh^2}{0.12}\right)^{11/6} D^{-1}(z)\exp\left(-\frac{M^{1/3}}{M_*^{1/3}}\right) \, ,
\end{align}
We see in Eq.~\eqref{eqn:sigma_def_small} that $f(M)\sim M^{-5/6}$ for small halo mass $M\ll M_*$. For heavy halos with $M\gg M_*$, the halo mass function is exponentially suppressed, and thus will not be able to provide enough statistics for measuring $f_{\mathrm{NL}}$. For analytic estimates, we will focus only on lighter halos for the rest of the paper and ignore the exponential factor in Eq.~\eqref{eqn:nbar_halo}, although it will be retained in the numerical results shown later.

Recall from Eq.~\eqref{eqn:b1_deriv} that the bias coefficients of halos with mass $M$ depend on $\sigma(M)$. Generally we find from Eq.~\eqref{eqn:sigma_def_small} that $\sigma(M)\gtrsim 1$ for light halos as $M$ moves further away from $M_*$, so that the bias coefficients in Eq.~\eqref{eqn:b1_deriv} are asymptotically constant 
\begin{align}\label{eqn:b1_single}
	b_1^E(M) &\sim 1-(1/\delta_c) \nonumber \\
	b_{\Phi}^E(M) &\sim -2  \, .
\end{align}
\begin{figure}
	\includegraphics[width=1\textwidth]{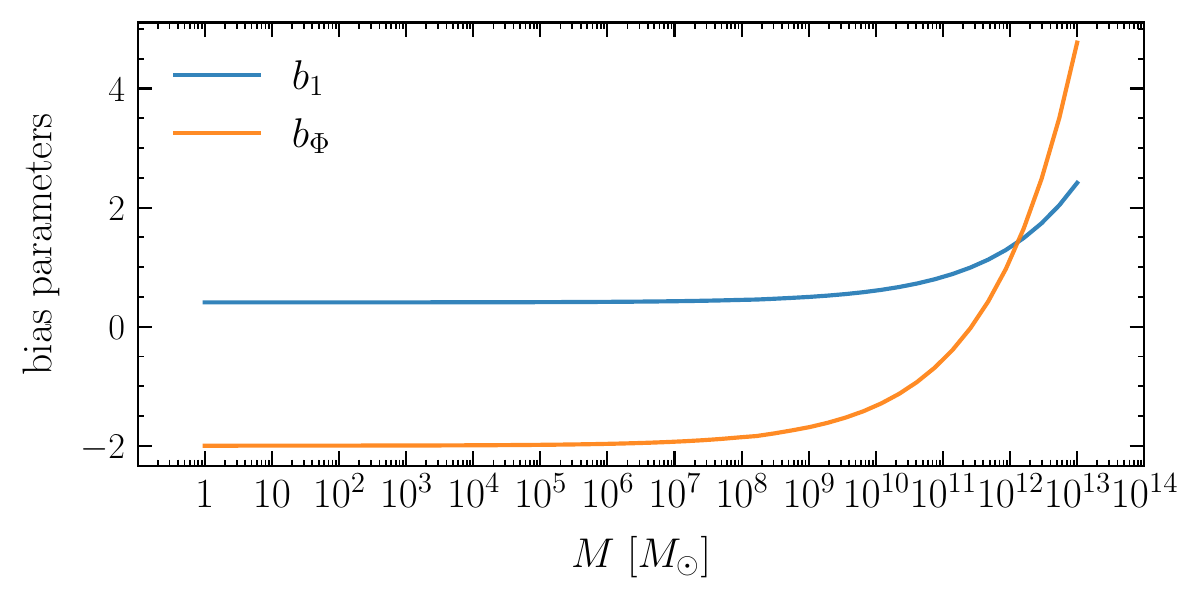} 
	\caption{Bias parameters in the Eulerian frame, $b_1$ and $b_{\Phi}$, for halos with mass $M$, computed using Eqs.~\eqref{eqn:b1_deriv}-\eqref{eqn:bphi_deriv} with exact expressions for $\sigma^2(M)$ in Eq.~\eqref{eqn:sigma_def} and Eq.~\eqref{eqn:sigma2_exact}. The numerical values of the cosmological parameters used are summarized in Table.~\ref{tab:parameters}. Note that the naive analytic estimates in Eq.~\eqref{eqn:b1_single} approximate both $b_1$ and $b_{\Phi}$ as constants, which is only true when $M$ is sufficiently small.}\label{fig:bias_parameters}
\end{figure}
In Fig.~\ref{fig:bias_parameters}, we numerically evaluate the bias parameters in Eqs.~\eqref{eqn:b1_deriv}-\eqref{eqn:bphi_deriv} using the exact expressions for $\sigma^2(M)$ in Eq.~\eqref{eqn:sigma2_exact} as a function of halo mass. We verify that the bias parameters are asymptotically constant as $M$ decreases, while large mass halos (but still smaller than $M_*$) receive an $\mathcal{O}(1)$ correction in the numerical values of the bias parameters. 

The key thing to observe at this point is that for smaller halos, the scale-dependent bias remains non-zero. This is important because this is the regime with the largest number densities. As a result, it will be the amount of information in the scale-dependent bias will not be limited by the vanishing of $b_\Phi$ at high single-to-noise. 

\begin{figure}[!ht]
	\includegraphics[width=\textwidth]{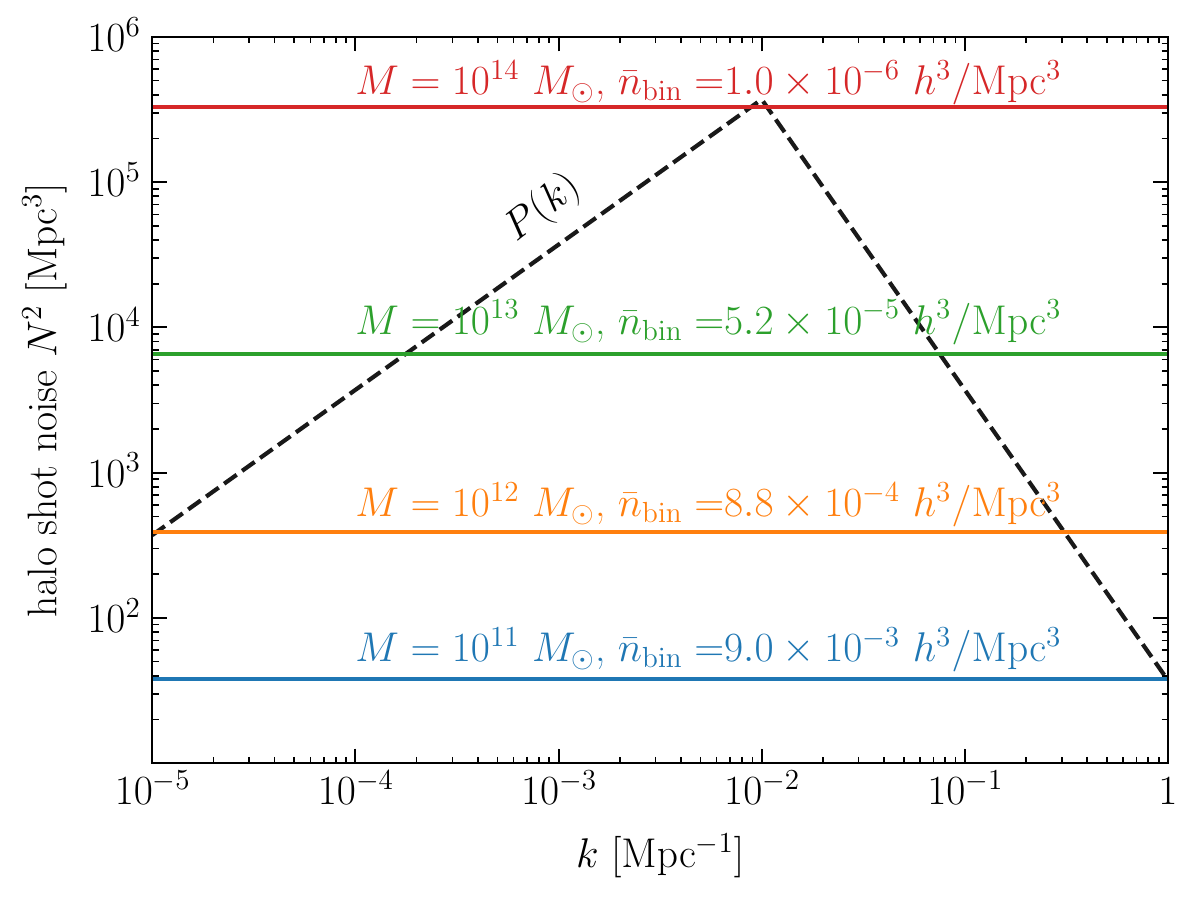} 
	\caption{Comparison between the matter power spectrum, $P(k)$, and the halo shot noise, $N^2$, for a selection of different halo masses $M$. Here $P(k)$ is defined in Eq.~\eqref{eqn:DM_density_field} and Eq.~\eqref{eqn:Pk_3}, while the halo shot noise $N^2$ is defined in Eq.~\eqref{eqn:poisson_noise_def}, is evaluated numerically using the exact expression of $f(M)$ in Eqs.~\eqref{eqn:hmf_convenient}-\eqref{eqn:hmf_differential_exact} and the Press-Schechter prediction of halo number density in Eq.~\eqref{eqn:Press-Schechter}. We also compute the mean halo number density in each mass bin, $\bar{n}_{\mathrm{bin}}\equiv N^{-2}$, for reference, assuming $h=0.7$~\cite{Planck:2018vyg}. Fiducial values of parameters chosen are listed in Table.~\ref{tab:parameters}.}\label{fig:power_spectrum_shot_noise}
\end{figure}

\begin{figure}[!ht]
\includegraphics[width=\textwidth]{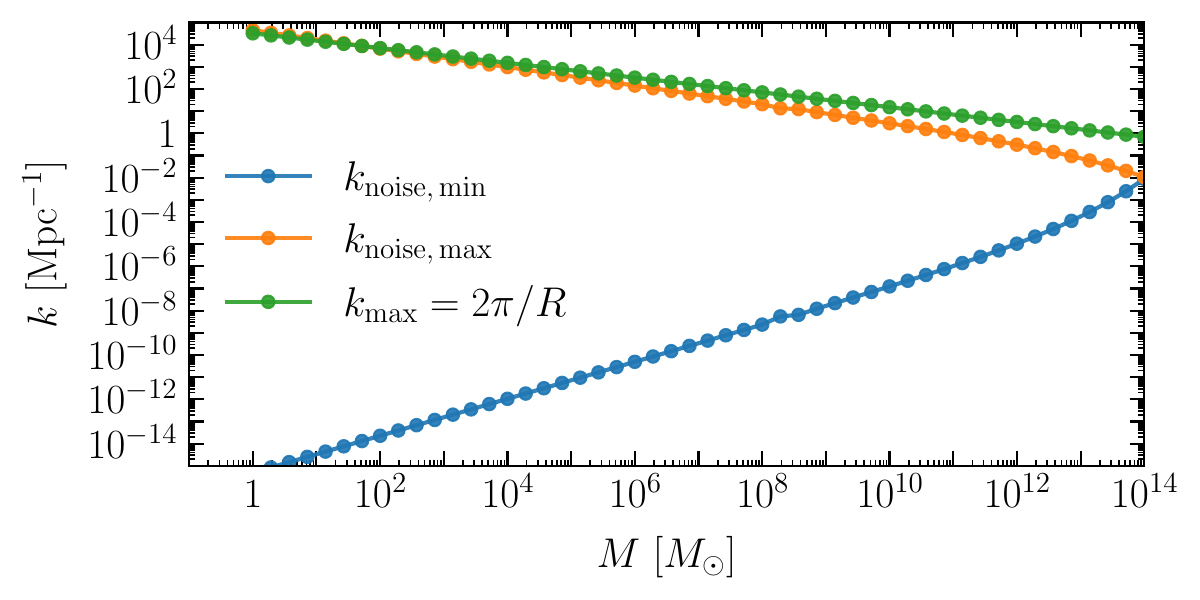} 
	\caption{Minimum and maximum wavenumber $k$ for different types of analysis. Here $k_{\min}\,\mathrm{noise}$ and $k_{\max}\,\mathrm{noise}$ are defined to be minimum and maximum $k$ such that $P(k)>N^2$ with $N^2$ being the halo shot noise. The expression of $P(k)$ is given in Eq.~\eqref{eqn:DM_spectrum} and Eq.~\eqref{eqn:Pk_3}, and $N^2$, defined in Eq.~\eqref{eqn:poisson_noise_def}, is evaluated numerically using the exact expression of $f(M)$ in Eqs.~\eqref{eqn:hmf_convenient}-\eqref{eqn:hmf_differential_exact} and the Press-Schechter prediction of halo number density in Eq.~\eqref{eqn:Press-Schechter}. We also show the smallest scale of a halo analysis, $2\pi/R$, with $R$ being the size of a halo with mass $M$. Fiducial values of parameters chosen are listed in Table.~\ref{tab:parameters}.}\label{fig:k}
\end{figure}

In order to quantify the amount of information in the halos, we will finally need the amplitude of Poisson noise associated with the different halo populations. To a good approximation, this is determined by the mean number of halos that are being measured. For the purpose of determining the maximum amount of information in halos, we will assume all the halos in a given mass range are observed. For example, if we are interested in measuring halos within a log mass bin centered at $\log M$ with width $\Delta(\log M)$, then the shot noise of the measurement is given by the inverse halo number density within the mass bin, \textit{i.e.} 
\begin{equation}\label{eqn:poisson_noise_def}
	N^2(M) \equiv \frac{1}{\bar{n}_h^L(M)M\Delta(\log M)} \, .
\end{equation}
For self-consistency, we will evaluate these in our critical collapse model using Eq.~\eqref{eqn:nbar_halo} so that
\begin{align}\label{eqn:poisson_noise}
	N^2(M)
	\sim 180\,\mathrm{Mpc}^3\,\left(\frac{M}{10^{12}\,M_{\odot}}\right)^{5/6}\left(\frac{A_s}{2\times 10^{-9}}\right)^{1/2} \left(\frac{k_{\mathrm{eq}}}{0.01\,\mathrm{Mpc}^{-1}}\right)^{3/2}\left(\frac{0.12}{\Omega_mh^2}\right)^{11/6}\frac{D(z)}{\Delta(\log M)}\, ,
\end{align}
for $M<M_*$, assuming a narrow log mass bin with $\Delta (\log M)\lesssim 1$. The majority of information on non-Gaussianity will be extracted from $k$-modes where the matter power spectrum exceeds the shot noise. Hence, it is useful to define $k_{\mathrm{noise},\,\min}$ and $k_{\mathrm{noise},\,\max}$ such that $P(k)\gtrsim N^2(M)$ when $k_{\mathrm{noise},\,\min}\lesssim k\lesssim k_{\mathrm{noise},\,\max}$. These can be computed by setting $P(k)=N^2(M)$ using Eq.~\eqref{eqn:DM_spectrum}, Eq.~\eqref{eqn:Pk_3}, and Eq.~\eqref{eqn:poisson_noise}
\begin{align}\label{eqn:k_noise}
	k_{\mathrm{noise},\,\min} 
	&\sim 4.9\times 10^{-6}\,\mathrm{Mpc}^{-1}\,\left(\frac{M}{10^{12}\,M_{\odot}}\right)^{5/3}\left(\frac{k_{\mathrm{eq}}}{0.01\,\mathrm{Mpc}^{-1}}\right)^{3/2} \nonumber \\
    &\times\left(\frac{2\times 10^{-9}}{A_s}\right)^{1/2}\left(\frac{\Omega_mh^2}{0.12}\right)^{1/6}\Delta (\log M)^{-1} D^{-1}(z)\nonumber \\
	k_{\mathrm{noise},\,\max} 
	&\sim 0.45\,\mathrm{Mpc}^{-1}\,\left(\frac{10^{12}\,M_{\odot}}{M}\right)^{5/12}\left(\frac{k_{\mathrm{eq}}}{0.01\,\mathrm{Mpc}^{-1}}\right)^{3/4} \nonumber \\
    &\times\left(\frac{A_s}{2\times 10^{-9}}\right)^{1/4}\left(\frac{0.12}{\Omega_mh^2}\right)^{1/12}\Delta (\log M)^{1/2}D^{1/2}(z) \, .
\end{align}
These scales will provide a useful analytic guide for how the scales where we can measure scale-dependent bias are affected by the number density of the samples. In Fig.~\ref{fig:power_spectrum_shot_noise}, we numerically evaluate the halo shot noise for several choices of halo mass using the definition of $N^2$ in Eq.~\eqref{eqn:poisson_noise_def} and the exact expressions for $f(M)$ given in Appendix~\ref{app:exact_expressions}, Eqs.~\eqref{eqn:hmf_convenient}–\eqref{eqn:hmf_differential_exact}. We also show the matter power spectrum, $P(k)$, for comparison. The figure shows the region where $P \gg N^2$ and indicates the corresponding values of $k_{\mathrm{noise},\,\min}$ and $k_{\mathrm{noise},\,\max}$. Since the Press–Schechter formalism predicts that more massive halos are rarer, the shot noise level is correspondingly higher, consistent with Fig.~\ref{fig:power_spectrum_shot_noise}. In particular, for our fiducial parameter choices, very massive halos with $M \sim 10^{14}\,M_{\odot}$ have an exponentially suppressed abundance (\textit{cf.} Eqs.~\eqref{eqn:M_star}–\eqref{eqn:nbar_halo}), such that $P(k)$ is subdominant to $N^2$ for all $k$. This indicates a rough upper limit on halo mass for which Fisher information on $f_{\mathrm{NL}}$ can still be meaningfully extracted. In addition, we also show in Fig.~\ref{fig:k} the numerically evaluated values of $k_{\mathrm{noise},\,\min}$ and $k_{\mathrm{noise},\,\max}$ as functions of halo mass $M$.


\subsection{Single-Tracer}

With the theoretical predictions for halo statistics in hand, we now move on to compute the information content on non-Gaussianity from halo measurements. We begin with a single-tracer analysis, in which we infer $f_{\mathrm{NL}}$ by measuring the halo power spectrum. This analysis is the easiest to perform on real data and thus is closest to the forecasts for many surveys.

Denoting the SNR computed from a single-tracer galaxy\footnote{At the level of approximation we are using, the distinction between halos and galaxies is unimportant. We only observe galaxies, so we will use that terminology moving forward.} analysis as $(S/N)_{g,\,s}$, we compute it in the large signal-to-noise ratio limit, \textit{i.e.} $P\gg N^2$, and using the expression for the transfer function $\mathcal{T}(k)$ in Eq.~\eqref{eqn:DM_density_field} (see Eq.~3.4 of~\cite{Green:2023uyz})
\begin{align}\label{eqn:SNR_galaxy}
	\left(\frac{S}{N}\right)_{g,\,s}^2 &\approx \frac{9b_{\Phi}^2\Omega_m^2H_0^4V_{\rm survey}}{2b_1^2D^2(z)}\int\frac{d^3\vec{k}}{(2\pi)^3}\frac{1}{k^4T^2(k)}  \nonumber \\
    &\approx \frac{9}{4\pi^2}\frac{b_{\Phi}^2}{b_1^2}\frac{1}{D^2(z)}\frac{\Omega_m^2H_0^4V_{\rm survey}}{k_{\min}} \, ,
\end{align}
where, in the second line, we assumed $k_{\max}<k_{\mathrm{eq}}$, so we can approximate the transfer function as $T(k)\approx 1$ in the entirety of the momentum integral. The fact that the integral is dominated by $k_{\rm min}$ justifies this approximation, and a more complete treatment shows that there is little information on $\fnlloc$ that is gained directly from $k > \keq$ (although these modes do break degeneracies between $b_1$ and $\fnlloc$).
We can compare this quantity with the matter bispectrum Fisher information, $(S/N)_m$. Recall the respective Fisher information from Eq.~\eqref{eqn:loc_triangle_3} and Eq.~\eqref{eqn:SNR_galaxy}, the ratio is thus given by 
\begin{align}\label{eqn:SNR_loc_single_halo_vs_matter_bispectrum}
	\frac{(S/N)^2_{g,\,s}}{(S/N)^2_{m}} 
	\sim 0.23 \,\left(\frac{2\times 10^{-9}}{A_s}\right)\left(\frac{10^{-3}\,\mathrm{Mpc}^{-1}}{k_{\min}}\right)\left(\frac{0.2\,\mathrm{Mpc}^{-1}}{k_{\max}}\right)^3 \quad \left(\frac{\log(200)}{\log\left(k_{\max}/k_{\min}\right)}\right)
    \frac{1}{D^2(z)} \, ,
\end{align}
which is less than one as expected. While it seems like that we can dial down $k_{\min}$ indefinitely to boost this ratio, note that we assumed signal $\gg$ noise in this estimate, which no longer holds if $k_{\min}$ is so small that the matter power spectrum dips below the noise level. With this in mind, we compute the ratio assuming $k_{\min}$ gets cut off at $k_{\mathrm{noise},\,\min}$ as in Eq.~\eqref{eqn:k_noise}. We will also require that $k_{\max}$ is bounded by the smallest length scale, which is related to the halo radius $k_{\max}\leq 2\pi/R$. This motivates us to set $k_{\min}\sim k_{\mathrm{noise},\,\min}$ as defined in Eq.~\eqref{eqn:k_noise}, and $k_{\max}\sim 2\pi/R$. Then the signal-to-noise ratio becomes
\begin{align}\label{eqn:SNR_loc_single_halo_vs_matter_bispectrum_theory}
	\frac{(S/N)^2_{g,\,s}}{(S/N)^2_{m}} 
	&\sim 4.2\times 10^{-3}\, \left(\frac{M}{10^{12}\,M_{\odot}}\right)^{1/6}\left(\frac{2\times 10^{-9}}{A_s}\right)^{1/2}\left(\frac{\Omega_mh^2}{0.12}\right)^{5/6}\frac{\Delta(\log M)}{D(z)} \, .
\end{align}
where we did not explicitly show the log-dependence of the parameters. Note that $\Delta (\log M)\leq 1$ for our formulas to make sense, since we integrated over the mass bin to estimate the number density of halos within the mass bin.
Eq.~\eqref{eqn:SNR_loc_single_halo_vs_matter_bispectrum_theory} thus concludes that a single halo analysis cannot win over the matter bispectrum information, as expected.

At this stage, it is important to notice that the single tracer signal to noise scales as $k_{\rm min}^{-1}$ while the bispectrum scales only as $\log (k_{\rm max}/k_{\min})$. The counter intuitive reason is that cosmic variance for single tracer means that the noise is smaller at low-$k$ because $P(k)\propto k \to 0$. As a result, we are pushed to larger scales to minimize the noise. Canceling cosmic variance can therefore affect the range of $k$ that contribute, not just the overall amplitude of noise.

\subsection{Multi-Tracer}

We saw in the previous scale that there is a large difference between the scaling of the signal to noise in the single tracer and bispectrum analyses. The main difference is that cosmic variance means the noise is $k$-dependent and is smallest on large scales. In a multi-tracer analysis at high number density, cosmic variance is canceled and shot noise is the main source of noise, which is approximately $k$-independent. For this reason, we will now show that multi-tracer analyses give qualitatively different results from a single tracer analysis.

For simplicity, we will consider two tracers binned according to their halo masses, $M_1$ and $M_2$. The shot noise for each halo measurement is given by the inverse mean number density of tracer in a mass bin, $N_i^2 =1/\bar{n}_{\mathrm{bin}}^{(i)}$. If both masses are far below the cutoff mass $M_*$, we can expand the bias parameters in Eq.~\eqref{eqn:b1_single} for $\sigma\gg 1$. Then the Fisher information from the double tracer analysis, denoted as $(S/N)_{g,\,d}$, is (\textit{cf.} Eq.~\eqref{eqn:Fisher_final_multi_high_P})
\begin{align}\label{eqn:Fisher_double}
	(S/N)^2_{g,\,d} &\sim V_{\rm survey}\frac{\left[b_1^{(1)}b_{\Phi}^{(2)}-b_1^{(2)}b_{\Phi}^{(1)}\right]^2}{\left[\left(b_1^{(1)}\right)^2N_2^2+\left(b_1^{(2)}\right)^2N_1^2\right]} \frac{9}{25}A_s \log\left(\frac{k_{\max}}{k_{\min}}\right) \nonumber \\
	&\sim V_{\rm survey}\frac{36}{25}\frac{\delta_c^6}{(\delta_c-1)^2}\frac{1}{N_1^2+N_2^2} \frac{\left[\sigma^2(M_1)-\sigma^2(M_2)\right]^2}{\sigma^4(M_1)\sigma^4(M_2)}A_s \log\left(\frac{k_{\max}}{k_{\min}}\right) \, .
\end{align}
Using the expression in Eq.~\eqref{eqn:sigma_def_small}, we can further simplify this to
\begin{align}\label{eqn:Fisher_double_2}
	(S/N)^2_{g,\,d} &\sim V_{\rm survey}\frac{3125\times 5^{1/2}}{32\pi^4\times 2^{2/3}\times 3^{1/2}}\frac{\delta_c^{7}}{(\delta_c-1)^2}\frac{(\Omega_mH_0^2)^{31/6}M_p^{1/3}\Delta(\log M)}{A_s^{3/2}k_{\mathrm{eq}}^{15/2}D^5(z)}\log\left(\frac{k_{\max}}{k_{\min}}\right) \nonumber \\
	&\times \frac{\left(M_1^{1/6}-M_2^{1/6}\right)^2\left(M_1^{1/6}+M_2^{1/6}\right)}{M_1^{2/3}-M_1^{1/2}M_2^{1/6}+M_1^{1/3}M_2^{1/3}-M_1^{1/6}M_2^{1/2}+M_2^{2/3}} \nonumber \\
	&\sim V_{\rm survey}\frac{3125\times 5^{1/2}}{64\pi^4\times 2^{2/3}\times 3^{1/2}}\frac{\delta_c^{7}}{(\delta_c-1)^2}\frac{(\Omega_mH_0^2)^{31/6}M_p^{1/3}\Delta(\log M)}{A_s^{3/2}k_{\mathrm{eq}}^{15/2}M_1^{1/6}D^5(z)}\log\left(\frac{k_{\max}}{k_{\min}}\right)  \, ,
\end{align}
where, in the last line, we assume $M_2\ll M_1$ without loss of generality. The SNR ratio between the halo double tracer analysis and the matter bispectrum analysis, for given $k_{\min}$ and $k_{\max}$, evaluated using Eq.~\eqref{eqn:Fisher_double_2} and Eq.~\eqref{eqn:loc_triangle_3}, becomes
\begin{align}\label{eqn:SNR_loc_double_halo_vs_matter_bispectrum}
	\frac{(S/N)^2_{g,\,d}}{(S/N)^2_{m}}  
	\sim 3\times 10^2\,\left(\frac{10^{12}\,M_{\odot}}{M_1}\right)^{1/6}\left(\frac{2\times 10^{-9}}{A_s}\right)^{5/2}\left(\frac{\Omega_mh^2}{0.12}\right)^{31/6}  \left(\frac{0.2\,\mathrm{Mpc}^{-1}}{k_{\max}}\right)^3 \frac{\Delta(\log M)}{D^5(z)} \, .
\end{align}
This turns out to be greater than one, which naively seems to violate the intuition that the matter bispectrum contains the maximum amount of information. The resolution to this apparent contradiction is that, similar to the single-tracer case above, a halo analysis implicitly contains high-$k$ information corresponding to the size of the objects being measured. Hence, a fair comparison requires us to take $k_{\max}$ for both analyses to be the corresponding wavenumber, which we set to $2\pi/R_1$. Then the ratio becomes
\begin{align}\label{eqn:SNR_loc_double_halo_vs_matter_bispectrum_2}
	\frac{(S/N)^2_{g,\,d}}{(S/N)^2_{m}}
	\sim 6.8\times 10^{-2}\,\left(\frac{M_1}{10^{12}\,M_{\odot}}\right)^{5/6}\left(\frac{2\times 10^{-9}}{A_s}\right)^{5/2}\left(\frac{\Omega_mh^2}{0.12}\right)^{25/6}\frac{\Delta(\log M)}{D^5(z)} \, ,
\end{align}
which now agrees with our expectation.

\begin{figure}[!ht]
	\includegraphics[width=1\textwidth]{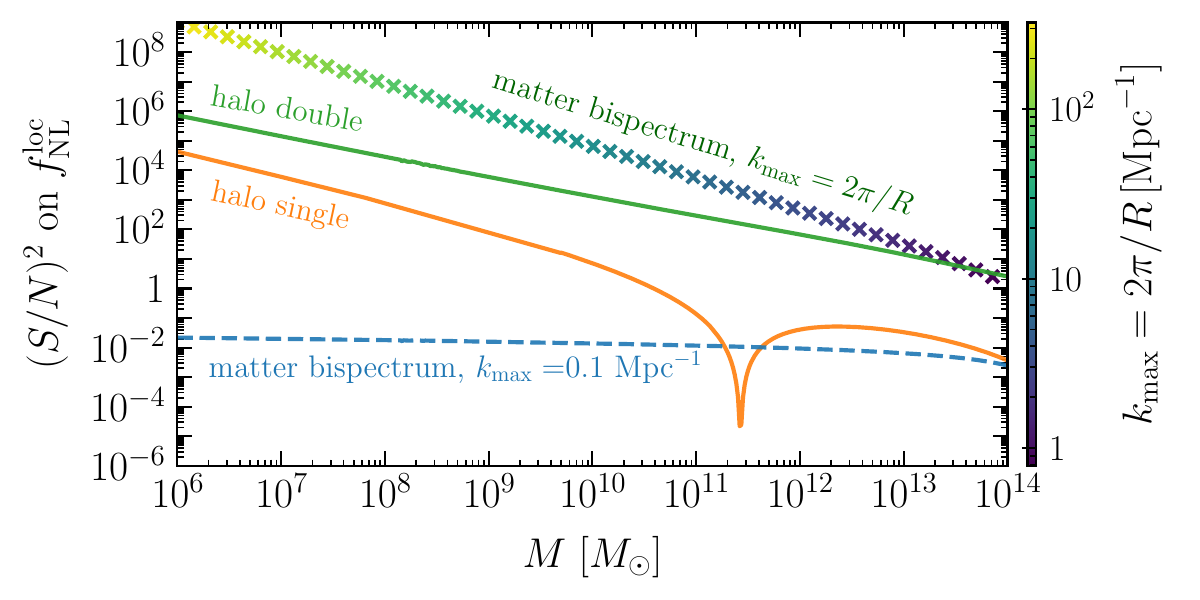} 
	\caption{Fisher information on $f_{\mathrm{NL}}^{\mathrm{loc}}$ for all analysis types considered in this section. The gradient-colored line denotes the Fisher information from a matter bispectrum analysis of the matter field, with $k_{\min}$ set by the minimum wavenumber for which $P(k)>N^2$ (\textit{i.e.} $k_{\mathrm{noise},\,\min}$), where $N^2$ is the shot noise for halos of mass $M$, and $k_{\max}$ set by $2\pi/R$, with $R$ the size of a halo of mass $M$. This line illustrates the maximum amount of information contained in halo objects of mass $M$. The Fisher information is given by the first line of Eq.~\eqref{eqn:SNR_expression_quote}. The expression for $P(k)$ is given in Eqs.~\eqref{eqn:DM_spectrum} and \eqref{eqn:Pk_3}, and $N$, defined in Eq.~\eqref{eqn:poisson_noise_def}, is evaluated numerically using the exact expression for $f(M)$ in Eqs.~\eqref{eqn:hmf_convenient}–\eqref{eqn:hmf_differential_exact} together with the Press–Schechter prediction of the halo number density in Eq.~\eqref{eqn:Press-Schechter}. The color of each point along the line indicates the numerical value of $k_{\max}=2\pi/R$. The blue dashed line corresponds to a matter bispectrum analysis with the same $k_{\min}$ but with a fixed, conservative choice of $k_{\max}=0.1$\,Mpc$^{-1}$. The orange line corresponds to a single-tracer analysis of halos of mass $M$, obtained by numerically integrating over $k$ using Eq.~\eqref{eq:Ffinal_galaxy}, with $k_{\min}$ and $k_{\max}$ set as above by $k_{\mathrm{noise},\,\min}$ and $2\pi/R$. Finally, the green line corresponds to a double-tracer analysis of two halo populations: heavier halos with mass $M_1=M$ and lighter halos with mass $M_2=1\,M_{\odot}$, obtained by numerically integrating over $k$ using Eq.~\eqref{eqn:Fisher_final_multi}, with the same choices of $k_{\min}$ and $k_{\max}$. We did not use any expressions that assume $P\gg N^2$ when computing the Fisher information. Fiducial parameter values are listed in Table~\ref{tab:parameters}. The survey volume is taken to be $V_{\mathrm{survey}}=10^{10}$~Mpc$^3$.
    }\label{fig:fisher_comparison}
\end{figure}

In Fig.~\ref{fig:fisher_comparison}, we show the Fisher information on $f_{\mathrm{NL}}^{\mathrm{loc}}$ for all analysis types considered in this work. The Fisher information from a matter bispectrum analysis, evaluated analytically using the first line of Eq.~\eqref{eqn:SNR_expression_quote}, assumes $k_{\min}=k_{\mathrm{noise},\,\min}$ and $k_{\max}=2\pi/R$ for halos of mass $M$. As argued above, this represents the maximum amount of information contained in halo objects of mass $M$ via the Cramér–Rao bound, which is supported by Fig.~\ref{fig:fisher_comparison}, since this curve exceeds any halo analyses shown. For comparison, we also show the corresponding Fisher information from the same analysis but with a conservatively chosen fixed value $k_{\max}=0.1$~Mpc$^{-1}$. As expected, fixing $k_{\max}$ to a value typical of realistic observations reduces the Fisher information relative to halo analyses, since the halo formation process itself encodes high-$k$ information associated with the halo size that impacts measurements of halo populations.

In Fig.~\ref{fig:fisher_comparison}, we also present the Fisher information from a single-tracer halo analysis with mass $M$, and from a double-tracer halo analysis with two mass bins: heavier halos with $M_1=M$ and lighter halos with $M_2=1\,M_{\odot}$. Both analyses set $k_{\min}$ and $k_{\max}$ by $k_{\mathrm{noise},\,\min}$ and $2\pi/R$, respectively, similarly to the matter bispectrum analysis for ease of comparison. The Fisher information is computed by numerically integrating over $k$ using the exact expressions in Eq.~\eqref{eq:Ffinal_galaxy} and Eq.~\eqref{eqn:Fisher_final_multi}, as derived in App.~\ref{app:fisher_information_derivation}, without using any analytic expressions that assume $P\gg N^2$. Note that the single-tracer Fisher information briefly dips to zero for halo masses around $M \sim 3\times 10^{11}\,M_{\odot}$, since the scale-dependent bias $b_{\Phi}$ vanishes at this mass for our fiducial parameters (\textit{cf.} Fig.~\ref{fig:bias_parameters}). As expected, the double-tracer analysis is more optimal than the single-tracer analysis across the full halo mass range, and both single-tracer and double-tracer halo analyses contain less Fisher information than the matter bispectrum analysis with $k_{\max}$ set by the halo size. Notably, for our choice of fiducial parameters, the double-tracer halo analysis saturates the Cramér–Rao bound associated with the matter bispectrum (assuming $\kmax$ is inferred from the highest halo mass, $M_1\approx 10^{14}\,M_{\odot}$). Although it may appear that the double-tracer analysis can exceed the bound at even higher halo masses, Figs.~\ref{fig:power_spectrum_shot_noise}-\ref{fig:k} shows that $P(k)$ falls below $N^2$ in this regime, indicating an exponential cutoff in the halo population as predicted by Press–Schechter (\textit{cf.} Eq.~\eqref{eqn:Press-Schechter} and Eqs.~\eqref{eqn:M_star}–\eqref{eqn:nbar_halo}). Consequently, no meaningful halo analyses can be performed for even more massive halos.

\subsection{Summary}

The key take-away from this section is that the matter bispectrum contains all the information relevant to measuring $\fnlloc$, if we include all the modes that affect the formation of the halos. However, because halos are sensitive to small scale modes, analyses with conservative choices of $\kmax$ can derive most of their constraining power from the scale-dependent bias.

The single tracer analysis is generally far from being an optimal measurement, as indicated by the different scaling with $\kmin$ from both the bispectrum and multi-tracer analyzes. In practice, the multi-tracer forecasts show the same scaling as the optimal measurement of the bispectrum and offer orders of magnitude more sensitivity than a conventional bispectrum analysis with $\kmax \approx 0.1 \, h \, {\rm Mpc}^{-1}$.

Naturally, this suggests that the restricting to moderate $\kmax$ is unnecessary for these surveys. Primordial non-Gaussianity couples short modes in a way that appears to violate the equivalence principle and therefore does not arise from nonlinear evolution. Several steps have been taken in this direction~\cite{Goldstein:2022hgr,Kvasiuk:2024gbz} and show promise for extending the sensitivity to $\fnlloc$.

\section{Future of Local and Equilateral Non-Gaussianity}\label{sec:fnleq}

While current limits on primordial non-Gaussianity are given by the CMB~\cite{Planck:2019kim}, targets for local and equilateral templates will require measurements from LSS that exceed the CMB by at least an order of magnitude~\cite{Alvarez:2014vva}. Reaching these goals requires control over the nonlinear effects that influence the distribution of matter and galaxies. Of course, PNG includes many more possible shapes than these two bispectrum templates. Nevertheless, these two templates are sufficient to understand the sensitivity for a much wider range of signals.

From Section~\ref{sec:likelihood}, we know the information about PNG is contained in the power spectrum and bispectrum. Yet, some of the information about the nonlinear evolution arises from $N>3$-point correlators. The field-level Cram\'er-Rao bound will determine how much information remains accessible accounting for all available information. In this section, we will show these kinds of field-level forecasts for an array of upcoming surveys.

\subsection{Forecasting Methodology}

In idealized forecasts that adopt aggressive choices of $k_{\rm max}$ motivated by halo-scale physics and assume controlled theoretical uncertainties, the bispectrum can therefore dominate the constraints on $f_{\rm NL}$. However, in realistic analyses the bispectrum must be restricted to a more conservative $k_{\rm max}$ owing to limitations from theoretical modeling, survey systematics, and data analysis. In this practical regime, the Fisher information on local PNG extracted from the power spectrum alone can become comparable to, or even exceed, that from the bispectrum.

The comparison in Sec.~\ref{sec:local} focused on the relative information content of the power spectrum and bispectrum under minimal marginalization. In realistic survey forecasts, however, constraints on $f_{\rm NL}$ are further degraded by degeneracies with galaxy bias parameters, cosmological parameters, and redshift-space effects. The primary purpose of this section is therefore not to reassess where the raw PNG information resides, but to quantify how the inclusion of the bispectrum mitigates these degeneracies and stabilizes constraints in realistic survey settings.

In this section, we consider three survey specifications, BOSS, DESI, and MegaMapper\footnote{We will use MegaMapper~\cite{Schlegel:2022vrv} as the design for Spec-S5~\cite{Spec-S5:2025uom} in order to make contact with existing forecasts~\cite{Cabass:2022epm}. }, and compare forecasts of marginalized $\sigma(f_{\rm NL})$ obtained from power-spectrum--only analyses with those from joint power-spectrum and bispectrum forecasts. In practical survey settings, the dominant role of the bispectrum is to break parameter degeneracies that are poorly constrained by the power spectrum alone. These degeneracies can inflate the uncertainty on local PNG by factors of a few, and on equilateral PNG by up to two orders of magnitude. Even a relatively modest bispectrum contribution is therefore sufficient to substantially reduce these degeneracy-driven degradations.

One essential input to our forecasts from the field-level likelihood is that we do not include bias parameters beyond quadratic order. This is a direct consequence of the field level CR bound for the bias parameters, which shows that Fisher information for the parameters $b_{n>2}$ arises from the $n+1$- through $2n$-point correlators. In contrast, the CR bound for $\fnl$, at the field level, is determined by the bispectrum and therefore is only degenerate with bias parameters to quadratic order. 

The forecasts presented in this section rely on the following assumptions:
\begin{enumerate}
    \item Field-level redshift-space distortions are modeled using the linear Kaiser approximation,
    $\delta_g(k)=(b_{1}+f\mu^2+\ldots)\dL(k)$. Additional redshift-space effects beyond linear Kaiser theory are     incorporated at the level of the power spectrum and bispectrum through nuisance parameters, as described below and in Sec.~\ref{subsection:rsd}.

    \item For BOSS and DESI, we adopt $k_{\rm max}=0.22 \, h\,\mathrm{Mpc}^{-1}$ for the power spectrum and $k_{\rm max}=0.11 \, h\,\mathrm{Mpc}^{-1}$ for the bispectrum. For MegaMapper, we use the $k_{\rm max}^{\rm 1L}$ values reported in Table~3 of Ref.~\cite{Braganca:2023pcp} for the power spectrum, and again take the bispectrum $k_{\rm max}$ to be half of that value. This conservative choice reflects the increased sensitivity of the bispectrum to nonlinear mode coupling and modeling uncertainties.

    \item We adopt $k_{\rm min}=0.001 \, h\,\mathrm{Mpc}^{-1}$ for DESI and MegaMapper. For BOSS, owing to its smaller survey volume, we instead adopt  $k_{\rm min}=0.006 \, h\,\mathrm{Mpc}^{-1}$.

    \item The fiducial values of all bias parameters are set to zero except for  $b_1$, which is taken to match the values in Tables~1–3 of Ref.~\cite{Braganca:2023pcp}. The characteristic length scale is fixed to  $R_*=2.66 \, h^{-1}\mathrm{Mpc}$, following Ref.~\cite{Green:2023uyz}.

    \item We assume that bias parameters are shared across redshift bins in our baseline forecast, which avoids
    introducing artificially degenerate nuisance directions in the Fisher
    matrix. For comparison, we also consider a more conservative scenario in which bias parameters are treated as independent degrees of freedom in each redshift bin. 
\end{enumerate}
As our baseline methodology, we will use a single set of bias parameters shared across all bins with known redshift evolution.  We will contrast theses results with common forecasting practice of marginalizing over nuisance parameters independently in each redshift bin. Treating the same physical nuisance parameters as independent degrees of freedom in each bin introduces multiple weakly constrained directions in the Fisher matrix, even when their fiducial values are identical across bins. When information from multiple redshift bins is combined, these additional nuisance directions can become significantly degenerate with $f_{\rm NL}$, leading to degraded constraints. Sharing bias parameters across redshift bins mitigates this effect and yields more stable constraints on primordial non-Gaussianity. This effect is particularly pronounced for equilateral PNG in power-spectrum-only forecasts, where treating bias parameters independently across redshift bins leads to a degradation of the constraint.

\subsection{Results}

The objective of our forecasts are to better understand the reach and limitations of future surveys to constrain $\fnl$. The limiting factor is the degeneracy with the uncertainties in the bias parameters and other aspects of the nonlinear formation of structure. Our goal is not only to quantify the sensitivity to $\fnlloc$ and $\fnleq$, but also to isolate the parameters that most affect these numbers.

\begin{figure}[!ht]
    \centering
    \includegraphics[height=0.253\textwidth]{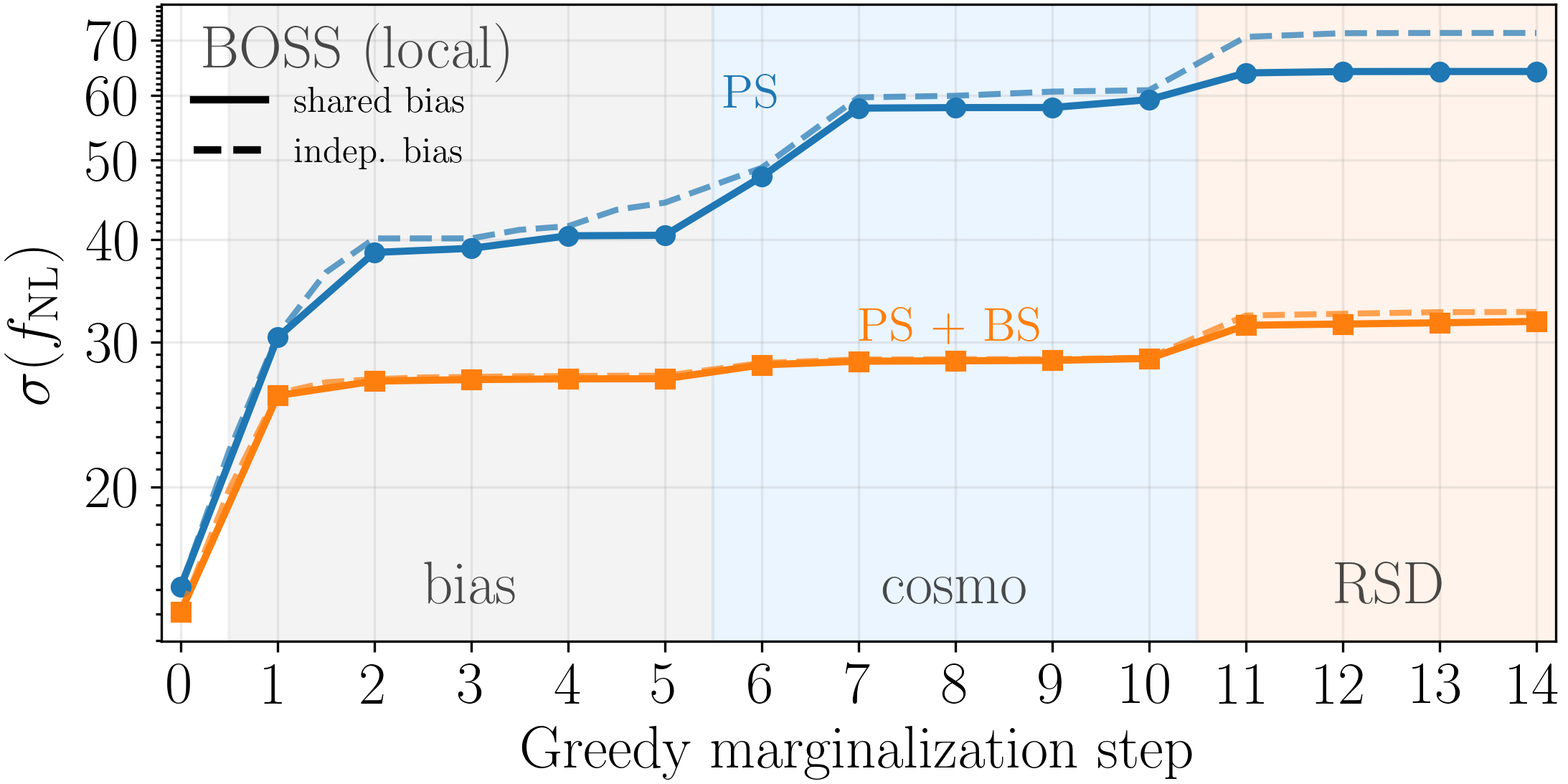}
    \hfill
    \includegraphics[height=0.253\textwidth]{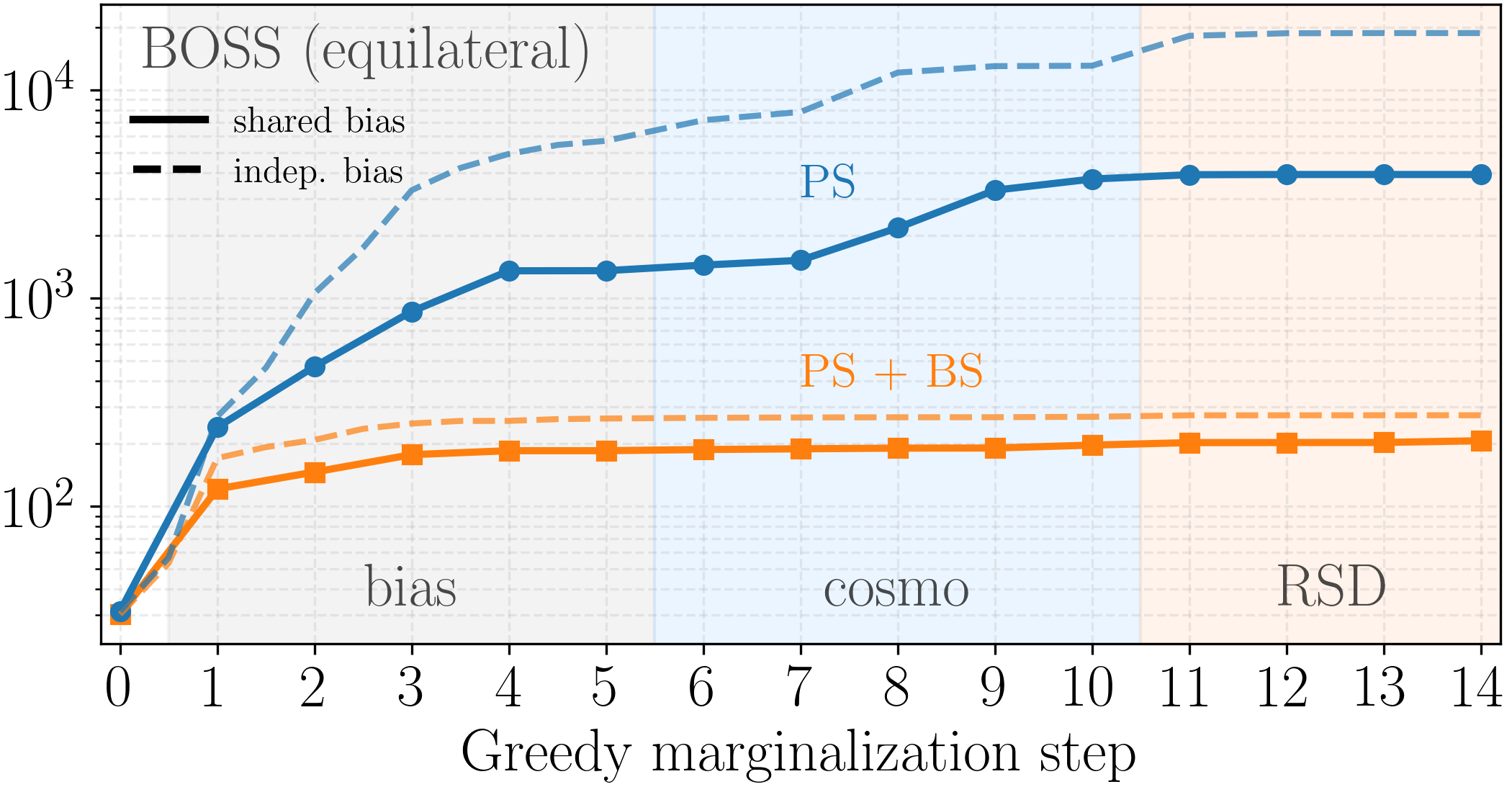}

    \caption{
   Forecasted marginalized uncertainties on $f_{\rm NL}$ for BOSS obtained using a greedy marginalization procedure.
The horizontal axis shows the cumulative number of marginalized parameters added by the greedy algorithm.
At each step, the parameter that maximally increases $\sigma(f_{\rm NL})$, conditional on the previously marginalized set, is added. The resulting greedy ordering differs between curves and is listed in Table.~\ref{tab:greedy-orderings}.
Solid curves assume bias parameters are shared across redshift bins, while dashed curves correspond to the conservative case in which bias parameters are treated independently in each bin.
Blue curves show power-spectrum-only (PS) constraints, while orange curves include both the power spectrum and bispectrum (PS+BS).
Shaded regions indicate broad parameter classes (bias, cosmological, and redshift-space nuisance parameters).
Left: $\sigma(f_{\rm NL}^{\rm loc})$ for local primordial non-Gaussianity .
Right: $\sigma(f_{\rm NL}^{\rm eq})$ for equilateral primordial non-Gaussianity .
}
    \label{fig:BOSS_png}
\end{figure}

\begin{figure}[!ht]
    \centering
    \includegraphics[height=0.253\textwidth]{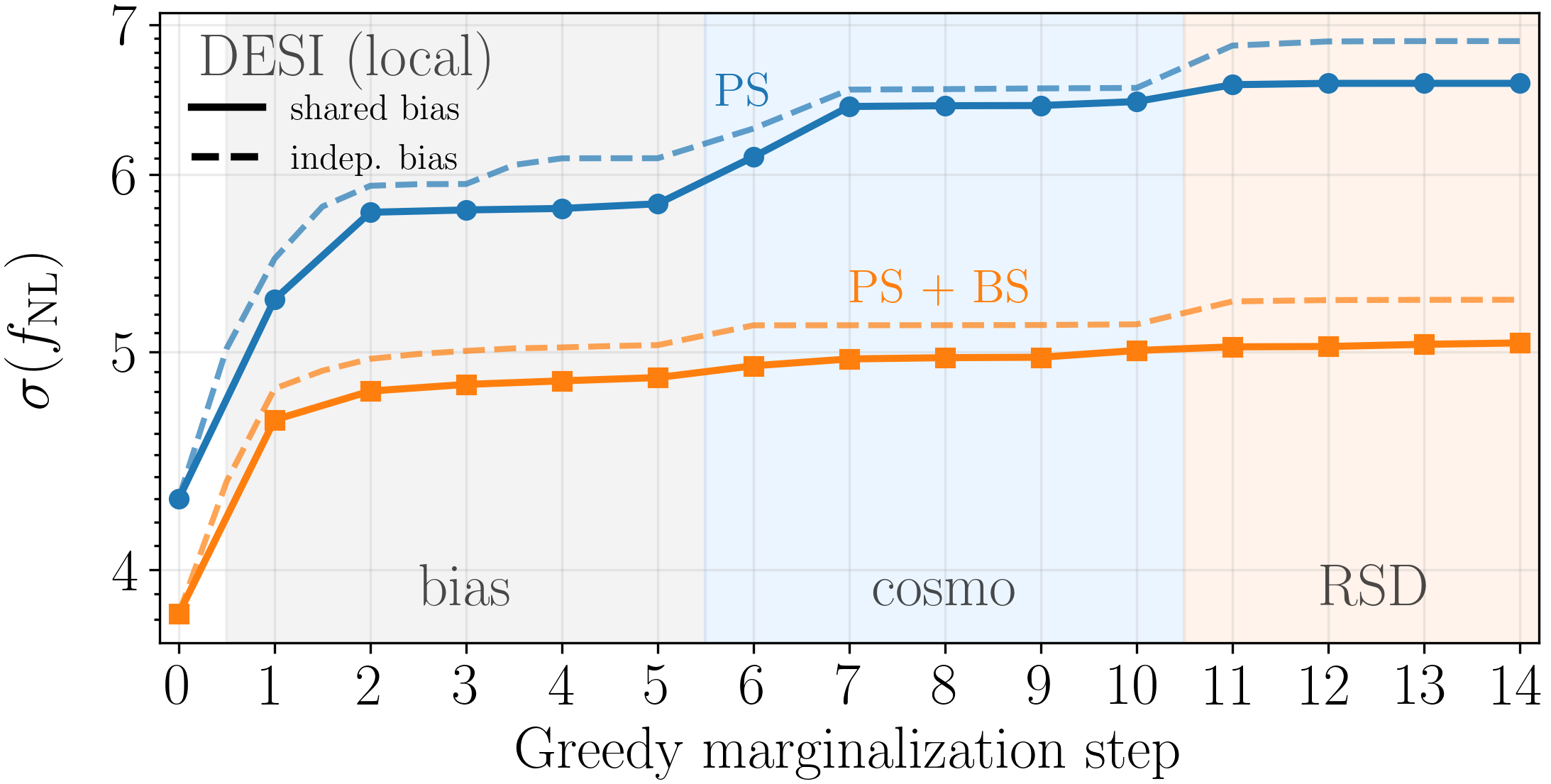}
    \hfill
    \includegraphics[height=0.253\textwidth]{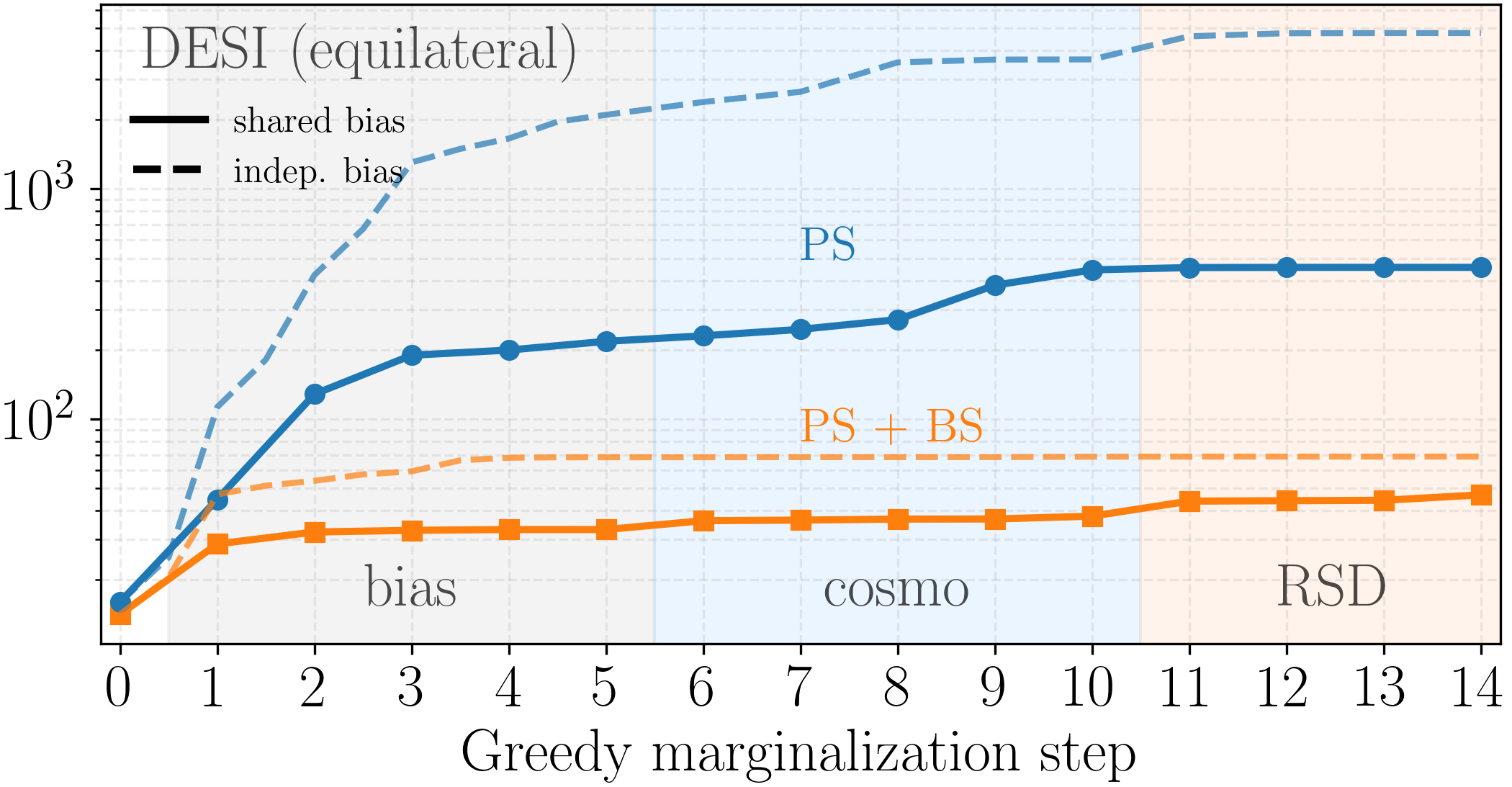}

    \caption{
    Same as Fig.~\ref{fig:BOSS_png}, but for DESI.
    }
    \label{fig:DESI_png}
\end{figure}

\begin{figure}[!ht]
    \centering
    \includegraphics[height=0.253\textwidth]{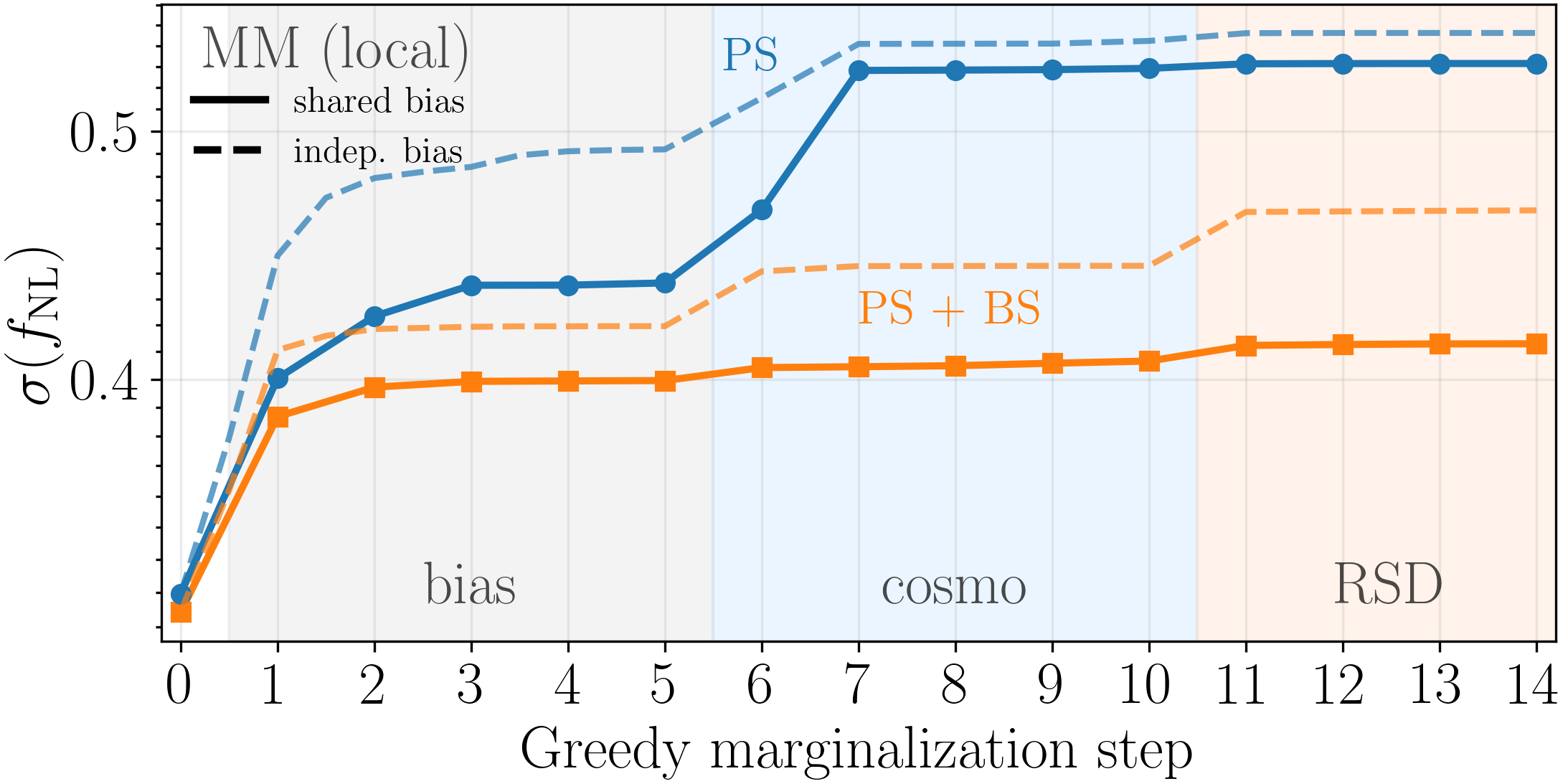}
    \hfill
    \includegraphics[height=0.253\textwidth]{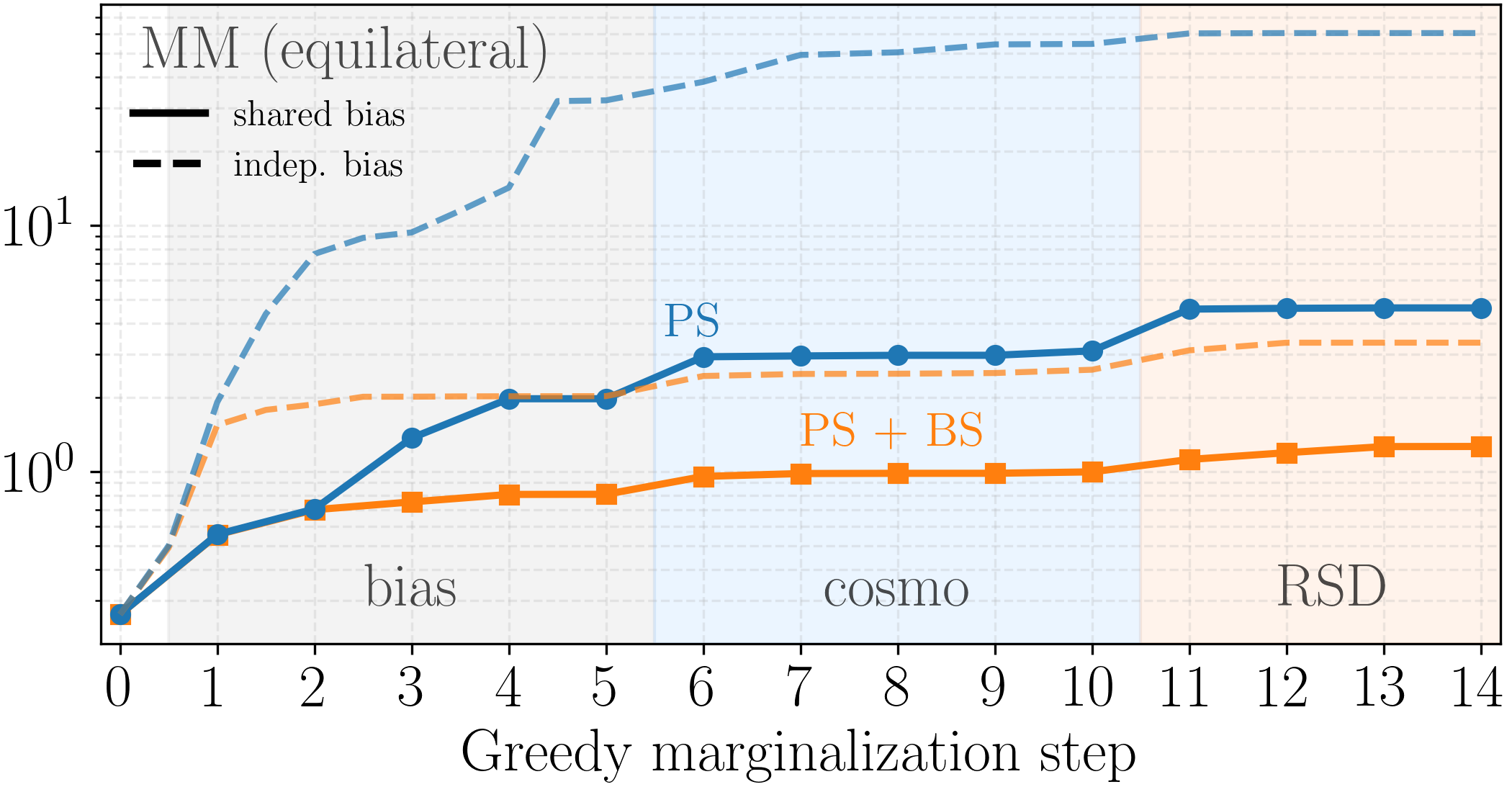}

    \caption{
    Same as Fig.~\ref{fig:BOSS_png}, but for MegaMapper.
    }
    \label{fig:MM_png}
\end{figure}

With this in mind, we will present the results in the greedy marginalization procedure shown below (see Table.~\ref{tab:greedy-orderings}). Parameters are added in stages corresponding to bias, cosmological, and redshift-space nuisance parameters. The results, shown in Figures~\ref{fig:BOSS_png},\ref{fig:DESI_png}, and~\ref{fig:MM_png}, are presented in terms of decreasing impact on the overall sensitivity.

The greedy marginalization shows that, qualitatively, marginalization over bias parameters has the largest impact on sensitivity. This is also consistent with recent results on the impact of simulation based priors on these kinds of measurements.~\cite{Ivanov:2024hgq} The details of which specific bias parameters have the most impact can be seen to vary both with the shape of non-Gaussianity and the sensitivity of the experiment, as seen in Table~\ref{tab:greedy-orderings}. It is noteworthy that in many cases it is the linear biasing terms (including gradients) that are the most impactful. These are the terms that are most easily measured by introducing additional data, or predicted from simulations or first principles.

In the literature, there is concern that forecasts based on simplified redshift-space modeling may be overly optimistic, particularly given the absence of a first-principles, survey-independent description of Finger-of-God (FoG) effects. Existing FoG treatments therefore rely on phenomenological parameterizations of small-scale velocity dispersion. At the forecast level, however, one can ask how strongly such effects impact the
inferred uncertainty on primordial non-Gaussianity. In Figures~\ref{fig:BOSS_png}, \ref{fig:DESI_png}, and \ref{fig:MM_png}, we include Gaussian FoG damping parameters for the power spectrum and bispectrum, denoted
by $\sigma_{v,P}$ and $\sigma_{v,B}$, respectively, following the treatment of Ref.~\cite{Gil-Marin:2016wya}. For the power spectrum, we additionally marginalize over higher-order redshift-space EFT counter-terms parameterized by
$\alpha_2$ and $\alpha_4$, following Ref.~\cite{Chen:2020fxs}. We find that marginalizing over these additional FoG and redshift-space nuisance parameters leads to only a mild further degradation of $\sigma(f_{\rm NL})$, subdominant compared to the degradation induced by marginalization over bias and cosmological parameters. This indicates that primordial non-Gaussianity is only weakly degenerate with these additional redshift-space nuisance parameters, or that their effects project predominantly onto degeneracy directions already spanned by the existing nuisance parameter set.


\begin{figure}[!ht]
    \centering
    \includegraphics[height=0.253\textwidth]{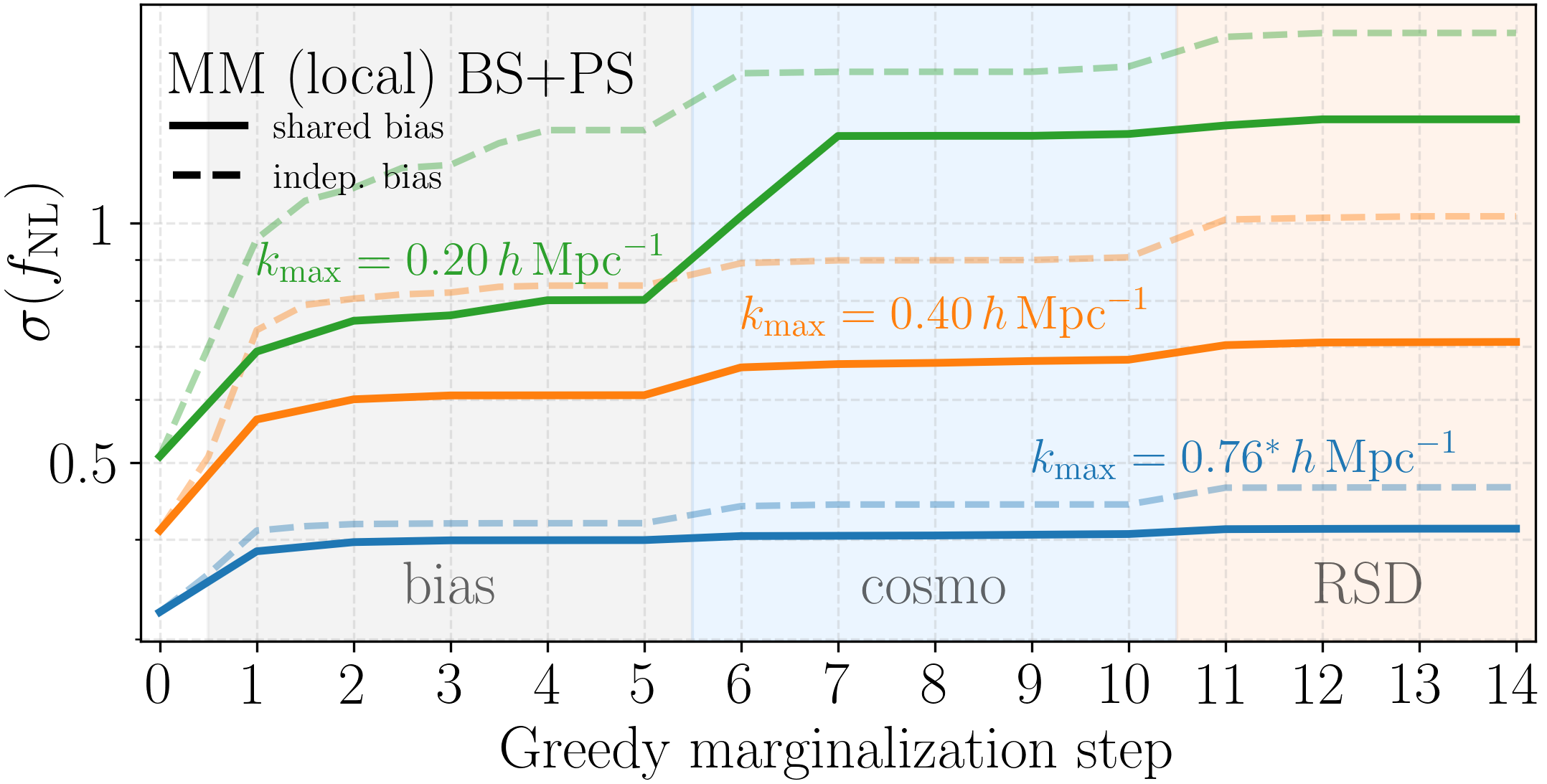}
\hfill
    \includegraphics[height=0.253\textwidth]{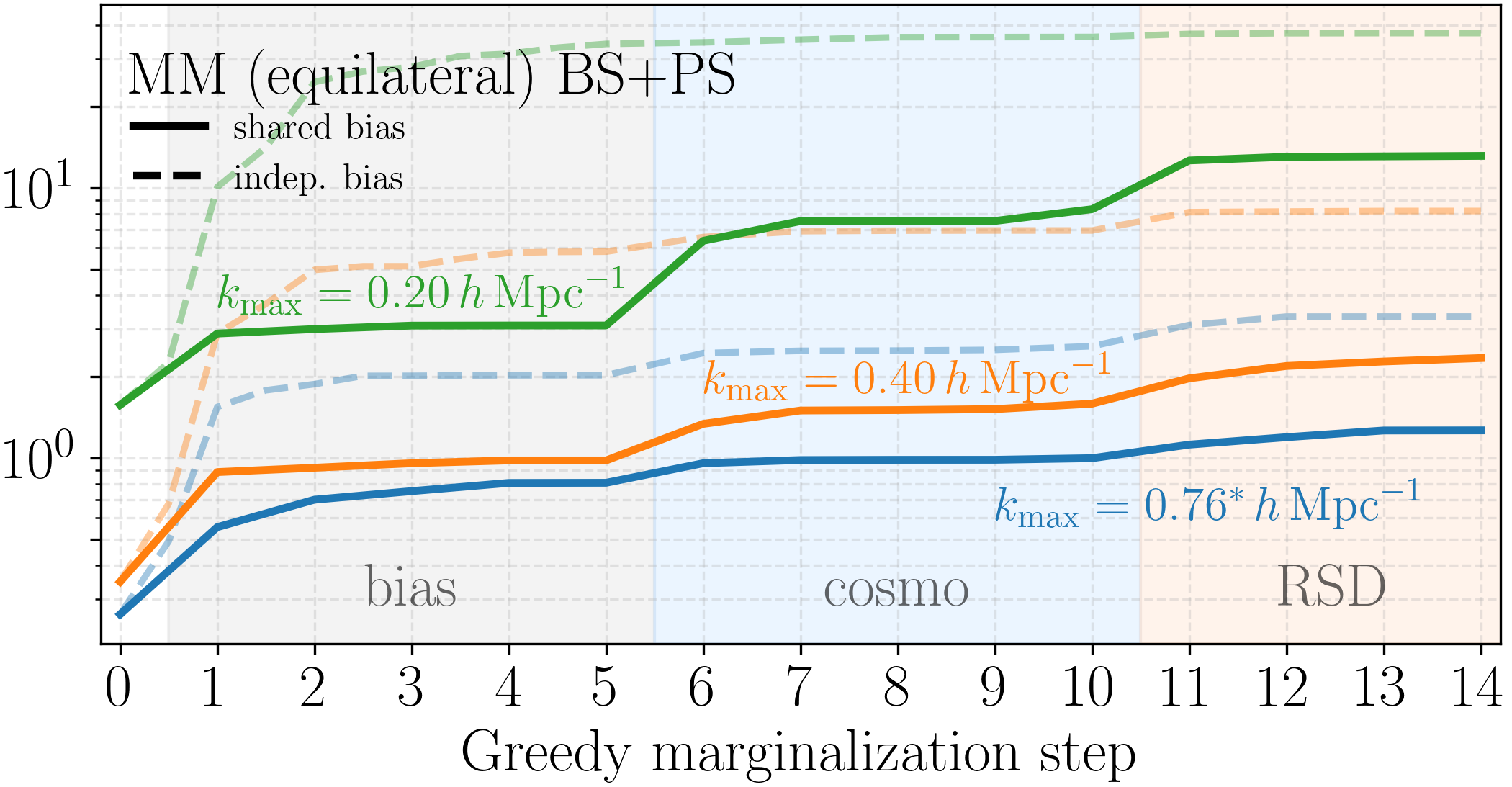}

    \caption{ Forecasted marginalized uncertainties on $f_{\rm NL}$ for 
   MegaMapper obtained using the greedy marginalization procedure as in Fig.~\ref{fig:BOSS_png}. Curves correspond to different choices of $k_{\rm max}$. The asterisk denotes the case where $k_{\rm max}$ values are used in two redshift bins, $k_{\rm max} = 0.36 \, h\,\text{Mpc}^{-1}$ and $k_{\rm max} = 0.76 \, h\,\text{Mpc}^{-1}$, following the choice made in \cite{Braganca:2023pcp}. 
    }
    \label{fig:MM_vary_k_max_png}
\end{figure}

\begin{table}[!ht]
\centering
\caption{Greedy marginalization orderings for all surveys and PNG shapes.
Within each cell, parameters are listed sequentially in the order they are added
during the greedy marginalization procedure. PS denotes the forecast with power spectrum only, while PS+BS includes both power spectrum and bispectrum. }
\label{tab:greedy-orderings}

\renewcommand{\arraystretch}{1.25}
\setlength{\tabcolsep}{6pt}

\begin{tabular}{l l p{5.1cm} p{5.1cm}}
\hline\hline
Survey & PNG shape & PS ordering & PS+BS ordering \\
\hline

\multicolumn{4}{l}{\textbf{BOSS}} \\[2pt]
&
Equilateral&
(none), $b_{k^2}$, $b_1$, $b_2$, $b_{k^4}$, $b_{s^2}$,
$\omega_{\rm cdm}$, $\ln(10^{10}A_s)$, $\omega_b$, $h$, $n_s$, $\sigma_{v,P}$, $\alpha_4$, $\alpha_2$, $\sigma_{v,B}$&
(none), $b_2$, $b_1$, $b_{k^2}$, $b_{s^2}$,$b_{k^4}$, 
$\omega_b$,$n_s$,$h$,  $\ln(10^{10}A_s)$,$\omega_{\rm cdm}$,$\alpha_2$, $\sigma_{v,B}$, $\alpha_4$, $\sigma_{v,P}$\\[4pt]

&
Local&
(none), $b_1$, $b_{s^2}$, $b_{k^2}$, $b_2$, $b_{k^4}$,
$n_s$, $\ln(10^{10}A_s)$, $h$, $\omega_{\rm cdm}$, $\omega_b$, $\sigma_{v,P}$,
$\alpha_4$, $\alpha_2$, $\sigma_{v,B}$&
(none), $b_1$, $b_2$, $b_{k^4}$, $b_{s^2}$, $b_{k^2}$,
$\ln(10^{10}A_s)$, $\omega_{\rm cdm}$, $h$, $\omega_b$, $n_s$, $\sigma_{v,P}$, $\sigma_{v,B}$,
$\alpha_4$, $\alpha_2$\\[6pt]

\multicolumn{4}{l}{\textbf{DESI}} \\[2pt]
&
Equilateral&
(none), $b_{k^2}$, $b_2$, $b_{k^4}$, $b_1$, $b_{s^2}$,
$\omega_{\rm cdm}$, $n_s$, $\omega_b$, $h$, $\ln(10^{10}A_s)$, 
$\alpha_2$,$\sigma_{v,P}$, $\alpha_4$, $\sigma_{v,B}$&
(none), $b_2$, $b_1$, $b_{s^2}$, $b_{k^4}$, $b_{k^2}$,
$\ln(10^{10}A_s)$, $n_s$, $\omega_{\rm cdm}$, $h$, $\omega_b$,
$\alpha_2$, $\sigma_{v,B}$, $\alpha_4$, $\sigma_{v,P}$\\[4pt]

&
Local&
(none), $b_1$, $b_{s^2}$, $b_2$, $b_{k^4}$, $b_{k^2}$,
$n_s$, $\ln(10^{10}A_s)$, $\omega_{\rm cdm}$, $\omega_b$,$h$, $\sigma_{v,P}$, $\alpha_2$, $\alpha_4$, $\sigma_{v,B}$&
(none), $b_1$, $b_{k^2}$, $b_{k^4}$, $b_2$, $b_{s^2}$,
$n_s$, $\ln(10^{10}A_s)$, $\omega_{\rm cdm}$, $h$, $\omega_b$, $\sigma_{v,B}$, $\sigma_{v,P}$, $\alpha_4$, $\alpha_2$\\[6pt]

\multicolumn{4}{l}{\textbf{MegaMapper}} \\[2pt]
&
Equilateral&
(none), $b_{k^4}$, $b_{k^2}$, $b_2$, $b_{s^2}$, $b_1$,
$\ln(10^{10}A_s)$, $n_s$, $\omega_b$, $\omega_{\rm cdm}$, $h$, 
$\alpha_2$, $\sigma_{v,P}$,  $\alpha_4$, $\sigma_{v,B}$&
(none), $b_{k^4}$, $b_{k^2}$, $b_2$, $b_1$, $b_{s^2}$,
$n_s$, $\ln(10^{10}A_s)$, $\omega_{\rm cdm}$, $h$, $\omega_b$, $\sigma_{v,B}$, $\alpha_2$, $\sigma_{v,P}$, $\alpha_4$\\[4pt]

&
Local&
(none), $b_1$, $b_2$, $b_{s^2}$, $b_{k^4}$, $b_{k^2}$,
$\ln(10^{10}A_s)$, $n_s$, $\omega_{\rm cdm}$, $h$, $\omega_b$, 
$\alpha_2$, $\sigma_{v,P}$,  $\alpha_4$, $\sigma_{v,B}$&
(none), $b_1$, $b_2$, $b_{k^2}$, $b_{s^2}$, $b_{k^4}$,
$\ln(10^{10}A_s)$, $n_s$, $\omega_{\rm cdm}$, $h$, $\omega_b$,  $\sigma_{v,B}$, $\alpha_2$, $\sigma_{v,P}$, $\alpha_4$\\

\hline\hline
\end{tabular}
\end{table}

Overall, the results in Fig.~\ref{fig:BOSS_png}, Fig.~\ref{fig:DESI_png}, and Fig.~\ref{fig:MM_png} illustrate the role played by the bispectrum in realistic forecasts. For unmarginalized constraints, the bispectrum contributes very little, as expected from the conservative choice of $k_{\rm max}$. Its primary utility arises when parameters are marginalized: the bispectrum partially breaks the degeneracies present in the power spectrum, reducing the degradation in $\sigma(f_{\rm NL})$, especially for the equilateral shape. Another trend visible in the comparison across surveys is that the relative impact of the bispectrum decreases at higher redshift. In surveys like MegaMapper, the relative impact of the bispectrum becomes smaller because the PNG signal in the power spectrum is substantially enhanced at high redshift. The scale-dependent bias induced by PNG scales as 
\begin{align}
    \Delta b \propto \frac{f_{\rm NL}}{k^2\mathcal{T}(k,z)} 
\end{align}
with $\mathcal{T}(k,z)\propto D(z)$, where $D(z)$ is the linear growth factor. Hence, $\Delta b\propto D^{-1}(z)$, and the PNG contribution to the power spectrum grows rapidly at high redshift, when all other parameters such as the shot noise, $k_{\rm min}$, and $k_{\rm max}$ are held fixed. At redshift $z\approx 0$, the Fisher information on $f_{\rm NL}$ from the power spectrum and the bispectrum can be comparable, but as redshift increases, the power spectrum contribution grows much faster than that of the bispectrum. As a result, the bispectrum's relative contribution to the total constraint diminishes at high redshift, making its impact on surveys like MegaMapper substantially weaker than in lower-redshift surveys like BOSS and DESI.

The forecasts for local and equilateral are qualitatively different, as we anticipated from the $k$-dependence of the signal. To see this different, notice that the $\sigma(\fnlloc)$ figures use a linear scale for the $y$-axis, while $\sigma(\fnleq)$ is a log scale. This is precisely what we could have anticipated from the previous section. For local non-Gaussianity, scale-dependent bias moves a lot of the information to the power spectrum. The bispectrum has some additional information, but significant improvements requires including modes well into the non-linear regime. The forecasts in Figure~\ref{fig:BOSS_png} and Figure~\ref{fig:DESI_png} are consistent with the measured constraints in~\cite{BOSS:2012vpn,Cabass:2022ymb,DAmico:2022osl} and~\cite{Chaussidon:2024qni,Chudaykin:2025vdh} respectively. In particular, DESI is competitive with Planck, and can improve on the CMB with simulation based priors. Forecasts with Megamapper are similar to those of SPHEREx~\cite{Bock:2025ijl} found elsewhere~\cite{Heinrich:2023qaa}. As noted in the previous section, significant improvements $\sigma(\fnlloc)$ can be achieved through cosmic variance cancellation using external data (such as CMB lensing~\cite{Schmittfull:2017ffw} or kSZ~\cite{Munchmeyer:2018eey}) or from multi-tracer analyses~\cite{Seljak:2008xr}. The purpose of the previous Section was the highlight the impact of these kinds of specialized analyses that are accessible with the conventional analyses from these surveys. Nevertheless, we have presented these results to highlight the impact of shape dependence in these surveys.

The forecasts for equilateral non-Gaussianity are our primary interest here. We see that forecasts vary by orders of magnitude depending on the assumptions and the data used. Previous work has shown that the power spectrum alone is not competitive with the bispectrum. The forecasts here reproduce those forecasts, but also Figures~\ref{fig:BOSS_png} and~\ref{fig:DESI_png} show there is an order of magnitude improvement available for BOSS and DESI if the biases in different redshift bins are correlated. This could be relevant for shapes that lie between local and equilateral.

From Figures~\ref{fig:DESI_png} and~\ref{fig:MM_png}, we see that reaching CMB-levels of sensitivity~\cite{Chang:2022lrw} lies somewhere between the capabilities of DESI and MegaMapper. The current sensitivity of Planck, $\sigma(\fnleq)=47$ is accessible with DESI with some knowledge of the redshift dependence of the bias parameters and more aggressive values of $\kmax$. MegaMapper shows similar behavior with an overall improvement in sensitivity by a factor of 50. In principle, MegaMapper has the sensitivity to approach the well-motivated target of $\sigma(\fnleq) \approx 1$~\cite{Baumann:2011su,Baumann:2014cja,Baumann:2015nta}.

The forecasts presented here present an optimistic view for the potential of future measurements of $\fnleq$ with next generation surveys. However, these forecasts still depend sensitively on the range of accessible scales and our knowledge of the bias parameter evolution. Figure~\ref{fig:MM_vary_k_max_png} shows how these forecasts change with more conservative choices of $\kmax$. We see almost two orders of magnitude in variation of the forecasts using the same observables in the same survey, only changing the $\kmax$ and the priors on redshift evolution of bias parameters. The practical implication is that, even assuming all the information encoded in higher-point statistics is used to break degeneracies (i.e.~the field-level Cram\'er-Rao bound), the range in $\fnleq$ in these forecasts is still equivalent to varying the volume of the survey by a factor of $10^4$. This highlights that the path to reaching our most ambitious goals will come from investment in theoretical and computational tools~\cite{Green:2022hhj}, rather than simply deeper and larger surveys.

\section{Conclusions}\label{sec:conclusions}

The search for primordial non-Gaussianity is one of the most theoretically compelling experimental endeavors in fundamental physics~\cite{Flauger:2022hie}. Unfortunately, the long-term potential for this program is limited as much by our ability to model the Universe~\cite{Green:2022hhj} as it is to design and build the next generation of surveys~\cite{Chang:2022lrw}. Our goal in this paper is to better understand the potential and limitations of the data the surveys will collect, and where the theoretical bottlenecks lie. Using the field-level likelihood, we determined the Cram\'er-Rao bound for the amplitude of both local and equilateral non-Gaussianity and expressed the result in terms of the correlators that dominate the result.

The long-term potential for these surveys is qualitatively different for equilateral and local non-Gaussianity, which are good proxies for generic (quasi-~\cite{Chen:2009zp}) single-field~\cite{Maldacena:2002vr,Chen:2006nt,Cheung:2007st} and multi-field~\cite{Lyth:2002my,Bernardeau:2002jy,Zaldarriaga:2003my} inflationary signals respectively. For local non-Gaussianity, we showed that the field-level Cram\'er-Rao bound is given by the matter bispectrum when we include modes with wavelengths as small to the size of the relevant halos. Fortunately, a multi-tracer analysis of the halo power spectrum alone is sufficiently close to this bound using only the long wavelength modes, that the long term potential for improved measurements of $\fnlloc$ looks quite robust.

The potential for equilateral non-Gaussianity, even using field-level inference, remains significantly more ambiguous. While field-level information reduces the sensitivity to biasing down to a small number of parameters, forecasts still vary by one- to two-orders of magnitude depending on the precise assumptions of the redshift dependence and range of accessible scales. A future MegaMapper survey has the statistical power to reach some of our most important theoretical thresholds, but this will rely on theoretical inputs for the bias parameters along with robust modeling well into the quasi-linear regime. 

More optimistically, our forecasts are organized to highlight the most critical opportunities for theory and simulations to improve our understanding of the inflationary era. Our level of knowledge of the bias parameters plays a large role in the potential sensitivity, far more than improved modeling of the redshift space distortions.~\cite{Sullivan:2024jxe} This further motivates investigations into simulation based priors~\cite{Kobayashi:2021oud,Ivanov:2025qie} or data driven techniques~\cite{Dalal:2025eve,Sullivan:2025fie} as opportunities for long term improvements in sensitivity.

\paragraph{Acknowledgments}
We are grateful to Daniel Baumann, Yun-Ting Cheng, Kyle Cranmer, Neal Dalal, Roland de Putter, Olivier Dor\'e, Jiashu Han, Austin Joyce, Tongyan Lin, Misha Ivanov,  Junyang Lu, Moritz Muenchmeyer, Raman Sundrum, Erwin Tanin, and Ben Wallisch for helpful discussions.
DG and EC are supported by the US~Department of Energy under grant~\mbox{DE-SC0009919}. DG thanks the KITP for hospitality while this work was being completed, and was supported in part by the National Science Foundation under PHY- 2309135. VL is supported by the Network for Neutrinos, Nuclear Astrophysics and Symmetries (N3AS) through the National Science Foundation Physics Frontier Center, Grant No. PHY-2020275.

\newpage
\appendix

\section{Marginal likelihood}\label{app:Marginal likelihood}

\subsection{Setup of the likelihood function}
The marginal likelihood function is 
\begin{equation}
   \mathcal{P}[\delta_g^{\rm obs}|b_i] = \int \mathcal{D}\dL \exp\left(-\int_k \left( \frac{|\dL(\k)|^2}{2\PL(k)}+\frac{|\delta_g(\k)-\delta^{\rm obs}_g(\k)|^2}{2N^2}\right)\right) \ .
\end{equation}
To compute this likelihood function, we need to first define the integrand as the joint likelihood $\mathcal{P}[\dL, b_i, \delta_g^{\rm obs}]$, find the saddle point $\hdL$ of the joint likelihood, and then integrate over $\dL$ around the saddle point.

\subsection{Saddle point}
The first derivative of the logarithm of joint likelihood is the following :
\begin{equation}
\begin{aligned}
     \frac{\partial}{\partial \dL (-k)} \ln \mathcal{P}
    =&
    -\frac{\dL(\k)}{\PL(k)} - \int_{k'}\frac{[\delta_g(\k')-\delta_g^{\rm obs}(\k')]}{N^2}\Big[b_{\rm lin}\delta_D(\text{-}\k'\text{+}\k)\\
    &+2\int_{k_1}[b_{\rm quad}(\k_1,\text{-} \k_1\text{-}\k')+b_{\rm lin}F_2(\k_1,\text{-}\k_1\text{-}\k')]\dL(\k_1)\delta_D(\k\text{-}\k_1\text{+}\k')  
    +\ldots\Big] \nonumber
\end{aligned}
\end{equation}
The maximum likelihood $\dL$, $\hdL$, corresponds to the density field where the first derivative of the joint likelihood is zero,
\begin{align}
    \frac{\partial}{\partial \dL(-\k)}\ln \mathcal{P}[\dL,\delta_g^{\rm obs},b_i]\Big|_{\dL  = \hdL} = 0
    \label{first derivative} \ .
\end{align}
Given that the nonlinear density field can be expanding in powers of the linear density, it similarly makes sense to expand the saddle point as 
\beq
\hdL = \hdLa+\hdLb+\hdLc+\ldots\ ,
\eeq
where the superscript indicates the order of $\delta_g^{\rm obs}$ of that term, $\hdL^{(n)} = {\cal O}(\delta_g^{\rm obs})^n$.
We can solve for the maximum likelihood field perturbatively, 
\begin{equation}
\begin{aligned}
     &\hdLa(\k) = \frac{b_{\rm lin}\PL(k)}{\PL b_{\rm lin}^2+N^2}\delta_g^{\rm obs}(\k)\\
    &\hdLb(\k) = -\int_{k_1}\frac{b_{\rm lin}\PL(k)b_{\rm quad}(\k_1,\k\text{-}\k_1)}{\PL(k)b_{\rm lin}^2+N^2} \hdLa(\k_1) \hdLa(\k\text{-}\k_1)\\
    &\hdLc(\k) = -\int_{k_1,k_2}b_{\rm cub}(\k_1,\k_2,\k\text{-}\k_1\text{-}\k_2)\frac{b_{\rm lin}(k)\PL(k)}{\PL b_{\rm lin}^2+N^2}\hdLa(\k_1)\hdLa(\k_2) \hdLa(\k\text{-}\k_1\text{-}\k_2)\\
    &+2\int_{k_1,k_2}b_{\rm quad}(\k_1,\k\text{-}\k_1)b_{\rm quad}(\k_2,\k\text{-}\k_1\text{-}\k_2)\\
    &\times \frac{b_{\rm lin}(k)\PL(k)}{\PL b_{\rm lin}^2+N^2}\frac{b_{\rm lin}(k\text{-}k_1)\PL(k\text{-}k_1)}{\PL b_{\rm lin}^2+N^2} \hdLa(\k_1)\hdLa(\k_2) \hdLa(\k\text{-}\k_1\text{-}\k_2)
    \end{aligned}
    \end{equation}

\subsection{Joint likelihood at the saddle point}
We can express the logarithm of the joint likelihood at the saddle point as follows.
\begin{equation}
   \begin{aligned}
&   \ln \mathcal{P}[\dL,\delta_g^{\rm obs}, b_i]\\
   & = R_0-\int_{k,k'}\delta_D(\k+\k')R_2(k)\dL'(\k)\dL'(\k')+\int_{k,k'}\Delta R_2(\k,\k')\dL'(\k)\dL'(\k')+\ldots
   \end{aligned}
\end{equation}
where $R_0\equiv \ln \mathcal{R}[\hat\L,\delta_g^{\rm obs},b_i]$, $R_2(k) \equiv \frac{\PL(k)b_{\rm lin}^2+N^2}{2\PL(k)N^2}$, and 
$$\Delta R_2(\k,\k') \equiv \frac{1}{2}\frac{\partial^2}{\partial \dL(\k)\partial \dL(\k')}\ln \mathcal{P}[\dL,\delta_g^{\rm obs},b_i]\Big|_{\dL = \hat \dL}+\delta_D(\k+\k')R_2(k)$$.

\subsection{Computation of the marginal likelihood}
At the saddle point, the path integral for the marginal likelihood is easier to compute.
\begin{align}
    \mathcal{P}[\delta_g^{\rm obs}|b_i] = \int \mathcal{D}\delta_L' e^{R_0-\int_k R_2(k)\delta_L'(\k)\delta_L'(-\k)-\int_{k,k'}\Delta R_2(\k,\k')\delta_L'(\k)\delta_L'(\k')}
\end{align}
All higher cumulants correspond to higher order in $\frac{N^2}{\PL b_{\rm lin}^2+N^2}$, so we omit all of them.
Because the $\Delta R_2$ term is not diagonal, we will first Taylor expand $e^{\int \Delta R_2 \delta_L'^2}$, and then do the Gauss integral term by term.
\begin{equation}
\begin{aligned}
    &\mathcal{P}[\delta_g^{\rm obs}|b_i] \\
    &= e^{R_0-\int_k \frac{(2\pi)^3\delta_D(0)}{2}\ln[\PL(k)b_{\rm lin}^2+N^2]}\\
    &\Big\{ 1-\int_k \frac{\Delta R_2(\k,\text{-}\k)\PL(k)N^2}{\PL(k)b_{\rm lin}^2+N^2} -\frac{1}{2}\Big[\int_k \frac{\Delta R_2(\k,\text{-}\k)\PL(k)N^2}{\PL(k)b_{\rm lin}^2+N^2}\Big]^2+\ldots\\
    &+\int_{k,k'}\frac{\Delta R_2(\k,\k')^2 N^4 \PL(k)\PL(k')}{[\PL(k)b_{\rm lin}^2+N^2][\PL(k')b_{\rm lin}^2+N^2]}+\ldots\Big\}
    \label{eq:marginal_likelihood}
\end{aligned}
\end{equation}

\subsection{\texorpdfstring{$R_0$}{R0}}
\label{app:R0}
Because $\delta_g[\hat \delta_L]-\delta_g^{\rm obs} \sim \mathcal{O}(\frac{N^2}{\PL b_{\rm lin}^2+N^2})$, $\int_k\frac{[\delta_g\text{-}\delta_g^{\rm obs}]^2}{2N^2}$ is also of order $\mathcal{O}(\frac{N^2}{\PL b_{\rm lin}^2+N^2})$. When we drop all the low signal to noise terms, the first term in the logarithm of the marginal likelihood $R_0$ becomes:
\begin{equation}
\begin{aligned}
    &R_0 = -\int_{k}\frac{\hat\delta_L(\k)\hat\delta_L(-\k)}{2\PL(k)}
    \\
  &=  -\int_k \frac{b_{\rm lin}^2\PL(k)}{2[\PL b_{\rm lin}^2+N^2]^2}P_g^{\rm obs}(k)(2\pi)^3\delta_D(0)\\
  &-\int_{k,k'}\frac{b_{\rm lin}^2\PL(k)}{2[\PL b_{\rm lin}^2+N^2]^2}b_{\rm quad}(\k',\k-\k')^2
  \\
 & \times 2 \Big[\frac{b_{\rm lin}\PL(k')}{\PL b_{\rm lin}^2+N^2}\Big]^2 \Big[\frac{b_{\rm lin}\PL(|\k\text{-}\k'|)}{\PL b_{\rm lin}^2+N^2}\Big]^2P_{g}^{\rm obs}(k')P_{g}^{\rm obs}(|\k\text{-}\k'|)(2\pi)^3\delta_D(0)
  \\
 & -4\int_{k,k'}\frac{b_{\rm lin}(k)}{2[\PL(k)b_{\rm lin}^2+N^2]}b_{\rm quad}(\k',\k\text{-}\k')b_{\rm quad}(\text{-}\k',\k)\Big[\frac{b_{\rm lin} \PL(k')}{\PL(k')b_{\rm lin}^2+N^2}\Big]^2
  \\
 & \times 2 \Big[\frac{b_{\rm lin} \PL(k')}{\PL(|\k-\k'|)b_{\rm lin}^2+N^2}\Big]^3P_g^{\rm obs}(|\k\text{-}\k'|)P_g^{\rm obs}(|\k\text{-}\k'|)(2\pi)^3\delta_D(0)
  \end{aligned}
\end{equation}

\subsection{Terms with \texorpdfstring{$\Delta R_2$}{Delta R2}}
\label{app:deltaR2}
\begin{equation}
\begin{aligned}
    &\Delta R_2(\k,\k') = \frac{[b_{\rm lin}(k')b_{\rm quad}(\k+\k',-\k)+b_{\rm lin}(k)b_{\rm quad}(\k+\k',-\k')]}{N^2}\hdL(\text{-}\k\text{-}\k')\\
    &+ \frac{2}{N^2}\int_{k_1}b_{\rm quad}(\k_1\text{-}\k,\k)b_{\rm quad}(\text{-}\k_1\text{-}\k',\k')\hdL(\k_1\text{-}\k)\hdL(\text{-}\k_1\text{-}\k') \\
  &  +\frac{3}{2N^2}\int_{k_1}\Big[ b_{\rm lin}(k')b_{\rm cub}(\k_1,\text{-}\k\text{-}\k'\text{-}\k_1,\k) +b_{\rm lin}(k)b_{\rm cub}(\text{-}\k_1,\text{-}\k\text{-}\k'\text{-}\k_1,\k') \Big]\hdL(\k_1)\hdL(\text{-}\k\text{-}\k'\text{-}\k_1)
\end{aligned}
\end{equation}
The first term in the logarithm of the marginal likelihood with $\Delta R_2$ is 
\begin{equation}
    \begin{aligned}
    &-\int_k \frac{\Delta R_2(k,-k)\PL(k)N^2}{\PL(k)b_{\rm lin}^2+N^2}\\
    &=-\int_k \frac{\PL(k)}{\PL(k)b_{\rm lin}^2+N^2}\Big\{ 2\int_{k_1} b_{\rm quad}(k_1\text{-}k,k)^2\Big[\frac{b_{\rm lin}\PL(k_1\text{-}k)}{\PL b_{\rm lin}^2+N^2}\Big]^2P_g^{\rm obs}(k_1\text{-}k)(2\pi)^3\delta_D(0)\Big\}\\
    &+\ldots
   \end{aligned}
\end{equation}

The second term is 
\begin{equation}
\begin{aligned}
    &\int_{k,k'}\frac{\Delta R_2(k,k')^2 N^4 \PL(k)\PL(k')}{[\PL(k)b_{\rm lin}^2+N^2][\PL(k')b_{\rm lin}^2+N^2]}
    \\
    &=\int_{k,k'} \frac{ \PL(k)\PL(k')[b_{\rm lin}(k')b_{\rm quad}(\k+\k',-\k)+b_{\rm lin}(k)b_{\rm quad}(\k+\k',-\k')]^2}{[\PL(k)b_{\rm lin}^2+N^2][\PL(k')b_{\rm lin}^2+N^2]}
    \\
    &\qquad \times \Big[\frac{b_{\rm lin}\PL(|\k+\k'|)}{\PL b_{\rm lin}^2+N^2}\Big]^2 P_g^{\rm obs}(|\k+\k'|)(2\pi)^3\delta_D(0)
\end{aligned}
\end{equation}
Next, we can put everything together.

\subsection{Marginal likelihood \texorpdfstring{$\ln \mathcal{P}[\delta_g^{\rm obs}\,|\,b_i]$}{ }}
 Substituting the results of App.~\ref{app:R0} and App.~\ref{app:deltaR2} into eq.~\ref{eq:marginal_likelihood}, we obtain
\begin{equation}
\begin{aligned}
   & \frac{1}{V_{\rm survey}}\ln \mathcal{P}[\delta_g^{\rm obs}|b_i]\\ 
   & =
  -\int_k \frac{1}{2}\ln[D(k)] -\int_k \frac{b_{\rm lin}^2\PL(k)}{2D^2(k)}P_g^{\rm obs}(k)\\
 &-\int_{k,k'}\frac{b_{\rm lin}^2\PL(k)}{D^2(k)}b_{\rm quad}(\k',\k\text{-}\k')^2
 \Big[\frac{b_{\rm lin}\PL(k')}{D(k')}\Big]^2 \Big[\frac{b_{\rm lin}\PL(|\k\text{-}\k'|)}{D(|\k\text{-}\k'|)}\Big]^2P_{g}^{\rm obs}(k')P_{g}^{\rm obs}(|\k\text{-}\k'|)\\
  & -\int_{k,k'}\frac{ 4b_{\rm quad}(\k',\k\text{-}\k')b_{\rm quad}(\text{-}\k',\k)}{\PL(k)}\Big[\frac{b_{\rm lin} \PL(k)}{D(k)}\Big]^3\Big[\frac{b_{\rm lin} \PL(k')}{D(k')}\Big]^2 \frac{b_{\rm lin} \PL(|\k\text{-}\k'|)}{D(|\k\text{-}\k'|)}
  \\
  &\hspace{11cm}
  \times P_g^{\rm obs}(k')P_g^{\rm obs}(|\k\text{-}\k'|)\\
&+\ln \Bigg\{
1  
-\int_k \frac{\PL(k)}{D(k)} \Bigg(2\int_{k_1} b_{\rm quad}(\k_1\text{-}\k,\k)^2\Big[\frac{b_{\rm lin}\PL(|\k_1\text{-}\k|)}{D(|\k_1\text{-}\k|)}\Big]^2P_g^{\rm obs}(|\k_1\text{-}\k|)(2\pi)^3\delta_D(0) 
\Bigg)\\
&    \quad+\int_{k,k'} \frac{ \PL(k)\PL(k')[b_{\rm lin}(k') b_{\rm quad}(\k\text{+}\k',\text{-}\k)+b_{\rm lin}(k)b_{\rm quad}(\k\text{+}\k',\text{-}\k')]^2}{D(k)D(k')}\\
  &  \qquad\times \Big[\frac{b_{\rm lin}\PL(|\k\text{+}\k'|)}{D(|\k\text{+}\k'|)}\Big]^2 P_g^{\rm obs}(|\k+\k'|)(2\pi)^3\delta_D(0)\Bigg\}\frac{1}{(2\pi)^3\delta_D(0)}
    \label{eq:marginalized likelihood}
\end{aligned}
\end{equation}

where $D(k)\equiv \PL(k)b_{\rm lin}^2+N^2$

The first two terms correspond to the Gaussian likelihood of the power spectrum, while the remaining terms arise from the nonlinear bias contributions and encode information from the bispectrum.

\subsection{Marginal Likelihood for Multi-tracer}
For multi-tracers, the marginal likelihood can be expressed as the following
\begin{equation}
    \begin{aligned}
        &\mathcal{P}[\delta^{\rm obs(i)}|b_{\rm lin}^{(i)}]=\int \mathcal{D} \dL \; \exp\Big[ -\int_k \frac{|\dL(\k)|^2}{2\PL(k)}-\int_k\sum_i \frac{|\delta_g^{(i)}(\k)-\delta_g^{\rm obs(i)}(\k)|^2}{2N_i^2}\Big]\\
    \end{aligned}
\end{equation}
if we only consider linear bias for now, $\delta_g^{(i)}= b_{\rm lin}^{(i)}(k)\dL(k)$. This allows us to rearrange the exponent and complete the Gauss integral.
\begin{equation}
    \begin{aligned}
         &\mathcal{P}[\delta^{\rm obs(i)}|b_{\rm lin}^{(i)}] 
         \\
         &= \exp\Bigg\{-\frac{V_{\rm survey}}{2}\int_k \; \Bigg(\ln\Big[\Big( \frac{1}{2\PL(k)}+\sum_i\frac{b_{\rm lin}^{(i)}(k)}{2N_i^2} \Big)\Big(\PL(k)\prod_i N_i^2\Big)\Big]\Bigg)\\
         &\qquad-\int_k\Big( \sum_i\frac{|\delta_g^{\rm obs (i)}(\k)|^2}{2N_i^2}-\sum_{ij}\frac{b_{\rm lin }^{(i)}(k)b_{\rm lin }^{(j)}(k)\PL(k)}{2N_i^2N_j^2[1+\sum_l\frac{b_{\rm lin}^{(l)}(k)^2\PL(k)}{N_l^2}]} \delta_g^{\rm obs(i)}(\k)\delta_g^{\rm obs(j)*}(\k)\Big)\Bigg\}
    \end{aligned}
\end{equation}
Notice that if we let 
\begin{align}
    C_{ij}(k)\equiv b_{\rm lin}^{(i)}b_{\rm lin}^{(j)}\PL(k)+\delta_{ij}N_i^2
\end{align}
\begin{align}
    \det[C(k)] = \prod_i N_i^2 \Big( 1 + \sum_j\frac{b_{\rm lin}^{(j)}(k)^2\PL(k)}{N_j^2}\Big)
\end{align}
and 
\begin{align}
    C^{-1}_{ij}(k) = \frac{\delta_{ij}}{N_i^2}-\frac{b_{\rm lin}^{(i)}b_{\rm lin}^{(j)}\PL(k)}{N_i^2\, N_j^2}\frac{1}{[1+\sum_l\frac{b_{\rm lin}^{(l)}(k)^2\PL(k)}{N_i^2}]}
\end{align}
Hence the likelihood can be rewritten as the following
\begin{equation}
    \begin{aligned}
        -\ln \mathcal{P}[\delta^{\rm obs(i)}|b_{\rm lin}^{(i)}]  = \frac{V_{\rm survey}}{2}\int_k\ln\Big(\det C(k)\Big)+\frac{1}{2}\int_k C^{-1}_{ij}(k)\delta_g^{\rm obs(i)}(k)\delta_g^{\rm obs(j)*}(k)
    \end{aligned}
    \label{multi-tracer likelihood}
\end{equation}

\section{Fisher Matrix}\label{app:Fisher matrix}
\subsection{The leading order \texorpdfstring{$F_{\rm quad,quad} $}{F-quad,quad}}
\label{app:leading_order_F_quad_quad}
In section \ref{CR bound at the field level}, we've seen two sections in the Fisher matrix. Here we will present the quadratic-quadratic section. 
\begin{equation}
\begin{aligned}
     &\frac{1}{V_{\rm survey}}F_{\rm quad,quad} 
     \\
     &= \frac{\text{-}1}{V_{\rm survey}}\frac{\partial^2}{\partial b_{\rm quad}^2}   \ln \mathcal{P}[\delta_g^{\rm obs}|b_i]\\
     &=2\int_{k,k'}\frac{b_{\rm lin}^2\PL(k)}{D^2(k)}K_2(\k',\k-\k')^2 \Big[\frac{b_{\rm lin}\PL(k')}{D(k')}\Big]^2 \Big[\frac{b_{\rm lin}\PL(|\k\text{-}\k'|)}{D(|\k\text{-}\k'|)}\Big]^2
     \\
     &\hspace{8cm}
     \times P_{g}^{\rm obs}(k')P_{g}^{\rm obs}(|\k\text{-}\k'|)\\
      & +\int_{k,k'}\frac{8K_2(\k',\k\text{-}\k')K_2(\text{-}\k',\k)}{\PL(k)}\Big[\frac{b_{\rm lin}\PL(k)}{D(k)}\Big]^3\Big[\frac{b_{\rm lin} \PL(k')}{D(k')}\Big]^2\frac{b_{\rm lin} \PL(|\k\text{-}\k'|)}{D(|\k\text{-}\k'|)}
      \\
      &\hspace{8cm}
      \times P_g^{\rm obs}(k')P_g^{\rm obs}(|\k\text{-}\k'|)\\
       &+\int_{k,k'} \frac{4\PL(k)}{D(k)}  K_2(\k'\text{-}\k,\k)^2\Big[\frac{b_{\rm lin}\PL(|\k\text{-}\k'|)}{D(|\k\text{-}\k'|)}\Big]^2P_g^{\rm obs}(|\k\text{-}\k'|)\\
       &-\int_{k,k'} \frac{ 2\PL(k)\PL(k')[b_{\rm lin}(k') K_2(\k'\text{-}\k,\k)+b_{\rm lin}(k)K_2(\k\text{-}\k',\k')]^2}{D(k)D(k')} \\
       &\hspace{8cm}
       \times\Big[\frac{b_{\rm lin}\PL(|\k\text{-}\k'|)}{D(|\k\text{-}\k'|)}\Big]^2 P_g^{\rm obs}(|\k\text{-}\k'|)
       \\[3pt]
    &\approx 2\int_{k,k'}\frac{K_2(\k',\k)^2 [\PL(k)b_{\rm lin}]^2[\PL(k')b_{\rm lin}]^2}{[\PL(k) b_{\rm lin}^2+N^2][\PL(k') b_{\rm lin}^2+N^2][\PL(|\k-\k'|) b_{\rm lin}^2+N^2]}\\
      & +4\int_{k,k'}\frac{K_2(\k-\k',\k')K_2(\k'-\k,\k)[\PL(k)b_{\rm lin}][\PL(k')b_{\rm lin}][\PL(|\k-\k'|)b_{\rm lin}]^2}{[\PL(k) b_{\rm lin}^2+N^2][\PL(k') b_{\rm lin}^2+N^2][\PL(|\k-\k'|) b_{\rm lin}^2+N^2]}
\end{aligned}
\end{equation}
This is identical to the conventional assumption of Gaussian distributed bispetrum,
\begin{align}
    &\ln \mathcal{P}\ni V_{\rm survey} \int_{k,k'} \frac{[B_g(k,k',|\k+\k'|)-B_g^{\rm obs}(k,k',|\k+\k'|)]^2}{12[\PL(k) b_{\rm lin}^2+N^2][\PL(k') b_{\rm lin}^2+N^2][\PL(k-k') b_{\rm lin}^2+N^2]}
\end{align}
\begin{equation}
    \begin{aligned}
        F_{\rm quad\;quad} =  V_{\rm survey} \int_{k,k'} \frac{\frac{\partial}{\partial b_{\rm quad}}B_g(k,k',|\k+\k'|)\frac{\partial}{\partial b_{\rm quad}}B_g(k,k',|\k+\k'|)}{6[\PL(k) b_{\rm lin}^2+N^2][\PL(k') b_{\rm lin}^2+N^2][\PL(k-k') b_{\rm lin}^2+N^2]}
    \end{aligned}
    \label{quadratic Fisher}
\end{equation}
\subsection{Fisher matrix at order \texorpdfstring{$\mathcal{O}(F_2^2)$}{O(F2 sqaured)}: the \texorpdfstring{$F_{\rm lin, lin}$}{F lin,lin} example}
\label{app:loop_term}
In typical large-scale structure applications, the nonlinear contribution to the galaxy density is expected to be subdominant,
$$
\int_{k'}b_{\rm quad}(\k',\k-\k')\dL(\k')\dL(\k-\k')\ll b_{\rm lin}(k)\dL(\k)
$$
This hierarchy does not need to be imposed explicitly when deriving the marginal likelihood, but it provides a useful ordering when organizing the Fisher matrix expansion. 
In particular, when derivatives with respect to the bias parameters act on the nonlinear bias coefficients, terms involving the quadratic kernels $K_2$ or $F_2$ can contribute at  the same order as the leading terms. That is the ordering adopted in the main text and in App.~\ref{app:leading_order_F_quad_quad}.

In the following, we compute the Fisher matrix including the subdominant  contributions and retain all terms up to order $\mathcal{O}(F_2^2)$.

The explicit expansion is lengthy, since derivatives acting on the nonlinear term generate many contributions. We display a few representative terms below,

\begin{equation}
\begin{aligned}
\label{eq:full_lin_lin_sec}
    &\frac{1}{V_{\rm survey}}F_{\rm lin,lin}
    \\
    &= \frac{\text{-}1}{V_{\rm survey}}\frac{\partial^2}{\partial b_{\rm lin}^2}   \ln \mathcal{P}[\delta_g^{\rm obs}|b_i]
    \\
    &= -\int_k\frac{b_{\rm lin}^2\PL(k)^2}{D^2(k)}K_1(k)^2
   \\
   &+3\int_k \frac{b_{\rm lin}^2\PL(k)^2}{D^3(k)}P_g^{\rm obs}(k)K_1(k)^2
    \\
    &+12\int_{k,k'}\frac{b_{\rm lin}^4 \PL(k)F_2(\k',\k\text{-}\k')^2}{D^2(k)} \frac{b_{\rm lin}^4\PL^4(k')}{D^4(k')}\Big[\frac{b_{\rm lin}\PL(|\k\text{-}\k'|)}{D(|\k\text{-}\k'|)}\Big]^2 
    \\
    & \hspace{8cm}
    \times P_g^{\rm obs}(k')P_g^{\rm obs}(|\k\text{-}\k'|) K_1(k')^2
    \\
   & +8\int_{k,k'}\frac{b_{\rm lin}^4 \PL(k)F_2(\k',\k\text{-}\k')^2}{D^2(k)} \frac{b_{\rm lin}^3\PL^3(k')}{D^3(k')}\frac{b_{\rm lin}^3\PL^3(|\k\text{-}\k'|)}{D^3(|\k\text{-}\k'|)} 
   \\
 & \hspace{8cm}
  \times P_g^{\rm obs}(k')P_g^{\rm obs}(|\k\text{-}\k'|) K_1(k')K_1(|\k\text{-}\k'|)
    \\
    &+\int_{k,k'}\frac{24b_{\rm lin}(k)^2F_2(\k',\k\text{-}\k')F_2(\k\text{-}\k',\k)}{D(k)}\frac{b_{\rm lin} ^4 \PL(k')^4K_1(k')^2}{D^4(k')}\frac{b_{\rm lin}^4 \PL(|\k\text{-}\k'|)^3}{D^3(|\k\text{-}\k'|)}
    \\
     & \hspace{8cm}
    \times P_g^{\rm obs}(k')P_g^{\rm obs}(|\k\text{-}\k'|)
    \\
   & +\int_{k,k'}\frac{24b_{\rm lin}(k)^2F_2(\k',\k\text{-}\k')F_2(\k\text{-}\k',\k)}{D(k)}\frac{b_{\rm lin}^6 \PL(|\k\text{-}\k'|)^5}{D^5(|\k\text{-}\k'|)}\frac{b_{\rm lin} ^2 \PL(k')^2}{D^2(k')}
   \\
   & \hspace{8cm}
  \times  P_g^{\rm obs}(k')P_g^{\rm obs}(|\k\text{-}\k'|)K_1(|\k\text{-}\k'|)^2
   \\
   &+\ldots
\end{aligned}
\end{equation}

The full expression contains 94 contributions at order $F_2^2$. The sum of the coefficients matches those appearing in the compact expression below.
\begin{equation}
\begin{aligned}
         \ln \mathcal{P} = -\int_k\frac{[P^{\rm 1-loop}(k)-P^{\rm obs}(k)]^2}{4[P(k)+N^2]^2}-\int_{k,k'}\frac{[B^{\rm tree}(k,k')-B^{\rm obs}(k,k')]^2}{12[P(k)+N^2][P(k')+N^2][P(k+k')+N^2]}\\
     -\int_{k,k'}\frac{[T^{\rm tree}(k,k',\text{-}k')-T^{\rm obs}(k,k',\text{-}k')][P^{\rm 1-loop}(k)-P^{\rm obs}(k)]}{6[P+N^2]^3}
\end{aligned}
\end{equation}
The power spectrum includes the one-loop correction, while the bispectrum and trispectrum appear at tree level. Note that the power spectra appearing in the denominators are the tree-level spectra, i.e. they do not include the loop corrections.

\section{Derivation for Fisher Information Expressions for the different surveys}\label{app:fisher_information_derivation}

The purpose of this work is to compare the Fisher information on $f_{\mathrm{NL}}$ for different analyses. In particular, $f_{\mathrm{NL}}$ can be either inferred directly from the matter field $\delta(\vec{x})$, or indirectly from the galaxy / halo field $\delta_g(\vec{x})$. In this appendix, we derive the expressions for the Fisher information in each analysis. We will denote the \textit{observed} galaxy field as $\delta_g(\vec{x})$ without the subscript ``obs" in this appendix.

\subsection{analysis on \texorpdfstring{$\delta(\vec{x})$}{delta(x)}}

We first assume that we have a map of measured $\delta(\vec{x})$, and we try to infer the value of $f_{\mathrm{NL}}$ from it. The likelihood for a given realization of $\delta(\vec{x})$ is
\begin{align}\label{eqn:likelihood}
	\mathcal{L} = \frac{1}{Z}\exp\Bigg\{ -\frac{1}{2}\int_k \frac{\delta(\vec{k})\delta(-\vec{k})}{P(k)} +\frac{f_{\mathrm{NL}}}{6}\int_{k_1,k_2,k_3}B_{t}(k_1,k_2,k_3)\frac{\delta(\vec{k}_1)\delta(\vec{k}_2)\delta(\vec{k}_3)}{P(k_1)P(k_2)P(k_3)}\Bigg\} \, ,
\end{align}
where $B_t \equiv \Bt$ is introduced as shorthand, and the normalization constant is defined as
\begin{equation}\label{eqn:Z_def}
	Z \equiv \int\mathcal{D}\delta \exp \left\{ -\frac{1}{2}\int_{k}\frac{\delta(\vec{k})\delta(-\vec{k})}{P(k)}+\frac{f_{\mathrm{NL}}}{6}\int_{k_1,k_2,k_3}B_{t}(k_1,k_2,k_3)\frac{\delta(\vec{k}_1)\delta(\vec{k}_2)\delta(\vec{k}_3)}{P(k_1)P(k_2)P(k_3)}\right\} 
	\, ,
\end{equation}
such that $\int \mathcal{D}\delta\,\mathcal{L}=1$. We can expand Eq.~\eqref{eqn:Z_def} for small $f_{\mathrm{NL}}$ and perform the path integral over $\delta$ at each order. The leading order contribution is
\begin{equation}\label{eqn:Z_evaluated}
    Z = Z_0\left\{1 + \frac{f_{\mathrm{NL}}^2}{12}V_{\rm survey}\int_{k_1,k_2,k_3}\frac{B^2_{t}(k_1,k_2,k_3)}{P(k_1)P(k_2)P(k_3)}+ \mathcal{O}(f_{\mathrm{NL}}^3)\right\} \, ,
\end{equation}
where we defined $Z_0\equiv Z|_{f_{\mathrm{NL}}\to 0}$ to be the Gaussian part of the partition function, and used the fact that the path integral over $\delta$ evalutes to zero where there is an odd number of $\delta$ in the integral. Here $V$ is the survey volume as defined in Table~\ref{tab:conventions}. 
This allow us to use Eq.~\eqref{eqn:likelihood} and Eq.~\eqref{eqn:Z_evaluated} to write the log-likelihood as
\begin{align}\label{eqn:log-likelihood}
	\log\mathcal{L} =  &-\frac{1}{2}\int_k\frac{\delta(\vec{k})\delta(-\vec{k})}{P(k)}+\frac{f_{\mathrm{NL}}}{6}\int_{k_1,k_2,k_3}B_{t}(k_1,k_2,k_3)\frac{\delta(\vec{k}_1)\delta(\vec{k}_2)\delta(\vec{k}_3)}{P(k_1)P(k_2)P(k_3)} \nonumber \\
	&-\frac{f_{\mathrm{NL}}^2}{12}V_{\rm survey}\int_{k_1,k_2,k_3}\frac{B^2_{t}(k_1,k_2,k_3)}{P(k_1)P(k_2)P(k_3)} + \mathcal{O}(f_{\mathrm{NL}}^3)  \, .
\end{align}
The optimal estimator, given by solving
\begin{equation}\label{eqn:optimal_estimator_def}
	\frac{d}{df_{\mathrm{NL}}}\log\mathcal{L}\Big|_{f_{\mathrm{NL}}=\hat{f}_{\mathrm{NL}}} = 0  \, .
\end{equation}
can be computed by differentiating Eq.~\eqref{eqn:log-likelihood}
\begin{align}\label{eqn:optimal_estimator}
	\hat{f}_{\mathrm{NL}} = \frac{\int_{k_1,k_2,k_3}B_t(k_1,k_2,k_3)\frac{\delta(\vec{k}_1)\delta(\vec{k}_2)\delta(\vec{k}_3)}{P(k_1)P(k_2)P(k_3)} }{V_{\rm survey}\int_{k_1,k_2,k_3}\frac{B^2_t(k_1,k_2,k_3)}{P(k_1)P(k_2)P(k_3)}  } \, .
\end{align}
The Fisher information in Eq.~\eqref{eqn:sigma_fNL} can be similarly derived
\begin{align}\label{eqn:SNR_formula}
	\left(\frac{S}{N}\right)^2 &= V_{\rm survey}\int_{k_1,k_2,k_3} \frac{B^2_t(k_1,k_2,k_3)}{6P(k_1)P(k_2)P(k_3)}  \nonumber \\
	&= V_{\rm survey}\int_{k_1,k_2,k_3} \frac{B^2_{\Phi,t}(k_1,k_2,k_3)}{6P_{\Phi}(k_1)P_{\Phi}(k_2)P_{\Phi}(k_3)}  \, .
\end{align}
This expression can also be derived from the marginal likelihood in Eq.~\eqref{quadratic Fisher}. We see that all dependence on the transfer functions in Eq.~\eqref{eqn:SNR_formula} drop out, which is as expected. To compute these momentum integrals, we employ the following trick (see App.~C of Ref.~\cite{Green:2022bre})
\begin{align}\label{eqn:k_integral_trick}
	&\int_{k_1,k_2,k_3}f(k_1,k_2,k_3)  \nonumber \\
    &= \frac{1}{8\pi^4}\int_{\Delta} dk_1dk_2dk_3\,(k_1k_2k_3)f(k_1,k_2,k_3) \nonumber \\
    &= \frac{3}{4\pi^4} \int_{k_{\min}}^{k_{\max}}dk_1 \int_{k_1/2}^{k_1}dk_2\int_{\max(k_1-k_2,k_{\min})}^{k_2}dk_3\,(k_1k_2k_3)f(k_1,k_2,k_3) \nonumber \\
    &\approx \frac{3}{4\pi^4} \int_{k_{\min}}^{k_{\max}}dk_1 \int_{k_1/2}^{k_1}dk_2\int_{k_1-k_2+k_{\min}}^{k_2}dk_3\,(k_1k_2k_3)f(k_1,k_2,k_3) \, ,
\end{align}
for any function $f(k_1,k_2,k_3)$, where $\int_{\Delta} dk_1dk_2dk_3$ in the first line denotes the triangle inequality, in which the sum of any two momentum magnitudes is required to be greater than the third, and in the second line we impose $k_1 \geq k_2 \geq k_3$ without loss of generality and compensate the phase space with a factor of 6. Here $k_{\min}$ and $k_{\max}$ are the minimum and maximum $k$-modes that an observation has access to. 

With this integral trick, the SNR for local and equilateral non-gaussianity are given by putting Eq.~\eqref{eqn:B_template} and Eq.~\eqref{eqn:Gaussian_power_spectrum} into Eq.~\eqref{eqn:SNR_formula} and Eq.~\eqref{eqn:k_integral_trick}
\begin{align}\label{eqn:loc_triangle_2}
	\left(\frac{S}{N}\right)_{\mathrm{loc}}^2 &\approx \frac{9A_sV_{\rm survey}}{25\pi^2}  \int_{2k_{\min}}^{k_{\max}} dk_1 \int_{k_1/2}^{k_1}dk_2 \int_{k_1-k_2+k_{\min}}^{k_2}dk_3 \, \frac{\left(k_1^3+k_2^3+k_3^3\right)^2}{k_1^2k_2^2k_3^2} \nonumber \\
	\left(\frac{S}{N}\right)_{\mathrm{eq}}^2 &\approx \frac{18225 V_{\rm survey}}{16\pi^{10}A_s}\int_{2k_{\min}}^{k_{\max}} dk_1 \int_{k_1/2}^{k_1}dk_2 \int_{k_1-k_2+k_{\min}}^{k_2}dk_3 \,  \frac{k_1^2k_2^2k_3^2}{(k_1+k_2+k_3)^6} \, .
\end{align}
These integrals can be performed analytically. Extracting only the leading order piece as $k_{\min}\to 0$, we find
\begin{align}\label{eqn:loc_triangle_3}
	\left(\frac{S}{N}\right)_{\mathrm{loc}}^2 &\approx \frac{12A_sV_{\rm survey}k_{\max}^3}{25\pi^2} \left[\log\left(\frac{k_{\max}}{k_{\min}}\right) + \mathcal{O}(1)\right] \nonumber \\
	\left(\frac{S}{N}\right)_{\mathrm{eq}}^2 &\approx \frac{189V_{\rm survey}k_{\max}^3}{2048\pi^{10}A_s}\left[1 + \mathcal{O}\left(\frac{k_{\min}}{k_{\max}}\right)\right] \, , 
\end{align}
where the numerical factor in the second line of Eq.~\eqref{eqn:loc_triangle_3} evaluates to $\approx 10^{-6}$.

\subsection{analysis on \texorpdfstring{$\delta_g(\vec{x})$}{delta g (x)}, single tracer}

We now move on to computing the Fisher information on $f_{\mathrm{NL}}$ from a single-tracer galaxy survey using the galaxy power spectrum. The galaxy power spectrum, $P_g(k)$, is defined by
\begin{equation}\label{eqn:halo_power_spectrum}
	\langle \delta_g(\vec{k})\delta_g(\vec{k}')\rangle = P_g(k)\,(2\pi)^3\delta^3_D(\vec{k}+\vec{k}') \, .
\end{equation}
We can relate the galaxy power spectrum with the matter power spectrum using the bias expansion in Eq.~\eqref{eqn:bias_expansion}, which is given in Eq.~\eqref{eqn:galaxy_power_spectrum}. Under the Gaussian assumption each Fourier mode $\delta_{g}(\mathbf k)$ is drawn from a zero‐mean Gaussian with variance $P_g(k)$.  In the continuum limit the joint likelihood is
\begin{equation}\label{eqn:galaxy_likelihood}
	\mathcal L[\delta_{g}]=\exp\!\left\{-\,\frac{1}{2}\int\frac{d^3\vec{k}}{(2\pi)^{3}}\left[
	\frac{\left\lvert\delta_{g}(\vec{k})\right\rvert^{2}}{P_{g}(k)}+V_{\rm survey}\ln\!\bigl(2\pi\,P_{g}(k)\bigr)\right]\right\} \, ,
\end{equation}
where the second term comes from the normalization constant. The Fisher information on $f_{\mathrm{NL}}$ in the galaxy measurement can be computed using the definition of Fisher information in Eq.~\eqref{eqn:sigma_fNL} and
taking the expectation value $\left\lvert\delta_{g}(\vec{k})\right\rvert^{2}=P(k)V_{\mathrm{survey}}$, and find 
\begin{equation}\label{eq:Fisher_galaxy}
	\left(\frac{S}{N}\right)^2 
    =\frac{V_{\rm survey}}{2}\int\frac{d^{3}k}{(2\pi)^{3}}\frac{\bigl[\partial_{f_{\rm NL}}P_{g}(k)\bigr]^{2}}{P_{g}^{2}(k)}\Bigg|_{f_{\rm NL}=0} \, .
\end{equation}
Finally, differentiating Eq.~\eqref{eq:Fisher_galaxy} using Eq.~\eqref{eqn:galaxy_power_spectrum}, we find 
\begin{equation}\label{eq:Ffinal_galaxy}
	F_{f_{\rm NL}f_{\rm NL}}=V_{\rm survey}\int\frac{d^{3}k}{(2\pi)^{3}}\frac{2b_1^2b_{\Phi}^2P^{2}(k)}
	{\mathcal T^{2}(k)\,\bigl[b_{1}^{2}P(k)+N^2\bigr]^{2}}\,,
\end{equation}
where we have taken $\Delta=0$ in the scale-dependent bias. In the $P\gg N^2$ limit, the Fisher information further simplifies
\begin{equation}\label{eq:Ffinal_galaxy_high_P}
	F_{f_{\rm NL}f_{\rm NL}} \approx V_{\rm survey}\left(\frac{2b_{\Phi}^2}{b_1^2}\right)\int\frac{d^{3}k}{(2\pi)^{3}}\frac{1}
	{\mathcal T^{2}(k)} \, .
\end{equation}
This is the expression quoted in Eq.~\eqref{eqn:SNR_expression_quote}. 

\subsection{analysis on \texorpdfstring{$\delta_g(\vec{x})$}{delta g(x)}, double tracer}

Finally, we compute the Fisher information on $f_{\mathrm{NL}}$ from a double-tracer galaxy survey using their respective power spectrum. We will focus on two subset of galaxies in this section. The bias expansion for each tracer $i=1,2$ is given by Eq.~\eqref{eqn:bias_expansion_multi}, which we repeat here for clarity
\begin{equation}\label{eqn:bias_expansion_multi_1}
	\delta_{g}^{(i)}(\vec{k})=\left[b_{1}^{(i)} + f_{\rm NL}\,\frac{b_{\Phi}^{(i)}}{\mathcal T(k)}\,(kR_*)^{\Delta}\right]\delta(\vec{k})+ \epsilon^{(i)}(\vec{k}),
\end{equation}
where the stochastic bias for each galaxy satisfies $\langle \epsilon^{(i)}(\vec{k})\epsilon^{(j)*}(\vec{k}')\rangle = (2\pi)^{3}\,\delta_{D}(\vec{k}+\vec{k}')\,\delta_{ij}N_i^2$.
The covariance matrix is simiarly given by Eq.~\eqref{eqn:covariance_matrix}
\begin{align}
	C_{ij}(k)&\equiv\langle \delta_{g}^{(i)}(\vec{k})\,\delta_{g}^{(j)*}(\vec{k})\rangle' \nonumber \\
	&=\left[b_{1}^{(i)} + f_{\rm NL}\,\frac{b_{\Phi}^{(i)}}{\mathcal T(k)}\,(kR_*)^{\Delta}\right]\left[b_{1}^{(j)} + f_{\rm NL}\,\frac{b_{\Phi}^{(j)}}{\mathcal T(k)}\,(kR_*)^{\Delta}\right]P(k)+ \delta_{ij}\,N_i^2 \, .
\end{align}
Treating $\vec{\delta}_g=(\delta_{g}^{(1)},\delta_{g}^{(2)})^T$ as a zero-mean Gaussian with covariance $C(k)$, the log-likelihood is given by
\begin{equation}\label{eqn:log-likelihood_galaxy_multi}
	\ln\mathcal L= -\frac{1}{2}\int \frac{d^3k}{(2\pi)^3}
	\left[V_{\rm survey}\,\ln\det C(k) +\vec{\delta}_g^{T}(\vec{k})\,C^{-1}(k)\,\vec{\delta}_g(\vec{k})\right] \,.
\end{equation}
Differentiating Eq.~\eqref{eqn:log-likelihood_galaxy_multi} with respect to $f_{\rm NL}$ gives (denoting $\partial\equiv \partial_{\mathrm{NL}}$)
\begin{equation}\label{eqn:d1L_multi}
	\frac{\partial\ln\mathcal L}{\partial f_{\rm NL}}= -\frac{1}{2}\int \frac{d^3\vec{k}}{(2\pi)^3}
	\left[V_{\rm survey}\,\mathrm{Tr}\left(C^{-1}(k)\,\partial C(k)\right)- \vec{\delta}_g^{T}(\vec{k})\,C^{-1}(k)\,\partial C(k)\,C^{-1}(k)\,\vec{\delta}_g(\vec{k})\right] \,,
\end{equation}
where we used $\partial\,C^{-1}=-C^{-1}\,\partial C\,C^{-1}$. Taking a second derivative gives
\begin{align}\label{eqn:second_derivative_multi}
	\frac{\partial^2 \ln \mathcal{L}}{\partial f_{\rm NL}^2}
	&= -\frac{1}{2} \int \frac{d^3\vec{k}}{(2\pi)^3}
	\Bigg\{
	V_{\rm survey}\,\mathrm{Tr}\left(
	- C^{-1}(k)\,\partial C(k)\,C^{-1}(k)\,\partial C(k)
	+ C^{-1}(k)\,\partial^2 C(k)
	\right) \nonumber \\
	&\quad\;-\;\vec{\delta}_g^{T}(\vec{k})
	\left[C^{-1}(k)\,\partial^2C(k)\,C^{-1}(k) - 2C^{-1}(k)\,\partial C(k)C^{-1}(k)\,\partial C(k)C^{-1}(k)\right]\vec{\delta}_g(\vec{k})\Bigg\}\,.
\end{align}
Setting the ensemble average $\langle \vec{\delta}_g(\vec{k})\,\vec{\delta}_g^T(\vec{k})\rangle = C(k)V_{\mathrm{survey}}$ and hence $\langle \vec{\delta}_g(\vec{k})\,M\,\vec{\delta}_g^T(\vec{k})\rangle = \mathrm{Tr}\left[C(k)M\right]$ for any matrix $M$. We compute the Fisher information as defined in Eq.~\eqref{eqn:sigma_fNL}
\begin{equation}\label{eqn:Fisher_multi}
	F_{f_{\rm NL}f_{\rm NL}} = \frac{V_{\rm survey}}{2}
	\int \frac{d^3\vec{k}}{(2\pi)^3}
	\mathrm{Tr}\left[
	C^{-1}(k)\,\partial C(k)\,C^{-1}(k)\,\partial C(k)
	\right] \Bigg|_{f_{\rm NL}=0}\,.
\end{equation}
We then explicitly work with the $2\times2$ matrix $C(k)$ in Eq.~\eqref{eqn:covariance_matrix} and evaluation of the trace in Eq.~\eqref{eqn:Fisher_multi} gives
\begin{align}\label{eqn:Fisher_final_multi}
		F_{f_{\rm NL}f_{\rm NL}}
		&= V_{\rm survey}\int \frac{d^3\vec{k}}{(2\pi)^3} \frac{P^2(k)}{\left\{\left[\left(b_1^{(1)}\right)^2N_2^2+\left(b_1^{(2)}\right)^2N_1^2\right]P(k)+N_1^2N_2^2\right\}^2\mathcal{T}^2(k)} \nonumber \\
		&\times \Big\{2\left[\left(b_1^{(1)}b_{\Phi}^{(1)}N_2^2\right)^2+\left(b_1^{(2)}b_{\Phi}^{(2)}N_1^2\right)^2\right]  + \left(b_1^{(1)}b_{\Phi}^{(2)}+b_1^{(2)}b_{\Phi}^{(1)}\right)^2N_1^2N_2^2   \nonumber \\
        &+ \left(b_1^{(1)}b_{\Phi}^{(2)}-b_1^{(2)}b_{\Phi}^{(1)}\right)^2\left(\left(b_1^{(1)}\right)^2N_2^2+\left(b_1^{(2)}\right)^2N_1^2\right)P(k)\Big\} \,,
\end{align}
where we have taken $\Delta=0$. In the limit where the power is much larger than the noise, $P\gg N_i^2$ for each $i$, this expression can be simplified as
\begin{align}\label{eqn:Fisher_final_multi_high_P}
	F_{f_{\rm NL}f_{\rm NL}} 
	= V_{\rm survey}\frac{\left[b_1^{(1)}b_{\Phi}^{(2)}-b_1^{(2)}b_{\Phi}^{(1)}\right]^2}{\left[\left(b_1^{(1)}\right)^2N_2^2+\left(b_1^{(2)}\right)^2N_1^2\right]} \int \frac{d^3\vec{k}}{(2\pi)^3}P_{\Phi}(k) \, ,
\end{align}
which is the expression quoted in Eq.~\eqref{eqn:SNR_expression_quote}.

\section{Details of Halo Model}\label{app:exact_expressions}

\subsection{Lagrangian versus Eulerian Bias}
In standard cosmology, halos form through the gravitational collapse of overdense regions in the early universe. Regions with higher matter density are therefore more likely to host halos, while underdense regions host fewer. The systematic relation between the matter density field and the halo number density, as mentioned in the main text, is known as the bias expansion. The relevant quantity here is the comoving halo number density per mass
\begin{equation}\label{eqn:n_def}
	n_h(M) \equiv \frac{d^2N_h}{dVdM}\, ,
\end{equation}
where $N_h$ is the physical halo number, $V$ is the volume, and $M$ is the halo mass. Here we can define $n_h(M)$ in either the Eulerian or Lagrangian frame, $n_h^E(\vec{x})$ and $n_h^L(\vec{q})$, where $\vec{x}$ is the comoving (Eulerian) position, and $\vec{q}$ is the initial (Lagrangian) position, which is later mapped to the final Eulerian position using the displacement field, $\Psi(\vec{q})$, by
\begin{equation}\label{eqn:Eulerian_vs_Lagrangian}
	\vec{x} = \vec{q} + \vec{\Psi}(\vec{q}) \, .
\end{equation}
For a review on Eulerian and Lagrangian dynamics, see Ref.~\cite{Bernardeau:2001qr}. 

We define the halo density contrast by
\begin{align}\label{eqn:halo_density_contrast_def}
	\delta^E_h(\vec{x}) &\equiv \frac{n_h^E(M,\vec{x}) - \bar{n}_h^E(M)}{\bar{n}_h^E(M)} \nonumber \\
    \delta^L_h(\vec{x}) &\equiv \frac{n_h^L(M,\vec{x}) - \bar{n}_h^L(M)}{\bar{n}_h^L(M)} \, ,
\end{align}
where $M$ is the halo mass, and $\bar{n}_h(M)$ is the cosmic mean, which is the comoving halo number density per mass when there is no over-density. 

The halo bias expansion relates the halo over density, $\delta_h$, with the matter over-density $\delta$. This can again be done in either the Langrangian or the Eulerian frame. Neglecting higher order terms, we write
\begin{align}\label{eqn:halo_bias}
	\delta^E_h(\vec{x}) &= b_1^E \delta(\vec{x}) + f_{\mathrm{NL}}b^E_{\Phi}\Phi(\vec{x}) + \epsilon^E(\vec{x}) \nonumber \\
    \delta^L_h(\vec{q}) &= b_1^L \delta(\vec{q}) + f_{\mathrm{NL}}b^L_{\Phi}\Phi(\vec{q}) + \epsilon^L(\vec{q}) \, ,
\end{align}
where $b_1^L$, $b_{\Phi}^L$ and $b_1^E$, $b_{\Phi}^E$ are the bias parameters in the Lagrangian and Eulerian frame, respectively, and $\Phi$ is the primordial potential. Here $\epsilon^L(\vec{q})$ and $\epsilon^E(\vec{x})$ encode the stochastic noise in the bias expansion, which set the size of measurement noise with their two-point function
\begin{align}\label{eqn:stochastic_bias}
	\langle \epsilon^E(\vec{k})\epsilon^E(\vec{k}')\rangle &= N^2\,(2\pi)^3\delta_D(\vec{k}+\vec{k}') \nonumber \\
    \langle \epsilon^L(\vec{k})\epsilon^L(\vec{k}')\rangle &= N^2\,(2\pi)^3\delta_D(\vec{k}+\vec{k}')\, .
\end{align}
Using conservation of halo number, we can relate the halo overdensities in the two frames by
\begin{equation}\label{eqn:halo_number_conservation}
	1+\delta_h^E(\vec{x}) = \int d^3\vec{q}\left[1+\delta_h^L(\vec{q})\right]\delta_D\left(\vec{x}-\vec{q}-\vec{\Psi}(\vec{q})\right) \, .
\end{equation}
On the other hand, conservation of matter mass between the two frames implies 
\begin{equation}\label{eqn:matter_mass_conservation}
	1+\delta(\vec{x}) = \int d^3\vec{q}\,\delta_D\left(\vec{x}-\vec{q}-\vec{\Psi}(\vec{q})\right) \, .
\end{equation}
Combining Eqs.~\eqref{eqn:halo_number_conservation}-\eqref{eqn:matter_mass_conservation} gives
\begin{equation}\label{eqn:halo_number_conservation_2}
	\delta_h^E(\vec{x}) = \delta(\vec{x})+\int d^3\vec{q}\,\delta_h^L(\vec{q})\delta_D\left(\vec{x}-\vec{q}-\vec{\Psi}(\vec{q})\right) \, .
\end{equation}
Taking Fourier transform on both sides of Eq.~\eqref{eqn:halo_number_conservation_2} and expanding for small displacement field, we find
\begin{equation}\label{eqn:halo_number_conservation_3}
	\delta_h^E(\vec{k}) = \delta(\vec{k})+\delta_h^L(\vec{k}) + \cdots \, ,
\end{equation}
to linear order in $\delta_h^L$ and $\vec{\Psi}$. By comparing Eq.~\eqref{eqn:halo_number_conservation_3} with the bias expansion in Eq.~\eqref{eqn:halo_bias}, we immediately see that the bias parameters $b_1$ in the two frames differ by $+1$, while $b_{\Phi}$ is identical in both frames, \textit{i.e.}
\begin{align}
	b_1^E &= b_1^L + 1 \nonumber \\
    b_{\Phi}^E &= b_{\Phi}^L \, ,
\end{align}
which is the relation that is used in the Sec.~\ref{sec:local} (\textit{cf.} Eq.~\eqref{eqn:b_Lagrangian_vs_Eulerian}). In addition, the bias coefficients can be computed by taking derivatives in the bias expansion in Eq.~\eqref{eqn:halo_bias}
\begin{align}
	b^L_1 &= \frac{\partial \delta_h}{\partial \delta}\Bigg|_{\delta=0,\Phi=0} = \frac{1}{\bar{n}_h^L} \frac{\partial n_h^L}{\partial \delta}\Bigg|_{\delta=0,\Phi=0} = \frac{\partial \log n_h^L}{\partial \delta}\Bigg|_{\delta=0,\Phi=0}\nonumber \\
	f_{\mathrm{NL}}b^L_{\Phi} &= \frac{\partial \delta_h}{\partial \Phi}\Bigg|_{\delta=0,\Phi=0} = \frac{1}{\bar{n}_h^L} \frac{\partial n_h^L}{\partial \Phi}\Bigg|_{\delta=0,\Phi=0}= \frac{\partial \log n_h^L}{\partial \Phi}\Bigg|_{\delta=0,\Phi=0}\, ,
\end{align}
and similarly for the Eulerian frame. Here we note that the bias coefficients depend on $M$. 

\subsection{Exact Formulas for Press-Schechter}

Here we summarize several exact expressions within the Press–Schechter analysis. Recall that we begin with a simple model of the matter power spectrum in the form of a broken power law
\begin{equation}\label{eqn:power_spectrum_ansatz}
    P(k) = \frac{8\pi^2}{25}\frac{A_sD^2(z)}{\Omega_mH_0^4}\times \begin{dcases*}
        k & if $k\lesssim k_{\mathrm{eq}}$ \\
        k_{\mathrm{eq}}^3/k^2 & if $k\gtrsim k_{\mathrm{eq}}$
    \end{dcases*} \, ,
\end{equation}
with $k_{\mathrm{eq}}$ being the pivot scale of the matter power spectrum. Using this form of the power spectrum, we can compute the variance for a halo with size $R$ using the top-hat filter function as in Eq.~\eqref{eqn:sigma_def} and Eq.~\eqref{eqn:top_hat}. It turns out that this integral can be computed exactly, and the result is
\begin{align}\label{eqn:sigma2_exact}
    \sigma^2 = \frac{3y}{40\pi^2x^2}\Bigg\{&-9-10(2-3\gamma_E)x^2+4\pi x^5+(9+2x^2-4x^4)\cos 2x \nonumber \\
    &-2x\left[15x\,\mathrm{Ci}(2x)-15x\log 2x-(9-x^2)\sin 2x+4x^4\,\mathrm{Si}(2x)\right]\Bigg\} \, ,
\end{align}
where Si and Ci are the sine and cosine integral functions, $\gamma_E\approx 0.577$ is the Euler's constant, and we defined dimensionless quantities
\begin{align}\label{eqn:xy_def}
    x &\equiv k_{\mathrm{eq}}R \nonumber \\
    y &\equiv  \frac{8\pi^2}{25}\frac{A_sD^2(z)}{\Omega_mH_0^4}\frac{1}{R^4} \, .
\end{align}
Here $x$ controls the shape of $\sigma^2$ as  a function of $R$, while $y$ sets the amplitude of the fluctuation. Expanding Eq.~\eqref{eqn:sigma2_exact} assuming $x\ll 1$ gives Eq.~\eqref{eqn:sigma_def_small}.

The halo mass function per logarithmic mass in Eq.~\eqref{eqn:Press-Schechter} can be written in the following convenient form
\begin{align}\label{eqn:hmf_convenient}
    f(M) = \frac{1}{\sqrt{18\pi}}\frac{\bar{\rho}}{M}\frac{\delta_c}{\sigma^3}\frac{d\sigma^2}{d\log R}\exp\left(-\frac{\delta_c^2}{2\sigma^2}\right) \, ,
\end{align}
where the differential can be computed exactly by directly differentiating Eq.~\eqref{eqn:sigma2_exact}
\begin{align}\label{eqn:hmf_differential_exact}
    \frac{d\sigma^2}{d\log R} = \frac{3y}{20\pi^2x^2}\Bigg\{&27+5(11-12\gamma_E)x^2-2\pi x^5-(27+x^2-2x^4)\cos 2x \nonumber \\
    &+x\left[60x\,\mathrm{Ci}(2x)-60x\log 2x-(54-x^2)\sin 2x+4x^4\,\mathrm{Si}(2x)\right]\Bigg\} \, .
\end{align}
Expanding Eq.~\eqref{eqn:hmf_differential_exact} for $x\ll 1$ and using Eq.~\eqref{eqn:hmf_convenient}, we can directly derive Eq.~\eqref{eqn:nbar_halo}.

\clearpage
\phantomsection
\addcontentsline{toc}{section}{References}
\small
\bibliographystyle{utphys}
\bibliography{Refs}

\end{document}